\documentclass[12pt]{article}
\usepackage{amsmath,amssymb,amsthm}
\usepackage[colorlinks,citecolor=blue,urlcolor=blue]{hyperref}
\usepackage{algorithm, algpseudocode}
\usepackage{natbib}
\usepackage{graphicx}
\usepackage{multirow}
\usepackage{dsfont}
\usepackage{epstopdf}
\usepackage{booktabs}
\usepackage{textcomp}
\usepackage{ulem}
\usepackage{color}
\usepackage{threeparttable}
\usepackage{comment}
\usepackage{float}
\usepackage{indentfirst}
\usepackage{arydshln}
\usepackage{cleveref}
\usepackage{soul}
\usepackage{scalerel}
\usepackage{subfigure}

\newcommand{\blind}{0}

\newtheorem{assumption}{Assumption}
\newtheorem{theorem}{Theorem}
\newtheorem{proposition}{Proposition}
\newtheorem{lemma}{Lemma}
\newtheorem{remark}{Remark}
\newtheorem{example}{Example}

\DeclareMathOperator*{\argmin}{arg\,min}

\newcommand{\ope}{\mathbb{E}}
\newcommand{\bs}{\boldsymbol}

\newcommand{\prob}{\mathbb{P}}
\newcommand{\ms}{\mathcal{S}}
\newcommand{\plim}{\operatorname{plim}}

\newcommand{\mk}{\mathcal{K}}
\newcommand{\HT}{\textnormal{HT}}
\newcommand{\sumi}{\sum_{i=1}^N}

\newcommand{\cfadj}{\textnormal{cf}}
\newcommand{\oraclecf}{\textnormal{oracle-cf}}
\newcommand{\adj}{\textnormal{adj}}

\newcommand{\oracleadj}{\textnormal{oracle-adj}}
\def\Var{\textnormal{Var}}
\def\cov{\textnormal{Cov}}
\newcommand{\indep}{\perp\!\!\!\perp}
\newcommand{\op}{o_{\mathbb{P}}}
\newcommand{\Op}{O_{\mathbb{P}}}
\newcommand{\BT}{\mathrm{BRE}}
\newcommand{\CR}{\mathrm{CRE}}
\newcommand{\SR}{\mathrm{SRE}}
\newcommand{\MP}{\mathrm{MPE}}
\newcommand{\setk}{\{k\}}

\def\spacingset#1{\renewcommand{\baselinestretch}
	{#1}\small\normalsize} \spacingset{1}

\begin{document}
	
\if1\blind
{
  \title{Conditional cross-fitting for unbiased machine-learning-assisted covariate adjustment in randomized experiments}
  \author{}
    \date{}
  \maketitle
}\fi
	
% \bf Conditional cross-fitting method for design-based inference\\ \ \\ 

\if0\blind
{
	\title{Conditional cross-fitting for unbiased machine-learning-assisted covariate adjustment in randomized experiments}
    \author{Xin Lu$^{1}$, Lei Shi$^{2}$, Hanzhong Liu$^{3}$, and Peng Ding$^{4}$\\ \\
    $^1$ Department of Statistics and Data Science, Washington University in St. Louis\\
    $^2$ Division of Biostatistics, University of California, Berkeley\\
    $^3$ Department of Statistics and Data Science, Tsinghua University\\
    $^4$ Department of Statistics, University of California, Berkeley}
	\date{}
	\maketitle
}\fi

\bigskip
\begin{abstract}
Randomized experiments are the gold standard for estimating the average treatment effect (ATE). While covariate adjustment can reduce the asymptotic variances of the unbiased Horvitz--Thompson estimators for the ATE, it suffers from finite-sample biases due to data reuse in both prediction and estimation. Traditional sample-splitting and cross-fitting methods can address the problem of data reuse and obtain unbiased estimators. However, they require that the data are independently and identically distributed, which is usually violated under the design-based inference framework for randomized experiments. To address this challenge, we propose a novel conditional cross-fitting method, under the design-based inference framework, where potential outcomes and covariates are fixed and the randomization is the sole source of randomness. We propose sample-splitting algorithms for various randomized experiments, including Bernoulli randomized experiments, completely randomized experiments, and stratified randomized experiments. Based on the proposed algorithms, we construct unbiased covariate-adjusted ATE estimators and propose valid inference procedures. Our methods can accommodate flexible machine-learning–assisted covariate adjustments and allow for model misspecification. 

\end{abstract}


\newpage
\spacingset{1.9}

\section{Introduction}
Randomized experiments are widely considered as the gold standard to estimate the average treatment effect (ATE) in various fields such as medicine, economics, and social sciences \citep{imbens2015causal}. Although Horvitz--Thompson estimators are unbiased, their efficiency can be further improved through covariate adjustment  (also known as regression adjustment) using a prediction model for the potential outcomes. \citet{Lin2013Agnostic, bloniarz2016lasso, liu2020regression, lei2021regression, liu2KRandomization2024, lu2024tyranny} considered linear models as the prediction models. \citet{cohen2020no, guo2023generalized, su2023decorrelation,qu2025randomization} considered generalized linear and nonparametric models as the prediction models. Recent advancements have underscored the potential of machine learning methods, e.g., neural networks or random forests, as the prediction models to reduce the asymptotic variance of ATE estimators \citep{wager2016high,chernozhukov2018double,farrell2021deep}. However, a persistent challenge remains in mitigating finite-sample biases, which arise from the reuse of data in prediction and estimation \citep{su2023decorrelation}. This issue is particularly pronounced with high-dimensional covariates \citep{lei2021regression,lu2023debiased,chang2024exact}. 

 Under a super-population framework with independent and identically distributed (i.i.d.) experimental units and treatment assignments, traditional approaches such as sample-splitting and cross-fitting help remove these biases by dividing the sample into independent subsets for prediction and estimation \citep{wager2016high,chernozhukov2018double, zrnic2024cross}. However, in randomized experiments, the i.i.d. assumptions on the data-generating process may not hold. For example, the i.i.d. assumption on experimental units fails when the experimental units are a substantial fraction of the target population \citep{abadie2020sampling}, and the i.i.d. assumption on treatment assignments fails under stratified randomization \citep{imbens2015causal}. Therefore, random splitting does not eliminate the dependence between the two subsamples \citep{su2023decorrelation}.  Developing sample-splitting or cross-fitting methods that are tailored to the dependence structure in randomized experiments remains an open problem.

To address this problem, this paper presents a conditional cross-fitting method. We consider a design-based inference framework, where the potential outcomes and covariates are fixed quantities, and the randomization of treatment assignments is the only reasoned basis for inference \citep{neyman1923,freedman2008regression,abadie2020sampling}. The contributions of this paper are threefold.

First, we introduce conditional sample-splitting algorithms designed to handle various experimental designs, including Bernoulli randomized experiments, completely randomized experiments, and stratified randomized experiments with matched-pairs experiments as special cases. These algorithms leverage the dependence structure induced by randomization. The main idea is to divide the randomized experiment into two conditionally independent ones by utilizing the conditional cross-fitting mechanism and use them separately for prediction and estimation.

Second, building on these algorithms, we propose novel cross-fitted covariate-adjusted ATE estimators and variance estimators to ensure valid and efficient inference. We provide theoretical guarantees for the proposed estimators, demonstrating their unbiasedness, asymptotic normality, inference validity, and efficiency gains under mild regularity conditions. Our theory covers both regression-based and more flexible machine-learning–assisted covariate adjustment methods. Moreover, we validate the practical applicability and robustness of our approach through  simulation studies, showcasing its superior performance compared with existing methods.

Third, we show that, among the sample-splitting algorithms, 
the optimal one (in the sense of achieving minimal asymptotic variance) ensures that the conditional probability of treatment assignment for each unit remains consistent across subsamples. This insight provides a principled criterion for designing efficient sample-splitting algorithms in randomized experiments.

The remainder of this paper is structured as follows: Section~\ref{sec:framework} introduces the framework and notation. Section~\ref{sec:crossfitting-algorithm} details the conditional cross-fitting algorithm, which is designed to derive an unbiased estimator. Section~\ref{sec:asymptotic-property-of-cross-fitted-estimator} discusses the asymptotic properties of the cross-fitted estimator, including its asymptotic limit and efficiency gain, and discusses the optimal 
cross-fitting algorithm. Section~\ref{sec:confidence-intervals} provides confidence intervals for general experimental designs such as Bernoulli randomized experiments, completely randomized experiments, and stratified randomized and matched-pairs experiments. Section \ref{sec:simulation} provides empirical support through simulation studies. The paper concludes with Section~\ref{sec:discussion}.

\vspace{4mm}

\noindent \textbf{Related literature}:  The covariate adjustment can ensure valid inference and enhance precision with
imperfect and potentially biased prediction. This robustness has been highlighted by the prediction-powered inference in the recent literature \citep{angelopoulos2023prediction,zrnic2024cross}.  These works focus on estimating the mean of potential outcomes under a single treatment group, whereas we consider the ATE, which involves means in both treatment and control groups. Moreover, they are based on the superpopulation framework and i.i.d. assumptions.

 \cite{su2023decorrelation} introduced a novel decorrelation procedure designed to remove bias in covariate adjustment for Bernoulli randomized experiments under the design-based inference framework. The decorrelation procedure constructs overlapping random subsets of data that exhibit desirable independence properties. However, the decorrelation procedure faces three challenges. Firstly, the process of creating overlapping subsets is complex and heavily depends on the specific characteristics of the prediction function class. Furthermore, the decorrelation algorithm occasionally discards portions of the sample while estimating the ATE, which can be less efficient in finite samples compared to utilizing the entire dataset. Secondly, this procedure imposes constraints on the selection of prediction functions, excluding many viable options, such as the no-harm calibration function \citep{cohen2020no} and the pooled regression function \citep{negi2021revisiting}. Thirdly, the decorrelation algorithm is tailored for Bernoulli randomized experiments. It is unclear how to extend this algorithm to other experimental designs, such as completely randomized experiments and stratified randomized experiments. We provide a more detailed discussion of the decorrelation method in the Supplementary Material. Our conditional cross-fitting method fills in all these gaps.

Another closely related approach is the leave-one-out method: the prediction function of one unit is estimated using the rest of the data. \citet{wu2018loop,wu2021design} studied this method under the design-based inference framework. However, to make this method work, we need the correlation between the fitted predictions at different units to decay sufficiently fast, a condition that may be violated in practice. \cite{chiang2023regression} used a leave-one-out method to reduce the bias of the linear regression-adjusted estimator when there are many covariates. However, their method does not ensure the unbiasedness of the produced estimator. %\liu{Xin: please cite and add some discussion on \cite{chiang2023regression}.}

% can be very restrictive

 Another line of work addresses bias not by creating independent subsamples but by subtracting the estimated bias. \cite{lei2021regression,lu2023debiased,chang2024exact} used this method to remove the bias of linear prediction functions. However, it is unclear how to apply this method to nonlinear prediction functions.

%Third, the implications of our methodology extend beyond theoretical contributions. 
\vspace{4mm}

\noindent \textbf{Notation}: For a set $\ms$ of finite numbers, let $|\ms|$ denote the cardinality of $\ms$. Let $I(\cdot)$ denote the indicator function. Let $[Q] = \{1,\ldots,Q\}$ denote the set of integers from $1$ to $Q$. Let $\rightsquigarrow$ denote convergence in distribution.

\section{Design-based inference framework}
\label{sec:framework}

Consider a randomized experiment with $N$ units, where each unit is assigned a binary treatment according to the experimental design. Let random variable $Z_i \in \{0,1\}$ denote the treatment indicator of unit $i\in [N]$. We define $Y_i(z)$, $z\in \{0,1\}$, as the potential outcome under treatment $z$. The observed outcome for unit $i$ is $Y_i = Z_i Y_i(1) + (1-Z_i) Y_i(0)$. We adopt the design-based inference framework, where the potential outcomes are fixed and the only randomness comes from the treatment indicators. We are interested in estimating the ATE: 
$\bar{\tau} = N^{-1} \sum_{i=1}^N \{Y_i(1)-Y_i(0)\}.$

Let $\bs{Z} = (Z_1,\ldots,Z_N)$ denote the treatment indicators for all units. An experimental design $\prob_{\bs{Z}}$ is a distribution over $\bs{Z}$. Some popular experimental designs are as follows:

\begin{example}[Bernoulli randomized experiment (BRE)]
\label{example:introduce-bernoulli-trial}
    $Z_1,\ldots,Z_N$ are i.i.d., following from $\operatorname{Bernoulli}(r_1)$, i.e., Bernoulli distribution with mean $r_1 \in (0,1)$.  We denote the corresponding $\prob_{\bs{Z}}$ as ${\BT(N,r_1)}$.
\end{example}

\begin{example}[Completely randomized experiment (CRE)]
\label{example:introduce-CR}
    Given $0 < N_1 < N$, $N_0=N-N_1$. For any $\bs{z} = (z_1,\ldots, z_N)\in \{0,1\}^N$ with $\sum_{i=1}^N z_i = N_1$, we have $\prob(\bs{Z} = \bs{z}) = N_1!(N-N_1)!/N!$; otherwise, $\prob(\bs{Z} = \bs{z})=0$. We denote the corresponding $\prob_{\bs{Z}}$ as ${\CR(N,N_1)}$.
\end{example}

\begin{example}[Stratified randomized experiment (SRE)]
\label{example:introduce-SR}
Suppose the experimental units are stratified into $K$ strata $\mk = \{1,\ldots,K\}$ based on baseline covariates, such as age and gender. For each unit $i\in [N]$, let  $A_i \in \mk$ denote the stratum indicator. Let $N_{\{k\}} = \sum_{i=1}^N I(A_i = k) $, for $k\in \mk$. Given $0 < N_{\{k\}1} < N_{\{k\}}$ and $N_{\{k\}0} = N_{\{k\}}-N_{\{k\}1}$, for any $\bs{z}\in \{0,1\}^N$ with $\sumi z_i I(A_i=k) = N_{\{k\}1}$, we have $\prob(\bs{Z} = \bs{z}) = \prod_{k\in \mk} N_{\{k\}1}!(N_{\{k\}}-N_{\{k\}1})!/N_{\{k\}}!$; otherwise, $\prob(\bs{Z} = \bs{z})=0$. We denote the corresponding $\prob_{\bs{Z}}$ as ${\SR(N, \bs{A},(N_{\{k\}1})_{k\in \mk})}$, where $\bs{A} = (A_1,\ldots,A_N)$. In the special case where $N_{\{k\}z} = 1$ for all $k\in \mk$ and $z\in \{0,1\}$, the SRE reduces to a matched-pairs experiment ($\MP$). We denote the corresponding $\prob_{\bs{Z}}$ as $\MP(N,\bs{A})$.
\end{example}

The Horvitz--Thompson (HT) estimator is a classic unbiased estimator of $\bar{\tau}$:
\[
\hat{\tau}_{\HT} = \frac{1}{N}\sumi \bigg\{ \frac{Z_i Y_i}{\prob( Z_i=1)} - \frac{(1-Z_i) Y_i}{\prob( Z_i=0)} \bigg\}.
\]
When baseline covariates $\bs{x}_i\in \mathbb{R}^d$ are available for unit $i$, it is common to use covariate adjustment to reduce the asymptotic variance of the HT estimator \citep{Lin2013Agnostic,liu2020regression,guo2023generalized,cohen2020no,su2023decorrelation}. Performing covariate adjustment is equivalent to finding a pair of prediction functions $f_z^\ast(\bs{x}_i): \mathbb{R}^d\rightarrow \mathbb{R}$, such as $f_z^\ast(\bs{x}_i) = \bs{x}_i^\top \beta_z $, $z\in \{0,1\}$, to approximate the potential outcomes. Given such functions, we can obtain an oracle-adjusted estimator for $\bar{\tau}$: $$\hat{\tau}_{\oracleadj}= \hat{\mu}_{\oracleadj}(1)- \hat{\mu}_{\oracleadj}(0),$$ where
    \begin{align*}
    &\hat{\mu}_{\oracleadj}(z) = \frac{1}{N} \sum_{i:Z_i=z} \frac{Y_i-f^\ast_z(\bs{x}_i)}{\prob(Z_i=z)} + \frac{1}{N} \sum_{i=1}^N f^\ast_z(\bs{x}_i),\quad z\in \{0,1\}. 
\end{align*}
In practice, $f^\ast_z$ is often not available, so we have to replace it with an estimator $\hat{f}_z$. Using a plug-in principle, we have the standard covariate-adjusted estimator:
    \begin{align*}
     &\hat{\tau}_{\adj}= \hat{\mu}_{\adj}(1)- \hat{\mu}_{\adj}(0),\quad \text{where} \quad \hat{\mu}_{\adj}(z) = \frac{1}{N} \sum_{i:Z_i=z} \frac{Y_i-\hat{f}_z(\bs{x}_i)}{\prob(Z_i=z)} + \frac{1}{N} \sum_{i=1}^N \hat{f}_z(\bs{x}_i), \quad z\in \{0,1\}.
\end{align*}
Here, $\hat{f}_z$ is obtained by using the subsamples $\{(Y_i,\bs{x}_i)\}_{i:Z_i=z}$. For this approach, the same sample is used twice: first, to estimate the prediction function $\hat{f}_z(\bs{x}_i)$, and then again to estimate the ATE by plugging in $\hat{f}_z(\bs{x}_i)$. This sample reuse introduces a non-negligible bias.

When $\{(Y_i,Z_i,\bs{x}_i)\}_{i=1}^N$ are i.i.d. samples from a super-population, we can apply sample-splitting and cross-fitting (see e.g., \cite{chernozhukov2018double}) to obtain an unbiased estimator. To illustrate the main idea, we first consider two-fold sample splitting, with the extension to $K$-fold splitting being analogous. Specifically, we can partition the index set $[N]$ into two disjoint subsets $\ms_{[1]}$ and $\ms_{[2]}$, such that $\ms_{[1]} \cup \ms_{[2]} = [N]$ and $\ms_{[1]} \cap \ms_{[2]} = \emptyset$. The corresponding subsamples $\{(Y_i,Z_i,\bs{x}_i)\}_{i\in \ms_{[1]}}$ and $\{(Y_i,Z_i,\bs{x}_i)\}_{i\in \ms_{[2]}}$ are used either for estimating the prediction function or for estimating the ATE, but not both. The validity of this method relies on the independence between the two subsamples.

In the design-based inference framework, $\bs{x}_i$'s are deterministic, and $Z_i$'s may be dependent due to the experimental design. Let $\bs{Z}_{[q]} := (Z_i)_{i\in \ms_{[q]}}$, $q=1,2$. The randomness of the subsample $\{(Y_i,Z_i,\bs{x}_i)\}_{i\in \ms_{[q]}}$ arises from $(\ms_{[q]}, \bs{Z}_{[q]})$. 
Since $\ms_{[1]}$ and $\ms_{[2]}$ are disjoint subsets for the classic cross-fitting method, this creates a dependence between them, as well as a dependence between the two subsamples. This dependence violates key assumptions and makes standard theory on sample-splitting and cross-fitting unsuitable for design-based inference in randomized experiments. To address this issue, we propose a conditional cross-fitting mechanism tailored to the design-based inference framework.

\section{Unbiased estimator via conditional cross-fitting}
\label{sec:crossfitting-algorithm}

 The conditional cross-fitting algorithm is outlined in \Cref{algorithm:conditional-cross-fitting-algorithm} below. Unlike the decorrelation procedure proposed by \cite{su2023decorrelation}, our algorithm partitions the index set $[N]$ into disjoint $\ms_{[1]}$ and $\ms_{[2]}$ such that $\ms_{[1]}\cup\ms_{[2]} = [N]$ and $\ms_{[1]}\cap\ms_{[2]} = \emptyset$. A key distinction from the classic cross-fitting method, which typically performs a random split, is that our algorithm is carefully designed to satisfy \Cref{a:conditional-independent-subsamples} below to ensure the unbiasedness of the machine-learning-assisted covariate-adjusted ATE estimators.
\begin{assumption}
\label{a:conditional-independent-subsamples}
    (i) $\bs{Z}_{[1]} \indep \bs{Z}_{[2]} \mid (\ms_{[1]},\ms_{[2]})$; (ii) $\prob(Z_i=z\mid \ms_{[1]},\ms_{[2]}) \ne 0$  for $z\in\{0,1\}$ and $i\in [N]$.
\end{assumption}

\Cref{a:conditional-independent-subsamples}(i) is a conditional independence assumption, which requires that, given the sample-splitting, the treatment assignment vectors in the two subsamples are independent. This guarantees that the two resulting subsamples $\{(Y_i,Z_i,\bs{x}_i)\}_{i\in \ms_{[1]}}$ and $\{(Y_i,Z_i,\bs{x}_i)\}_{i\in \ms_{[2]}}$ are independent conditional on $(\ms_{[1]},\ms_{[2]})$. \Cref{a:conditional-independent-subsamples}(i) does not require $Z_i$ to be i.i.d. For dependent treatment assignments such as complete randomization and stratified randomization, we will develop algorithms in the next section that satisfy \Cref{a:conditional-independent-subsamples}(i). \Cref{a:conditional-independent-subsamples}(ii) requires that, given the sample-splitting, every unit has a positive possibility of being assigned to the treatment and control groups. \Cref{a:conditional-independent-subsamples} ensures that the conditional cross-fitting algorithm yields an unbiased estimator.

\begin{algorithm}
\caption{Conditional Cross-Fitting Algorithm}
\label{algorithm:conditional-cross-fitting-algorithm}
\begin{algorithmic}[1]
\State \textbf{Input:} Dataset $ \{( Y_i,Z_i,\bs{x}_i)\}_{i=1}^N$.
\State Randomly split the population $[N]$ into two disjoint subsets $\ms_{[1]}$ and $\ms_{[2]}$ under probability measure $\prob(\ms_{[1]},\ms_{[2]}\mid \bs{Z})$ such that $\bs{Z}_{[1]} \indep \bs{Z}_{[2]} \mid (\ms_{[1]},\ms_{[2]})$. Let $N_{[q]} = |\ms_{[q]}| $ and $\ms_{[-q]} = [N]\backslash\ms_{[q]}$, $q=1,2$. 
\For {$q = 1, 2$}
    \State  Use $\{( Y_i,Z_i,\bs{x}_i)\}_{i\in\ms_{[-q]}}$ to estimate the prediction function $\hat{f}_{[-q]z}(\bs{x}_i)$, $z\in \{0,1\}.$
    \State 
   Use $\{( Y_i,Z_i,\bs{x}_i)\}_{i\in\ms_{[q]}}$ to estimate the ATE:
        \begin{align*}
     &\hat{\tau}_{\cfadj, [q]}= \hat{\mu}_{\cfadj, [q]}(1)- \hat{\mu}_{\cfadj, [q]}(0),\\
    &\hat{\mu}_{\cfadj, [q]}(z) = \frac{1}{N_{[q]}} \sum_{i:Z_i=z,i\in \ms_{[q]}} \frac{Y_i-\hat{f}_{[-q]z}(\bs{x}_i)}{\prob(Z_i=z\mid \ms_{[1]},\ms_{[2]})} + \frac{1}{N_{[q]}} \sum_{i\in \ms_{[q]}} \hat{f}_{[-q]z}(\bs{x}_i). 
\end{align*}
\EndFor
\State \textbf{Combination:} Combine the two ATE estimators:
\[
\hat{\tau}_{\cfadj} = \sum_{q\in \{1,2\}} \frac{N_{[q]}}{N}\hat{\tau}_{\cfadj, [q]}.
\]
\State \textbf{Return:} $\hat{\tau}_{\cfadj}$.
\end{algorithmic}
\end{algorithm}

\begin{theorem}\label{thm:unbiased}
Under \Cref{a:conditional-independent-subsamples}, we have $   \ope(\hat{\tau}_{\cfadj} \mid \ms_{[1]},\ms_{[2]}) = \bar{\tau}$, where $\hat{\tau}_{\cfadj}$ is the cross-fitted ATE estimator defined in \Cref{algorithm:conditional-cross-fitting-algorithm}.
\end{theorem}

By \Cref{thm:unbiased}, $\hat{\tau}_{\cfadj}$ is conditionally unbiased under \Cref{a:conditional-independent-subsamples}, which further implies that it is unbiased by the law of total expectation. Importantly, the unbiased property imposes no requirements on the precision of the estimated prediction function. To apply the algorithm and achieve unbiasedness, we need to obtain feasible sample-splitting algorithms $\prob(\ms_{[1]},\ms_{[2]}\mid \bs{Z})$ that satisfy \Cref{a:conditional-independent-subsamples}.

\subsection{Sample-splitting algorithm for canonical experiments}
\label{sec:sample-splitting-algorithm}

 Below, we provide sample-splitting algorithms that satisfy \Cref{a:conditional-independent-subsamples} for the designs outlined in Section \ref{sec:framework}. Let $\prob_{\bs{Z}_{[q]}\mid \ms_{[1]},\ms_{[2]}}$ denote the conditional distribution of $\bs{Z}_{[q]}$ given $ (\ms_{[1]},\ms_{[2]})$. Since \Cref{algorithm:conditional-cross-fitting-algorithm} requires evaluating $\prob(Z_i = z \mid \ms_{[1]}, \ms_{[2]})$  to compute $\hat{\tau}_{\cfadj}$, characterizing $\prob_{\bs{Z}_{[q]} \mid \ms_{[1]}, \ms_{[2]}}$ is essential. We provide explicit expressions for these conditional distributions together with the corresponding algorithms.

\begin{example}[Sample-splitting algorithm in BRE]
\label{example:algorithm-bernoulli-trial}
For $\prob_{\bs{Z}}=\BT(N,r_1)$, we specify $\pi \in (0,1)$ and generate i.i.d $(R_i)_{i=1}^N$ from $\operatorname{Bernoulli}(\pi)$, independent of $(Z_i)_{i=1}^N$. Let $\ms_{[1]} = \{i:R_i=1\}$ and $\ms_{[2]} = \{i:R_i=0\}$.
\end{example}

\begin{proposition}\label{prop:BT}
Suppose that $\prob_{\bs{Z}}=\BT(N,r_1)$. The algorithm in \Cref{example:algorithm-bernoulli-trial} satisfies \Cref{a:conditional-independent-subsamples} with $\prob_{\bs{Z}_{[q]}\mid \ms_{[1]},\ms_{[2]}} = {\BT(N_{[q]},r_1)}$, $q=1,2$. 
\end{proposition}

By \Cref{prop:BT}, conditional on $(\ms_{[1]},\ms_{[2]})$, the two subsamples in \Cref{example:algorithm-bernoulli-trial} correspond to two independent BREs. Here, $\pi$ is the expected proportion of units assigned to subset $\ms_{[1]}$ and $\pi = 0.5$ corresponds to the default choice of equal split.

\begin{example}[Split-by-treatment algorithm in CRE]
\label{example:algorithm-split-by-treatment}
 For $\prob_{\bs{Z}}=\CR(N,N_1)$, (i) we specify $N_{[q]z}$, $z=0,1$, $q=1,2$, such that $N_z > N_{[q] z} > 0$ and $N_{[1] z}+N_{[2] z} = N_z$, (ii) randomly split $\ms_z$ into $\ms_{[1] z}$ and $\ms_{[2] z}$ such that $|\ms_{[1] z}| = N_{[1] z}$ and $|\ms_{[2] z}| = N_{[2] z}$, and (iii) let $\ms_{[1]} = \ms_{[1]1}\cup\ms_{[1]0}$ and $\ms_{[2]} = \ms_{[2]1}\cup\ms_{[2]0}$.
\end{example}

\begin{proposition}\label{prop:CR}
Suppose that $\prob_{\bs{Z}}=\CR(N,N_1)$. The algorithm in \Cref{example:algorithm-split-by-treatment} satisfies \Cref{a:conditional-independent-subsamples} with $\prob_{\bs{Z}_{[q]}\mid \ms_{[1]},\ms_{[2]}} = {\CR(N_{[q]},N_{[q]1})}$, $q=1,2$.
\end{proposition}

By \Cref{prop:CR}, conditional on $(\ms_{[1]},\ms_{[2]})$, the two subsamples in \Cref{example:algorithm-split-by-treatment} correspond to two independent CREs. Interestingly, although $({Z}_i)_{i=1}^N$ are dependent, we can obtain two conditionally independent subsamples through the conditional cross-fitting mechanism in \Cref{example:algorithm-split-by-treatment}.

SREs conduct independent CREs in each stratum. Therefore, we can apply the split-by-treatment algorithm in each stratum.

\begin{example}[Split-by-treatment algorithm in SRE]
\label{example:algorithm-split-by-treatment-sr}
For $\prob_{\bs{Z}}=\SR(N,\bs{A},(N_{\setk 1})_{k \in \mk})$, (i) we specify $N_{\{k\}[q] z}$, $k\in \mk$, $q=1,2$, $z=0,1$, 
such that $N_{\setk z} > N_{\{k\}[q] z} > 0$ and $N_{\{k\}[1] z}+N_{\{k\}[2] z} = N_{\{k\}z}$; (ii) let $\ms_{\{k\}z} = \{i:A_i=k,Z_i=z\}$ and randomly split $\ms_{\{k\}z}$ into $\ms_{\{k\} [1] z}$ and $\ms_{\{k\} [2] z}$ such that $|\ms_{\{k\} [q] z}| = N_{\{k\}[q] z}$; and (iii) let $\ms_{[q]} = \cup_{k\in \mk}\cup_{z\in\{0,1\}}\ms_{\{k\}[q]z}$, $q=1,2$.
\end{example}

\begin{proposition}\label{prop:sr-1}
Suppose that $\prob_{\bs{Z}}=\SR(N,\bs{A},(N_{\{k\}1})_{k\in \mk})$. For $q=1,2$, let $N_{\{k\}[q]} = N_{\{k\}[q]1} + N_{\{k\}[q]0}$ and $\bs{A}_{[q]} = (A_i)_{i\in \ms_{[q]}}$. The algorithm in \Cref{example:algorithm-split-by-treatment-sr} satisfies \Cref{a:conditional-independent-subsamples} with $\prob_{\bs{Z}_{[q]}\mid \ms_{[1]},\ms_{[2]}} = {\SR(N_{[q]},\bs{A}_{[q]},(N_{\{k\}[q]1})_{k\in \mk})}$.
\end{proposition}

\Cref{prop:sr-1} demonstrates that the split-by-treatment algorithm partitions the original SRE into two independent SREs conditional on the sample split. The split-by-treatment algorithm requires that $N_{\{k\}z} \geq 2$, $z\in \{0,1\}$, $k\in \mk$, to ensure that $\prob(Z_i=z\mid \ms_{[1]},\ms_{[2]}) > 0$ in \Cref{a:conditional-independent-subsamples}. However, this is violated in MPEs where $N_{\{k\}z} = 1$. In this case, we have to apply the split-by-stratum algorithm below.

\begin{example}[Split-by-stratum algorithm in SRE]
\label{example:split-by-stratum-sr}
  For $\prob_{\bs{Z}}=\SR(N,\bs{A},(N_{\setk 1})_{k \in \mk})$,  (i) we specify $K_{[1]},K_{[2]}$, such that $0 < K_{[1]},K_{[2]} < K$ and $K_{[1]} + K_{[2]} = K$, (ii) randomly split the $K$ strata into two parts: $\mathcal{K}_{[1]}\cup\mathcal{K}_{[2]} = [K]$, $\mathcal{K}_{[1]}\cap\mathcal{K}_{[2]} = \emptyset$, and (iii) let $\ms_{[q]} = \cup_{k\in \mathcal{K}_{[q]}}\ms_{\{k\}}$ with $\ms_{\{k\}}=\{i:A_i = k\}$.
\end{example}

\begin{proposition}\label{prop:sr-2}
    Suppose that $\prob_{\bs{Z}}=\SR(N,\bs{A},(N_{\setk 1})_{k \in \mk})$. The algorithm in \Cref{example:split-by-stratum-sr} satisfies  \Cref{a:conditional-independent-subsamples} with
 $\prob_{\bs{Z}_{[q]}\mid \ms_{[1]},\ms_{[2]}} = {\SR(N_{[q]},\bs{A}_{[q]},(N_{\{k\}1})_{k\in \mk_{[q]}})}$, $q=1,2$.
\end{proposition}

\Cref{prop:sr-2} demonstrates that the split-by-stratum algorithm also partitions the original SRE into two independent SREs conditional on the sample split.

\begin{remark}
    If the design includes a mix of some large and many small strata \citep{pashley2021insights,tian2025stratified}, a hybrid strategy combining \Cref{example:algorithm-split-by-treatment-sr} and \Cref{example:split-by-stratum-sr} can be applied.
\end{remark}

\subsection{Estimation of prediction function}
\label{sec:estimation-prediction}
Let $\mathcal{F}_z$ be a function class to which $f^\ast_z$ belongs. The prediction function is often defined by
\[
f^\ast_z = \argmin_{f\in \mathcal{F}_z} \frac{1}{N}\sumi \ell_{i,z}\big(Y_i(z),\bs{x}_i, f\big),
\]
where $\ell_{i,z}\big(Y_i(z),\bs{x}_i, f\big)$ is the loss function evaluated at unit $i$. For example, \cite{su2023decorrelation} uses $\ell_{i,z}(Y_i(z),\bs{x}_i, f\big) = \{Y_i(z)-f(\bs{x}_i)\}^2$ for a given parametric or nonparametric function class $\mathcal{F}_z$; logistic regression uses $\ell_{i,z}(Y_i(z),\bs{x}_i, f\big) = Y_i(z)\bs{x}_i^\top\bs{\theta}-\log\{ 1+\exp{ (\bs{\theta}^\top\bs{x}_i)}\}$ and $\mathcal{F}_z = \{f: f(\bs{x}_i) = \{1+\exp(-\bs{\theta}^\top \bs{x}_i)\}^{-1}$, $ \bs{\theta} \in \mathbb{R}^d\}$. We can also use machine learning function classes, such as the function classes of neural networks and random forests, as $\mathcal{F}_z$.
We obtain the estimator $\hat{f}_{[-q]z}$ by using the empirical loss:
\begin{align}
\label{eq:prediction-function-formula}
    \hat{f}_{[-q]z} = \argmin_{f\in \mathcal{F}_z} \frac{1}{N}\sumi \frac{I(i\in \ms_{[-q]},Z_i=z)}{\prob(i\in \ms_{[-q]}, Z_i=z)}\ell_{i,z}\big(Y_i(z),\bs{x}_i, f\big).
\end{align}

\begin{remark}
    If $f^\ast_0$ and $f^\ast_1$ share some parameters, as in the pooled regression, we can define them accordingly:
\[
(f_0^\ast,f_1^\ast)  = \argmin_{(f_1,f_0)\in \mathcal{F}} \frac{1}{N}\sumi \sum_{z\in \{0,1\}}\ell_{i,z}\big(Y_i(z),\bs{x}_i, f_z\big),
\]
where $\mathcal{F}$ is the joint function class of $(f_1,f_0)$. Then we can estimate $(f_0^\ast,f_1^\ast)$ jointly by
\begin{align*}
    &(\hat{f}_{[-q]0},\hat{f}_{[-q]1}) = \argmin_{(f_0,f_1)\in \mathcal{F}} \frac{1}{N}\sumi \sum_{z\in \{0,1\}}\frac{I(i\in \ms_{[-q]},Z_i=z)}{\prob(i\in \ms_{[-q]}, Z_i=z)}\ell_{i,z}\big(Y_i(z),\bs{x}_i, f_z\big).
\end{align*}
\end{remark}

\begin{remark}
      We can incorporate stratification information to obtain $\hat{f}_{[-q]z}$ for stratified randomized experiments. For example, under SREs, we can include stratum indicators as covariates to account for the stratum effects in linear regression. However, in this case, the weights $\prob(i\in \ms_{[-q]}, Z_i=z)^{-1}$ do not necessarily ensure that $\hat{f}_{[-q]z}$ is a consistent estimator of $f^\ast_z$ when the number of strata tends to infinity. The reason is that if we include stratum indicators as covariates, the number of parameters to be estimated in the prediction function tends to infinity with the sample size, leading to a loss of degrees of freedom. We address this issue by adjusting the weights to account for the loss of degrees of freedom. For MPEs, instead of regressing the outcome on covariates, we can regress the outcome difference on covariates, as proposed by \cite{fogarty2018regression}. Further details on regression adjustments for SREs and MPEs can be found in the Supplementary Material.
\end{remark}

\section{Asymptotic property of $\hat{\tau}_{\cfadj}$}
\label{sec:asymptotic-property-of-cross-fitted-estimator}
In this section, we study the asymptotic property of $\hat{\tau}_{\cfadj}$, including its asymptotic normality and efficiency gain.
\subsection{Asymptotic normality of $\hat{\tau}_{\cfadj}$}
\label{sec:cfadj-and-oracleadj}
  \Cref{a:conditional-experimental-design-bounded-variance} below is a regularity condition on $\prob_{\bs{Z}_{[q]}\mid \ms_{[1]},\ms_{[2]}}$. 
% \begin{align*}
%     \hat{\tau}_{\cfadj}-\hat{\tau}_{\oraclecf} =M_{[1]1} - M_{[1]0} + M_{[2]1} - M_{[2]0},
% \end{align*}
% where
% \begin{align*}
%     M_{[q]z} = \frac{1}{N}\sum_{i\in \ms_{[q]}} \Big(1-\frac{I(Z_i=z)}{\prob(Z_i=z\mid \ms_{[1]},\ms_{[2]})}\Big)(\hat{f}_{[-q]z}(\bs{x}_i)-f^\ast_{z}(\bs{x}_i))
% \end{align*}
\begin{assumption}
\label{a:conditional-experimental-design-bounded-variance}
For $q \in \{1,2\}$, suppose that $\prob_{\bs{Z}_{[q]}\mid \ms_{[1]},\ms_{[2]}}$ satisfies: for any finite population quantities $\{a_i\}_{i=1}^N$, there exists a constant $C_z$ independent of $N$, such that
    \[
    \Var\Big(\frac{1}{N}\sum_{i\in \ms_{[q]}} \frac{I(Z_i=z)}{\prob(Z_i=z\mid \ms_{[1]},\ms_{[2]})}a_i \mid \ms_{[1]},\ms_{[2]}\Big) \leq \frac{C_z}{N^2}\sum_{i\in \ms_{[q]}} a_i^2, \quad \textnormal{for $z\in \{0,1\}$.}
    \]
\end{assumption}

%As indicated by \Cref{prop:assumption-2-holds} below, \Cref{a:conditional-experimental-design-bounded-variance} holds for all the sample-splitting algorithms mentioned in \Cref{sec:sample-splitting-algorithm}.

\begin{proposition}
\label{prop:assumption-2-holds}
    For $q=1,2$, $z=0,1$, $k \in \mk$, a constant $c \in (0,0.5)$, we have (i) $\prob_{\bs{Z}_{[q]}\mid \ms_{[1]},\ms_{[2]}}=\BT(N_{[q]},r_1)$ satisfies \Cref{a:conditional-experimental-design-bounded-variance}, (ii) if $ N_{[q]z}/N \in (c,1-c)$, then $\prob_{\bs{Z}_{[q]}\mid \ms_{[1]},\ms_{[2]}}=\CR(N_{[q]},N_{[q]1})$ satisfies \Cref{a:conditional-experimental-design-bounded-variance}, (iii) if $N_{\setk[q]z}/N_{\setk} \in (c,1-c)$, then $\prob_{\bs{Z}_{[q]}\mid \ms_{[1]},\ms_{[2]}}=\SR(N_{[q]},\bs{A}_{[q]},(N_{\{k\}[q]1})_{k\in \mk})$ satisfies \Cref{a:conditional-experimental-design-bounded-variance}, and (iv) if $N_{\setk z}/N_{\setk} \in (c,1-c)$, $\prob_{\bs{Z}_{[q]}\mid \ms_{[1]},\ms_{[2]}}=\SR(N_{[q]},\bs{A}_{[q]},(N_{\setk 1})_{k\in \mk_{[q]}})$ satisfies \Cref{a:conditional-experimental-design-bounded-variance}.
\end{proposition}

Recalling Propositions \ref{prop:BT}--\ref{prop:sr-2}, \Cref{prop:assumption-2-holds} shows that \Cref{a:conditional-experimental-design-bounded-variance} holds for all the sample-splitting algorithms mentioned in \Cref{sec:sample-splitting-algorithm} in general if the proportion of units assigned to each treatment group in any subsample is positive.

\begin{assumption}[Stability]
\label{a:negligible-estimateion-error-for-prediction-function}
    For any $z \in \{0,1\}$, $q\in \{1,2\}$, we have $\|\hat{f}_{[-q]z}-f^\ast_{z}\|_N = \op(1)$, where
$
    \|\hat{f}_{[-q]z}-f^\ast_{z}\|_N^2 = N^{-1} \sumi \{\hat{f}_{[-q]z}(\bs{x}_i)-f^\ast_{z}(\bs{x}_i)\}^2 .
$
\end{assumption}

\Cref{a:negligible-estimateion-error-for-prediction-function} is a stability assumption, requiring the estimated prediction function to be close to the oracle prediction function. Similar assumptions have been used in \cite{cohen2020no} and \cite{su2023decorrelation}. \Cref{a:negligible-estimateion-error-for-prediction-function} requires the estimated prediction function based on subsample $\{(Y_i(z),\bs{x}_i)\}_{i:\ i\in \ms_{[-q]},Z_i=z}$ is close to that computed on $\{(Y_i(z),\bs{x}_i)\}_{i=1}^N$. The indicators $(I(i\in \ms_{[-q]},Z_i=z))_{i=1}^N$  follow i.i.d. Bernoulli distribution and $\CR(N,N_{[-q]z})$ for the sample-splitting algorithm in BRE and Split-by-treatment algorithm in CRE, respectively. \cite{su2023decorrelation} and \cite{cohen2020no} have justified the stability assumptions for the estimated prediction functions based on a subsample from BREs and CREs, respectively. We justify the stability assumption of linear prediction functions under SREs and MPEs in the Supplementary Material.

Define $\hat{\tau}_{\oraclecf}$ the same as $\hat{\tau}_{\cfadj}$ with $\hat{f}_{[-q]z}(\bs{x}_i)$ replaced by $f^\ast_{z}(\bs{x}_i)$. \Cref{a:nondegenerate-variance} below requires the variance to be non-degenerate.
\begin{assumption}
\label{a:nondegenerate-variance}  $N\Var(\hat{\tau}_{\oraclecf})$ has a positive limit.
\end{assumption}

\begin{theorem}
\label{thm:valid-confidence-interval}
    Under Assumptions~\ref{a:conditional-independent-subsamples}--\ref{a:nondegenerate-variance},  
    $N^{1/2}(\hat{\tau}_{\oraclecf}-\hat{\tau}_{\cfadj}) = \op(1)$.
   Thus, if we further assume $(\hat{\tau}_{\oraclecf}-\bar{\tau})/\Var(\hat{\tau}_{\oraclecf})^{1/2} \rightsquigarrow \mathcal{N}(0,1)$,
%    \begin{align}
%    \label{eq:clt-for-oracle-cf}
%        &(\hat{\tau}_{\oraclecf}-\bar{\tau})/\Var(\hat{\tau}_{\oraclecf})^{1/2} \rightsquigarrow \mathcal{N}(0,1),
%    \end{align}
 then $(\hat{\tau}_{\cfadj}-\bar{\tau})/\Var(\hat{\tau}_{\oraclecf})^{1/2} \rightsquigarrow \mathcal{N}(0,1)$.
 %and the confidence interval $    [\hat{\tau}_{\cfadj} - z_{1-\alpha/2}\hat{V}_{\cfadj}^{1/2}, \hat{\tau}_{\cfadj} + z_{1-\alpha/2}\hat{V}_{\cfadj}^{1/2}]$ has an asymptotic coverage rate of at least $1-\alpha$.
\end{theorem}

\Cref{thm:valid-confidence-interval} shows that, under appropriate conditions, $\hat{\tau}_{\cfadj}$ has the same asymptotic distribution as $\hat{\tau}_{\oraclecf}$. In \Cref{sec:confidence-intervals}, we provide central limit theorems for $\sqrt{N}(\hat{\tau}_{\oraclecf}-\bar{\tau})$ under the mentioned experimental designs and sample-splitting and cross-fitting algorithms, which further implies the asymptotic normality of $\hat{\tau}_{\cfadj}$.

\subsection{Optimal sample-splitting algorithm}
\label{sec:optimal-algorithm}

We can check that all the algorithms in \Cref{sec:sample-splitting-algorithm} satisfy \Cref{a:a-class-of-design} below. \begin{assumption}
\label{a:a-class-of-design}
$\prob(Z_i=z\mid \ms_{[1]},\ms_{[2]}) = \sum_{q=1}^2\prob(Z_i=z\mid i\in \ms_{[q]})I(i\in \ms_{[q]})$ and $\prob(i\in \ms_{[q]}\mid \bs{Z}) = \prob(i\in \ms_{[q]}\mid Z_i)$.
%\begin{align*}
%         &\prob(Z_i=z\mid \ms_{[1]},\ms_{[2]}) = \sum_{q=1}^2\prob(Z_i=z\mid i\in \ms_{[q]})I(i\in \ms_{[q]}),\quad \prob(i\in \ms_{[q]}\mid \bs{Z}) = \prob(i\in \ms_{[q]}\mid Z_i).
%\end{align*}
\end{assumption}

%Contrary to \Cref{a:constant-treatment-assignment-probability}, 
\Cref{a:a-class-of-design} indicates that the treatment probability of unit $i$ may depend on the subsample to which it belongs. The second part of \Cref{a:a-class-of-design} assumes that the conditional probability of assigning unit $i$ to subsample $\ms_{[q]}$ depends only on its own treatment assignment, independent of the treatment assignments of other units.

A practical question is: across all the feasible cross-fitting algorithms (algorithms satisfying \Cref{a:conditional-independent-subsamples}), which algorithm is optimal in the sense of achieving the smallest asymptotic variance? To answer this question, we need \Cref{a:constant-treatment-assignment-probability} below.

\begin{assumption}
\label{a:constant-treatment-assignment-probability}
    $
    \prob(Z_i = z\mid \ms_{[1]},\ms_{[2]}) = \prob(Z_i = z),
    $ $z=0,1.$
\end{assumption}

\Cref{a:constant-treatment-assignment-probability} assumes that the treatment probability of unit $i$ is independent of the sample split, and therefore, independent of the subsample to which it belongs. Taking the split-by-treatment algorithm for example, by Propositions \ref{prop:CR}--\ref{prop:sr-1}, \Cref{a:constant-treatment-assignment-probability} is equivalent to $N_{[1]1}/N_{[1]} = N_{[2]1}/N_{[2]}$ for CRE and $N_{\{k\}[1]1}/N_{\{k\}[1]} = N_{\{k\}[2]1}/N_{\{k\}[2]}$, $k\in \mk$, for SRE. For the BRE algorithm and split-by-stratum algorithm, \Cref{a:constant-treatment-assignment-probability} always holds.

\begin{theorem}
\label{thm:constant-probability-optimal}
    Under \Cref{a:a-class-of-design}, we have
    $
    \Var(\hat{\tau}_{\oraclecf}) \geq \Var(\hat{\tau}_{\oracleadj}).
    $
    The equality is achieved when \Cref{a:constant-treatment-assignment-probability} holds.
\end{theorem}

\Cref{thm:constant-probability-optimal} demonstrates that for the algorithms satisfying \Cref{a:a-class-of-design}, the algorithms satisfying \Cref{a:constant-treatment-assignment-probability} are optimal. Other conditions might also ensure that the asymptotic variance reaches the lower bound, but \Cref{a:constant-treatment-assignment-probability} is simple and gives us the principled criterion for designing the optimal sample-splitting algorithm. Moreover, under \Cref{a:constant-treatment-assignment-probability}, we have $\hat{\tau}_{\oracleadj} = \hat{\tau}_{\oraclecf}$, and thus, by \Cref{thm:valid-confidence-interval}, $\hat{\tau}_{\cfadj}$ recovers the oracle-adjusted estimator $\hat{\tau}_{\oracleadj}$ with asymptotically negligible difference.

%\lx{[Delete: The first part of \Cref{a:a-class-of-design} assumes that the conditional treatment probability for unit $i$ depends solely on its own membership and not on the membership of any other units. However, this treatment probability may depend on which subsample it belongs to. The second part of \Cref{a:a-class-of-design} assumes that the conditional probability of assigning unit $i$ to subsample $\ms_{[q]}$ depends only on its own treatment assignment, independent of the treatment assignments of other units. However, the conditional probability of assigning $i$ to subsample $\ms_{[q]}$ may vary based on the value $Z_i$.   \liu{I cannot understand this sentence--what is the difference between this sentence and the previous one?}. We can check that all the algorithms mentioned in \Cref{sec:sample-splitting-algorithm} satisfy \Cref{a:a-class-of-design}. ]}

\subsection{Efficiency gain of $\hat{\tau}_{\cfadj}$}

The asymptotic efficiency gain of the cross-fitting procedure equals $\Var(\hat{\tau}_{\HT})-\Var(\hat{\tau}_{\oraclecf})$. Under \Cref{a:constant-treatment-assignment-probability}, the asymptotic efficiency gain then equals $\Var(\hat{\tau}_{\HT})-\Var(\hat{\tau}_{\oracleadj})$, depending on the potential outcomes, experimental design, and the estimation procedure of prediction function.  Given the experimental design, we say that an estimation procedure has the do-no-harm property if the asymptotic efficiency gain is nonnegative for any potential outcomes \citep{cohen2020no}. A well-known result is that the ordinary least squares (OLS) regression adjustment provides a do-no-harm property for CRE \citep{Lin2013Agnostic}. In general, commonly used estimation procedures based on nonlinear functions do not provide this property and thus may have negative asymptotic efficiency gain. \cite{cohen2020no} proposed a no-harm calibration procedure under CRE for nonlinear estimation procedures. The resulting estimation procedure has the do-no-harm property. We show how to adapt the no-harm calibration procedure to our conditional cross-fitting algorithms under CREs in the Supplementary Material. Under other experimental designs, such as SREs and MPEs, OLS may not guarantee efficiency gains. To guarantee efficiency gains, we need to incorporate paring and stratification information into the linear regression; see the Supplementary Material for further details.

\section{Variance estimation and confidence interval}
\label{sec:confidence-intervals}
\subsection{Cross-fitted variance estimation}
We now turn to the inference problem for $\bar{\tau}$. Inference based on $\hat{\tau}_{\cfadj}$ is challenging due to the complex dependencies among treatment indicators induced by the experimental design. Even in the absence of cross-fitting, the variance estimators of the unadjusted HT estimators 
require careful construction, depending on the specific experimental design. The introduction of cross-fitting further complicates the dependence structure among the terms contributing to $\hat{\tau}_{\cfadj}$. In this section, we propose a novel cross-fitted variance estimator, leveraging the key insight that the two subsamples generated by the sample-splitting algorithm can be viewed as conditionally independent randomized experiments, given the sample split.

Consider the easier task of estimating the variance of the HT estimator in the absence of cross-fitting, namely, $\Var(\hat{\tau}_{\HT})$. Let $\hat{V}(\cdot;\prob_{\bs{Z}})$ denote a generic 
conservative variance estimator for the 
experimental design $\prob_{\bs{Z}}$, which takes the observed outcomes and treatment indicators as input. Based on the observed data $(Y_i)_{i=1}^N$ and $\bs{Z}$, $\hat V((Y_i)_{i=1}^N,\bs{Z};\prob_{\bs{Z}})$ is used as an conservative estimator for $\Var(\hat{\tau}_{\HT})$. 
  $\hat{V}(\cdot;\prob_{\bs{Z}})$ should be constructed case by case for different experimental designs. For examples of $\hat{V}(\cdot;\prob_{\bs{Z}})$, see \cite{neyman1923} for $\prob_{\bs{Z}} = \CR(N,N_1)$; \cite{liu2020regression} for $\prob_{\bs{Z}} = \SR(N,\bs{A},(N_{\{k\}1})_{k\in \mk})$; \cite{fogarty2018regression} for $\prob_{\bs{Z}} = \MP(N,\bs{A})$. In our approach, we will utilize $\hat{V}(\cdot;\prob_{\bs{Z}})$ as the basic estimator to construct our proposed cross-fitted variance estimators for $\hat{\tau}_{\cfadj}$.

\label{sec:variance-and-inference}
Since $\hat{\tau}_{\cfadj}$ has the asymptotic variance $\Var(\hat{\tau}_{\oraclecf})$, we need a conservative estimator of $\Var(\hat{\tau}_{\oraclecf})$ for the inference of $\bar{\tau}$ using $\hat{\tau}_{\cfadj}$. By the law of total variance, we have
  \begin{align*}
    \Var(\hat{\tau}_{\oraclecf}) = \ope\Big\{\sum_{q\in \{1,2\}}\frac{N_{[q]}^2}{N^2}\Var(\hat{\tau}_{\oraclecf[q]}\mid \ms_{[1]},\ms_{[2]})\Big\}.
\end{align*}
We hope that
\[
\sum_{q\in \{1,2\}}\frac{N_{[q]}^2}{N^2}\Var(\hat{\tau}_{\oraclecf[q]}\mid \ms_{[1]},\ms_{[2]}) \approx \ope\Big\{\sum_{q\in \{1,2\}}\frac{N_{[q]}^2}{N^2}\Var(\hat{\tau}_{\oraclecf[q]}\mid \ms_{[1]},\ms_{[2]})\Big\},
\]
such that it suffices to estimate $\Var(\hat{\tau}_{\oraclecf[q]}\mid \ms_{[1]},\ms_{[2]})$. 

Define $\varepsilon_i(z) := Y_i(z)-f^\ast_z(\bs{x}_i)$ and denote $\varepsilon_i = Z_i \varepsilon_i(1) + (1 - Z_i) \varepsilon_i(0)$.  We observe that
\[
\Var(\hat{\tau}_{\oraclecf[q]}\mid \ms_{[1]},\ms_{[2]}) = \Var\Big(\frac{1}{N_{[q]}}\sum_{i\in \ms_{[q]}} \frac{Z_i\varepsilon_i}{\prob(Z_i=1\mid \ms_{[1]},\ms_{[2]})} - \frac{(1-Z_i)\varepsilon_i}{\prob(Z_i=0\mid \ms_{[1]},\ms_{[2]})} \mid \ms_{[1]},\ms_{[2]}\Big).
\]
Viewing $\Var(\hat{\tau}_{\oraclecf[q]}\mid \ms_{[1]},\ms_{[2]})$ as the variance of a HT estimator under the experimental design $\prob_{\bs{Z}_{[q]}\mid \ms_{[1]},\ms_{[2]}}$, we can construct an estimator of $\Var(\hat{\tau}_{\oraclecf[q]}\mid \ms_{[1]},\ms_{[2]})$, using $\hat{V}(\cdot;\prob_{\bs{Z}_{[q]}\mid \ms_{[1]},\ms_{[2]}})$:
%\begin{align*}
%  \hat{V}_{\oraclecf, [q]} =  \hat V\Big(( \varepsilon_i)_{i\in \ms_{[q]}},\bs{Z}_{[q]};\prob_{\bs{Z}_{[q]}\mid \ms_{[1]},\ms_{[2]}} \Big).
%\end{align*}
%This estimator is infeasible since $\varepsilon_i$ is not observed. Replacing $\varepsilon_i$ with the cross-fitted residual
%\[
%\hat{\varepsilon}_{i} = Y_i-\sum_{q\in \{1,2\}}\sum_{z\in \{0,1\}}\hat{f}_{[-q]z}(\bs{x}_i)I(Z_i = z,i\in\ms_{[q]}),
%\]
%we obtain a feasible variance estimator:
\begin{align*}
  &\hat{V}_{\cfadj [q]} =  \hat V\Big((\hat{\varepsilon}_{i})_{i\in \ms_{[q]}},\bs{Z}_{[q]};\prob_{\bs{Z}_{[q]}\mid \ms_{[1]},\ms_{[2]}} \Big),
\end{align*}
where $\hat{\varepsilon}_{i}$ is the cross-fitted residual:
\[
\hat{\varepsilon}_{i} = Y_i-\sum_{q\in \{1,2\}}\sum_{z\in \{0,1\}}\hat{f}_{[-q]z}(\bs{x}_i)I(Z_i = z,i\in\ms_{[q]}).
\]
For example, $\prob_{\bs{Z}_{[q]}\mid \ms_{[1]},\ms_{[2]}} = \CR(N_{[q]}, N_{[q]1})$ under the Split-by-treatment algorithm in CREs, and $\hat{V}_{\cfadj [q]}$ can be constructed by the variance estimator under CREs. Combining the variance estimators from the two subsamples, we obtain an estimator of $\Var(\hat{\tau}_{\oraclecf})$:
\[
\hat{V}_{\cfadj} = \sum_{q\in \{1,2\}} \frac{N_{[q]}^2}{N^2}\hat{V}_{\cfadj [q]}.
\]
In Sections~\ref{sec:sub-BRE}--\ref{sec:sub-MPE}, we give specific examples of $\hat{V}_{\cfadj}$ of different cross-fitting algorithms.

We now demonstrate the validity of the corresponding confidence intervals.
Let $z_{1-\alpha/2}$ denote the $(1-\alpha/2)$-th quantile of a standard normal distribution. Let $\plim$ denote the probability limit.

\begin{theorem}
\label{thm:valid-confidence-interval-2}
    Under Assumptions~\ref{a:conditional-independent-subsamples}--\ref{a:nondegenerate-variance}, if 
    \begin{align}
    \label{eq:conservative-variance}
       & \plim_{N \rightarrow \infty} N \{\hat{V}_{\cfadj} -\Var(\hat{\tau}_{\oraclecf})\} \geq 0, \\
    \label{eq:clt-for-oracle-cf}
        &(\hat{\tau}_{\oraclecf}-\bar{\tau})/\Var(\hat{\tau}_{\oraclecf})^{1/2} \rightsquigarrow \mathcal{N}(0,1),
    \end{align}
 then the confidence interval $    [\hat{\tau}_{\cfadj} - z_{1-\alpha/2}\hat{V}_{\cfadj}^{1/2}, \hat{\tau}_{\cfadj} + z_{1-\alpha/2}\hat{V}_{\cfadj}^{1/2}]$ has an asymptotic coverage rate of at least $1-\alpha$.
\end{theorem}

\Cref{thm:valid-confidence-interval-2} demonstrates that the confidence interval is asymptotically conservative under some regularity conditions. Since we have verified Assumptions~\ref{a:conditional-independent-subsamples}--\ref{a:nondegenerate-variance} in previous sections, it suffices to verify Conditions \eqref{eq:conservative-variance}--\eqref{eq:clt-for-oracle-cf}. We will verify these conditions under BRE, CRE, SRE, and MPE below. 

\subsection{Bernoulli randomized experiments}
\label{sec:sub-BRE}

For BREs, let 
 $D_i = Z_i Y_i/r_1 - (1-Z_i) Y_i/(1-r_1).$
The HT estimator and its variance estimator are
\[
\hat{\tau}_{\HT} =\frac{1}{N} \sumi D_i,\quad \hat{V}((Y_i)_{i=1}^N,(Z_i)_{i=1}^N;{\BT(N,r_1)}) = \frac{1}{N^2}\sum_{i=1}^N (D_i - \hat{\tau}_{\HT})^2.
\]
Consider the sample-splitting algorithm in BRE in \Cref{example:algorithm-bernoulli-trial}. Let 
  \[
  \hat{D}_{\varepsilon,i} =  \frac{Z_i \hat{\varepsilon}_i}{r_1} - \frac{(1-Z_i) \hat{\varepsilon}_i}{1-r_1},\quad  \hat{\bar{D}}_{\varepsilon,[q]} = \frac{1}{N_{[q]}}\sum_{i\in \ms_{[q]}}\hat{D}_{\varepsilon,i}
  \]
 By \Cref{prop:BT}, we have $\prob_{\bs{Z}_{[q]}\mid \ms_{[1]},\ms_{[2]}} = \BT(N_{[q]},r_1)$ and then
 \[
 \hat{V}_{\cfadj [q]} = \frac{1}{N_{[q]}^2} \sum_{i\in \ms_{[q]}} (\hat{D}_{\varepsilon,i} - \hat{\bar{D}}_{\varepsilon,[q]})^2.
 \]

% \begin{assumption}
% \label{a:constant-sample-split-proportion}
%     For $q=1,2$, $N_{[q]}/N$ tends to a constant in $(0,1)$. 
% \end{assumption}
% \Cref{a:constant-sample-split-proportion} requires that there are enough units in the two subsamples.

\begin{assumption}
    \label{a:regularity-condition-bernoulli-trial}
    For $z=0,1$, $N^{-1} \sumi \{\varepsilon_i(z)\}^4 = O(1)$.
\end{assumption}

\Cref{a:regularity-condition-bernoulli-trial} is a bounded fourth moment condition on $\varepsilon_i(z)$, which is widely used for deriving finite sample asymptotic theory \citep[see, e.g.,][]{Lin2013Agnostic,guo2023generalized,cohen2020no,su2023decorrelation}.
% \Cref{a:regularity-condition-bernoulli-trial} is a regularity condition for the oracle residuals Which is implied by the bounded fourth-moment condition used by \cite{su2023decorrelation}. 
Let 
$$
\tau_{\varepsilon, i} = \varepsilon_i(1) - \varepsilon_i(0), \quad \bar{\tau}_{\varepsilon} = \frac{1}{N}\sumi \{  \varepsilon_i(1) -  \varepsilon_i(0)  \}, \quad \Delta = \frac{1}{N-1}\sumi (\tau_{\varepsilon,i} - \bar{\tau}_{\varepsilon})^2.
$$
\Cref{prop:valid-inference-bt} below justifies Conditions \eqref{eq:conservative-variance}--\eqref{eq:clt-for-oracle-cf} under BREs.

\begin{proposition}
\label{prop:valid-inference-bt}
   Under BREs and Assumptions \ref{a:negligible-estimateion-error-for-prediction-function}, \ref{a:nondegenerate-variance}, and \ref{a:regularity-condition-bernoulli-trial}, Conditions \eqref{eq:conservative-variance}--\eqref{eq:clt-for-oracle-cf} hold for the sample-splitting algorithm in BRE in \Cref{example:algorithm-bernoulli-trial}. In particular,
   $
       N\{\hat{V}_{\cfadj} - \Var(\hat{\tau}_{\oraclecf})\} -\Delta =  \op(1).
   $
\end{proposition}

\subsection{Completely randomized experiments}
\label{sec:example-CR}
Let ${\bar{Y}}_{z} = N_z^{-1} \sum_{i:Z_i=z} Y_i$ be the sample mean of $Y_i(z)$. Under CREs, we have $\hat{\tau}_{\HT} = \bar{Y}_1 - \bar{Y}_0$ and
\begin{align*}
% &\hat{\tau}_{\HT} = \sum_{i=1}^N \Big\{\frac{Z_i Y_i}{N_1} - \frac{(1-Z_i) Y_i}{N_0}\Big\};\\
&\hat{V}((Y_i)_{i=1}^N,(Z_i)_{i=1}^N;{{\CR(N,N_1)}}) =\sum_{i=1}^N \Big\{\frac{Z_i (Y_i - {\bar{Y}}_{1})^2}{N_1(N_1-1)} + \frac{(1-Z_i) (Y_i - {\bar{Y}}_{0})^2}{N_0(N_0-1)}\Big\}.
\end{align*}
Since  $\prob_{\bs{Z}_{[q]}\mid \ms_{[1]},\ms_{[2]}} = \CR(N_{[q]},N_{[q]1})$ by \Cref{prop:CR}, we have
\begin{align*}
       &\hat{V}_{\cfadj[q]} = \sum_{i\in \ms_{[q]}} \frac{Z_i (\hat{\varepsilon}_i - \hat{\bar{\varepsilon}}_{[q]1})^2}{N_{[q]1}(N_{[q]1}-1)} + \frac{(1-Z_i) (\hat{\varepsilon}_i - \hat{\bar{\varepsilon}}_{[q]0})^2}{N_{[q]0}(N_{[q]0}-1)},\quad \hat{\bar{\varepsilon}}_{[q]z} = \frac{1}{N_{[q]z}}\sum_{i:i\in \ms_{[q]},Z_i=z} \hat{\varepsilon}_i. 
\end{align*}

% \begin{assumption}
% \label{a:regularity-condition-for-outcome}
% For $z=0,1$,   $N^{-1} \sumi \{\varepsilon_i(z)  - \bar{\varepsilon}(z)  \}^2 = O(1)$ and $N^{-1/2}\max_{i=1}^N |\varepsilon_i(z) \liu{ - \bar{\varepsilon}(z) } | = o(1)$, where $\bar{\varepsilon}(z) = N^{-1} \sum_{i=1}^N \varepsilon_i(z)$.
% \end{assumption}

% \Cref{a:regularity-condition-for-outcome} is a standard assumption for deriving the finite-population central limit theory \citep{li2017general,Li9157,lu2024tyranny}.

\begin{assumption}
\label{a:constant-treated-probability}
   There exists a constant $c \in (0,0.5)$, such that $N_{[q]z}/N \in (c,1-c)$ for $q=1,2$ and $z=0,1$.
\end{assumption}

\Cref{a:constant-treated-probability} ensures that there are enough units in both subsamples, and we have enough units in both treatment arms $z\in\{0,1\}$ in each subsample. \Cref{prop:valid-inference-cr} below justifies Conditions \eqref{eq:conservative-variance}--\eqref{eq:clt-for-oracle-cf} under CREs.

\begin{proposition}
\label{prop:valid-inference-cr}
 Under CREs and Assumptions \ref{a:negligible-estimateion-error-for-prediction-function} and \ref{a:nondegenerate-variance}--\ref{a:constant-treated-probability}, if $\Delta$ (Defined in \Cref{prop:valid-inference-bt}) and $N
 \cov\big\{(N^{-1}N_{[q]}\hat{\mu}_{\oraclecf,[q]}(z))_{q=1,2;z=0,1}\big\}$ have finite limits, then Conditions \eqref{eq:conservative-variance}--\eqref{eq:clt-for-oracle-cf} hold for the split-by-treatment algorithm in \Cref{example:algorithm-split-by-treatment-sr}. In particular, 
 $
         N\{\hat{V}_{\cfadj} - \Var(\hat{\tau}_{\oraclecf})\} -\Delta =  \op(1).
 $
\end{proposition}

\subsection{Stratified randomized experiments}
\label{sec:sub-SRE}
Let $\bar{Y}_{\setk z} = N_{\setk z}^{-1}\sum_{i:Z_i=z, A_i=k} Y_i$. Under SREs with $N_{\setk z} \geq 2$, we have $\hat{\tau}_{\HT} =  \sum_{k=1}^K ({N_{\setk}}/{N})(\bar{Y}_{\setk 1} - \bar{Y}_{\setk 0})$ and
\begin{align*}
   &\hat{V}((Y_i)_{i=1}^N,(Z_i)_{i=1}^N;\SR(N,\bs{A},(N_{\setk 1})_{k \in \mk})) \\
   =& \sum_{k=1}^K\frac{N_{\setk}^2}{N^2}\sum_{i:A_i = k} \Big\{\frac{Z_i (Y_i - {\bar{Y}}_{\setk 1})^2}{N_{\setk 1}(N_{\setk 1}-1)} + \frac{(1-Z_i) (Y_i - {\bar{Y}}_{\setk 0})^2}{N_{\setk 0}(N_{\setk 0}-1)}\Big\}.
\end{align*}
Consider the split-by-treatment algorithm with $N_{\setk [q] z} \geq 2$, $k\in \mk$, $q=1,2$, $z=0,1$. Since  $\prob_{\bs{Z}_{[q]}\mid \ms_{[1]},\ms_{[2]}} = \SR(N_{[q]},\bs{A}_{[q]},(N_{\setk [q] 1})_{k \in \mk})$ by \Cref{prop:CR}, we have
\begin{align*}
  & \hat{V}_{\cfadj,[q]} = \sum_{k=1}^K\frac{N_{\setk [q]}^2}{N_{[q]}^2}\sum_{i:A_i = k} \Big\{\frac{Z_i (\hat{\varepsilon}_i - \hat{\bar{\varepsilon}}_{\setk [q] 1})^2}{N_{\setk [q] 1}(N_{\setk [q] 1}-1)} + \frac{(1-Z_i) (\hat{\varepsilon}_i - \hat{\bar{\varepsilon}}_{\setk [q] 0})^2}{N_{\setk [q] 0}(N_{\setk [q] 0}-1)}\Big\},\\
  & \hat{\bar{\varepsilon}}_{\setk [q] z} = \frac{1}{N_{\setk [q] z}}\sum_{i: A_i=k, i\in \ms_{[q]}, Z_i=z} \hat{\varepsilon}_i.
\end{align*}

% \begin{assumption}
% \label{a:regularity-condition-for-outcome-stratified-randomized-experiments}
% For $z=0,1$,   $N^{-1}\sum_{k=1}^K \sum_{i: A_i=k} \{\varepsilon_i(z)  - \bar{\varepsilon}_{\setk}(z) \}^2 = O(1)$ and $N^{-1/2}\max_{k=1}^K$ $\max_{i:A_i=k} |\varepsilon_i(z)  - \bar{\varepsilon}_{\setk}(z) | = o(1)$, where $\bar{\varepsilon}_{\setk}(z) = N_{\setk}^{-1} \sum_{i:A_i=k} \varepsilon_i(z)$.
% \end{assumption}

\begin{assumption}
\label{a:constant-probability-stratified-randomized-experiments}
    For $k\in \mk$, $q=1,2$, $z=0,1$, we have $N_{\setk 
[q] z}\geq 2$ and there exists a constant $c\in (0,0.5)$, such that $N_{\setk [q] z}/N_{\setk} \in (c,1-c)$.
\end{assumption}

\Cref{a:constant-probability-stratified-randomized-experiments} requires that \Cref{a:constant-treated-probability} holds for each stratum. Define
\[
\Delta_{\SR} = \sum_{k=1}^K \frac{N_{\setk}}{N(N_{\setk}-1)}\sum_{i:A_i=k}(\tau_{\varepsilon,i} - \bar{\tau}_{\varepsilon,\setk})^2, \quad \bar{\tau}_{\varepsilon,\setk} = \frac{1}{N_{\setk}} \sum_{i\in \setk}\tau_{\varepsilon,i}.
 \]
\Cref{prop:valid-inference-sr} below justifies Conditions \eqref{eq:conservative-variance}--\eqref{eq:clt-for-oracle-cf} under SREs. 

\begin{proposition}
\label{prop:valid-inference-sr}
Under SREs, suppose that Assumptions \ref{a:negligible-estimateion-error-for-prediction-function}, \ref{a:nondegenerate-variance}--\ref{a:regularity-condition-bernoulli-trial}, and \ref{a:constant-probability-stratified-randomized-experiments} hold, and $\Delta_{\SR}$ and $N
 \cov\big\{( N^{-1} N_{[q]}\hat{\mu}_{\oraclecf,[q]}(z))_{q=1,2;z=0,1}\big\}$ have finite limits. Then Conditions \eqref{eq:conservative-variance}--\eqref{eq:clt-for-oracle-cf} hold for the split-by-treatment algorithm in \Cref{example:algorithm-split-by-treatment-sr}. In particular, $N\{\hat{V}_{\cfadj} - \Var(\hat{\tau}_{\oraclecf})\} - \Delta_{\SR} =  \op(1)$.
\end{proposition}

\subsection{Mathced-pairs experiments}
\label{sec:sub-MPE}

Let $\hat{\tau}_{\setk} = \sum_{i:A_i=k}\{Z_i Y_i-(1-Z_i)Y_i\}$.
%$2k-1$ and ${2k}$ denote the two units in the $k$th pair, $k=1,\ldots, N/2$, $E_k = I(Z_{2k-1}=1)$, and $\hat{\tau}_{\setk} = (Y_{2k-1}-Y_{2k})E_{k} + (Y_{2k}-Y_{2k-1})(1-E_{k})$. 
Then we have $\hat{\tau}_{\HT} = (2/N) \sum_{k=1 }^{N/2}\hat{\tau}_{\setk}$ and
\begin{align*}
   \hat{V}((Y_i)_{i=1}^N,(Z_i)_{i=1}^N;\MP(N,\bs{A})) = \frac{4}{(N-2)N}\sum_{k=1}^{N/2}(\hat{\tau}_{\setk}-\hat{\tau}_{\HT})^2. 
\end{align*}
Under the split-by-stratum algorithm, $\prob_{\bs{Z}_{[q]}\mid \ms_{[1]},\ms_{[2]}} = \MP(N_{[q]},\bs{A}_{[q]})$ by \Cref{prop:sr-2}. Therefore,
\begin{align*}
    & \hat{V}_{\cfadj [q]} = \frac{4}{(N_{[q]}-2)N_{[q]}}\sum_{k\in \mk_{[q]}}(\hat{\tau}_{\varepsilon, \setk}-\hat{\tau}_{\varepsilon, [q]})^2,\\
    & \hat{\tau}_{\varepsilon, \setk} = \sum_{i:A_i=k}\big\{Z_i\hat{\varepsilon}_i-(1-Z_i)\hat{\varepsilon}_i\big\}, \quad  \hat{\tau}_{\varepsilon, [q]} = \frac{2}{N_{[q]}}\sum_{k\in \mk_{[q]}}\hat{\tau}_{\varepsilon, \setk}.
\end{align*}

% \begin{assumption}
% \label{a:regularity-condition-for-outcome-matched-pair}
% $(2N^{-1})\sum_{k=1}^{N/2}\sum_{i:A_i=k} \varepsilon_i(z)^4 = O(1)$.
% \end{assumption}

\begin{assumption}
\label{a:regularity-condition-for-matched-pair}
    There exists a constant $c\in (0,0.5)$, such that $c < N_{[q]}/N < 1-c$.
\end{assumption}

\Cref{prop:valid-inference-matched-pair} below justifies Conditions \eqref{eq:conservative-variance}--\eqref{eq:clt-for-oracle-cf} under MPEs. 
\begin{proposition}
\label{prop:valid-inference-matched-pair}
Under MPEs, suppose that 
   $
  \Delta_{\MP} = \{ 4/ (N-2) \} \sum_{k=1}^{N/2} (\bar{\tau}_{\varepsilon,\setk} - \bar{\tau}_{\varepsilon})^2
   $
has a finite limit. Then, Assumptions \ref{a:negligible-estimateion-error-for-prediction-function}, \ref{a:nondegenerate-variance}--\ref{a:regularity-condition-bernoulli-trial}, and \ref{a:regularity-condition-for-matched-pair} imply Conditions \eqref{eq:conservative-variance}--\eqref{eq:clt-for-oracle-cf} for the split-by-stratum algorithm in \Cref{example:split-by-stratum-sr}. In particular, $N\{\hat{V}_{\cfadj} - \Var(\hat{\tau}_{\oraclecf})\} -\Delta_{\MP} =  \op(1)$.
\end{proposition}

\section{Simulation}\label{sec:simulation}
\subsection{Linear regression adjustment for ATE estimation}
In this section, we conduct a simulation to compare different linear regression adjustment methods in BREs. For the data-generating process, we follow \cite{su2023decorrelation} to generate finite populations with potential outcomes from linear models on the covariates:
\begin{align*}
    Y_i(1) = \bs{x}_i^\top \bs{\theta}_1 + \varepsilon_i(1), \quad Y_i(0) =  \varepsilon_i(0), \quad i = 1, \dots, N, 
\end{align*}
where $\bs{x}_i\in\mathbb{R}^d$ is generated by centering a set of independent random vectors $\tilde{\bs{x}}_i$ with i.i.d. elements from the Student's $t$-distribution with degree-of-freedom $2$. $\bs{\theta}_1 \in \mathbb{R}^d$ is a vector of all ones normalized by $\sqrt{d}$, i.e., $\bs{\theta} = (1,\dots,1)^\top/\sqrt{d}$. We fix the sample size $N = 1,500$ and vary the dimension $d = \lfloor N^\gamma \rfloor$, where $\gamma \in \{0.50, 0.55, 0.60, 0.65, 0.70, 0.75\}$. 
For $\varepsilon_i(1)$, we generate the noise in bias-maximizing style, following \citet[][Appendix D]{su2023decorrelation} and \citet[][Section 4.4]{lei2021regression}. For $\varepsilon_i(0)$, we simply generate random Gaussian noise from $N(0, 0.01^2)$. For each population, we fix the potential outcomes and randomize the treatment $Z_i\sim\operatorname{Bernoulli}(0.5)$ for $1,000$ rounds.

We compare five different methods: (i) Cross-fitting, which is the method proposed in the current work. We follow Example \ref{example:introduce-bernoulli-trial} and set $\pi = 0.5$ to get a balanced split over the data. (ii) Decorrelation method, proposed by \cite{su2023decorrelation}. We follow the proposal therein to construct the variance estimator and tune the hyperparameters.  (iii) Difference-in-means, which is the simple estimator that does not adjust for pretreatment covariates. (iv) Debiased estimator, which is proposed by \cite{lei2021regression}. We also apply the variance estimator proposed therein. (v) Lin's regression, which is the interacted regression adjustment approach proposed by \citet{Lin2013Agnostic}. For these methods, we compare their performance based on their mean squared error (MSE), $95\%$ coverage rate, and the ratio between the estimated variance and the true variance (the variance inflation ratio). We run simulations over $50$ random seeds and report the median of these metrics across these populations. 

\begin{figure}[ht!]
    \centering
    \includegraphics[width=1.0\textwidth]{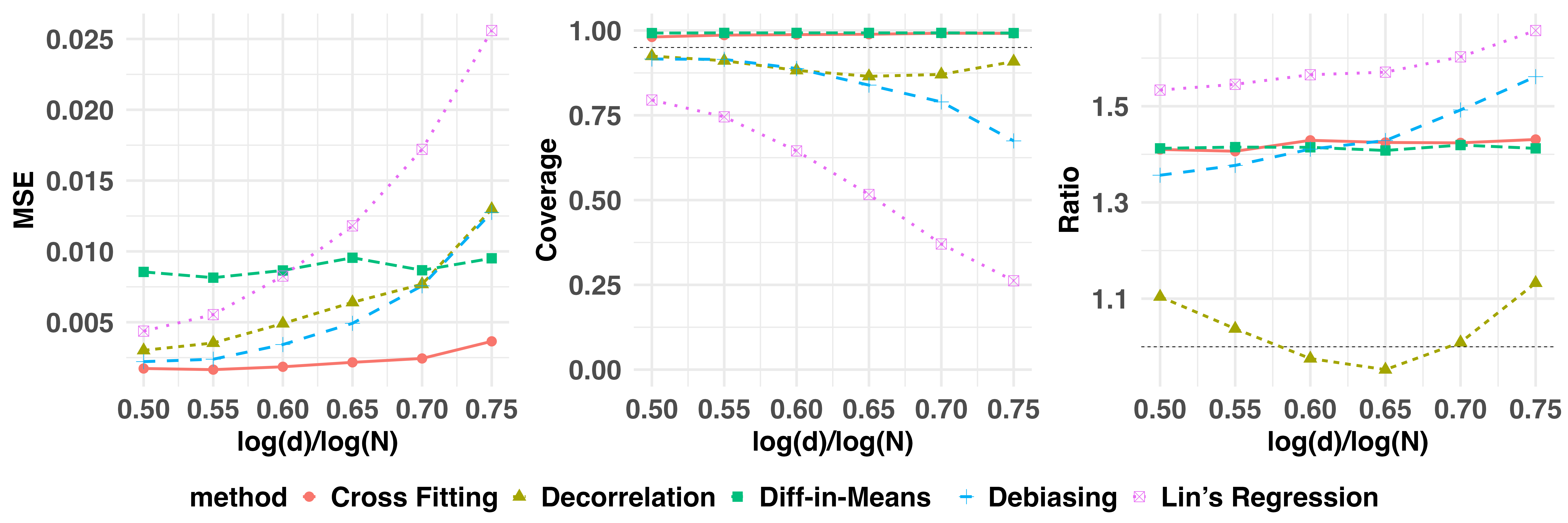}
    \caption{Simulation for linear regression adjustment for ATE estimation. From left to right, the panels report MSE, 95\% coverage rate, and variance inflation ratio for the five methods.   }
    \label{fig:lm}
\end{figure}

From Figure \ref{fig:lm}, we can see that under the given sample size, as the dimension of the covariates gets higher, Lin's regression eventually leads to higher MSE than the simple difference-in-means estimator. The debiasing method and the decorrelation method both perform relatively well for a moderate dimension, yet their MSEs still increase over difference-in-means when the dimension $d$ gets higher than $n^{0.7}$. The cross-fitting method, on the contrary, achieves the smallest MSE across all the compared dimensions. For type I error control, both difference-in-means and cross-fitting maintain validity across all dimensions. The decorrelation method leads to a slight under-coverage under the implemented sample size. Such under-coverage is mostly due to an under-estimation of the variance of the estimator in finite sample, as validated by the right panel of Figure \ref{fig:lm}, while all the other methods achieve conservative estimation of the variance with similar extent of inflation. However, Lin's regression and the debiasing method both fail to control type I error when $d$ is large compared to $N$.  
% \begin{figure}[!ht]
%     \centering
%     \subfigure[MSE]{
%         \includegraphics[width=0.47\textwidth]{lm_mse.png}
%         \label{fig:lm-mse}
%     }
%     \subfigure[Coverage]{
%         \includegraphics[width=0.47\textwidth]{lm_cov.png}
%         \label{fig:lm-cov}
%     }
%     \subfigure[Variance ratio]{
%         \includegraphics[width=0.6\textwidth]{lm_ratio.png}
%         \label{fig:lm-ratio}
%     }
%     \caption{}
%     \label{fig:lm}
% \end{figure}

\subsection{Nonlinear/nonparametric adjustment for ATE estimation}

\begin{figure}[ht!]
    \centering
    \includegraphics[width=1.0\textwidth]{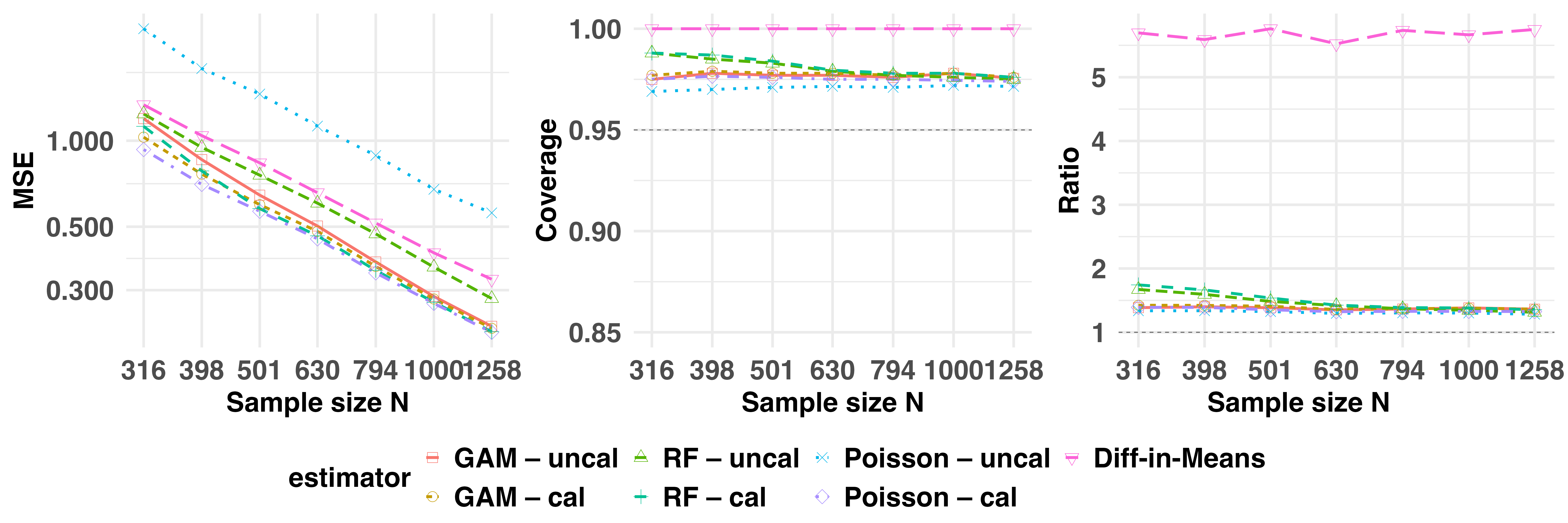}
    \caption{Simulation results for nonlinear/nonparametric adjustment for ATE estimation. From left to right, the panels report MSE, 95\% coverage rate, and variance inflation ratio for the methods.}
    \label{fig:np}
\end{figure}

In this section, we conduct a simulation to show the performance of nonlinear/nonparametric adjustment with conditional cross-fitting in BREs. We also illustrate the benefits of combining the strategy with the no-harm calibration procedure as proposed in the Supplementary Material. The simulation setup follows that of \cite{cohen2020no}. Specifically, we generate pretreatment covariates $x_i$'s from a uniform distribution on $[-5, 5]$. Then the potential outcomes are generated independently from the following model: 
\begin{align*}
    Y_i(1) \sim \textup{Poisson}\{\exp(x_i)\}, 
    \quad 
    Y_i(0) \sim \textup{Poisson}\{72 - 0.45\exp(x_i)\}. 
\end{align*}
The treatments are assigned as a Bernoulli randomization with mean $0.8$. We evaluate four different methods: (i) Difference-in-means; (ii) Generalized additive model (GAM); (iii) Random forest (RF); and (iv) Poisson regression. The potential outcome generation process guarantees a correct Poisson regression model for the treatment group but a wrong model for the control group. For (ii)-(iv), we apply conditional cross-fitting and compare the methods with/without using no-harm calibration. We vary the sample size $N = \lfloor 10^\gamma \rfloor$ with $\gamma \in \{2.5, 2.6, 2.7, 2.8, 2.9, 3.0, 3.1\}$, run simulations over $50$ random seeds, and report the median of these metrics across these populations. The results are reported in Figure \ref{fig:np}.

For MSE, all methods follow a root-$n$ convergence rate. Without calibration, Poisson regression and random forest lead to a large MSE that is higher than or comparable with the difference-in-means estimator. In contrast, the no-harm calibration serves as a remedy for decreasing the MSE. The GAM model works relatively well without calibration mostly because it applies an additive B-spline regression that works similarly to linear models, thus saving the efforts of calibration. In addition, all methods lead to valid type I error control, with the difference-in-means estimator being the most conservative. This is because the variance estimator is overly conservative without adjusting for covariates, as validated from the right panel of Figure \ref{fig:np}.  
% \begin{figure}[!ht]
%     \centering
%     \subfigure[MSE]{
%         \includegraphics[width=0.47\textwidth]{np_mse.png}
%         \label{fig:np-mse}
%     }
%     \subfigure[Coverage]{
%         \includegraphics[width=0.47\textwidth]{np_cov.png}
%         \label{fig:np-cov}
%     }
%     \subfigure[Variance ratio]{
%         \includegraphics[width=0.6\textwidth]{np_ratio.png}
%         \label{fig:np-ratio}
%     }
%     \caption{}
%     \label{fig:np}
% \end{figure}

\subsection{Optimal sample-splitting algorithm under CREs}
\begin{figure}[ht!]
    \centering
    \includegraphics[width=0.9\linewidth]{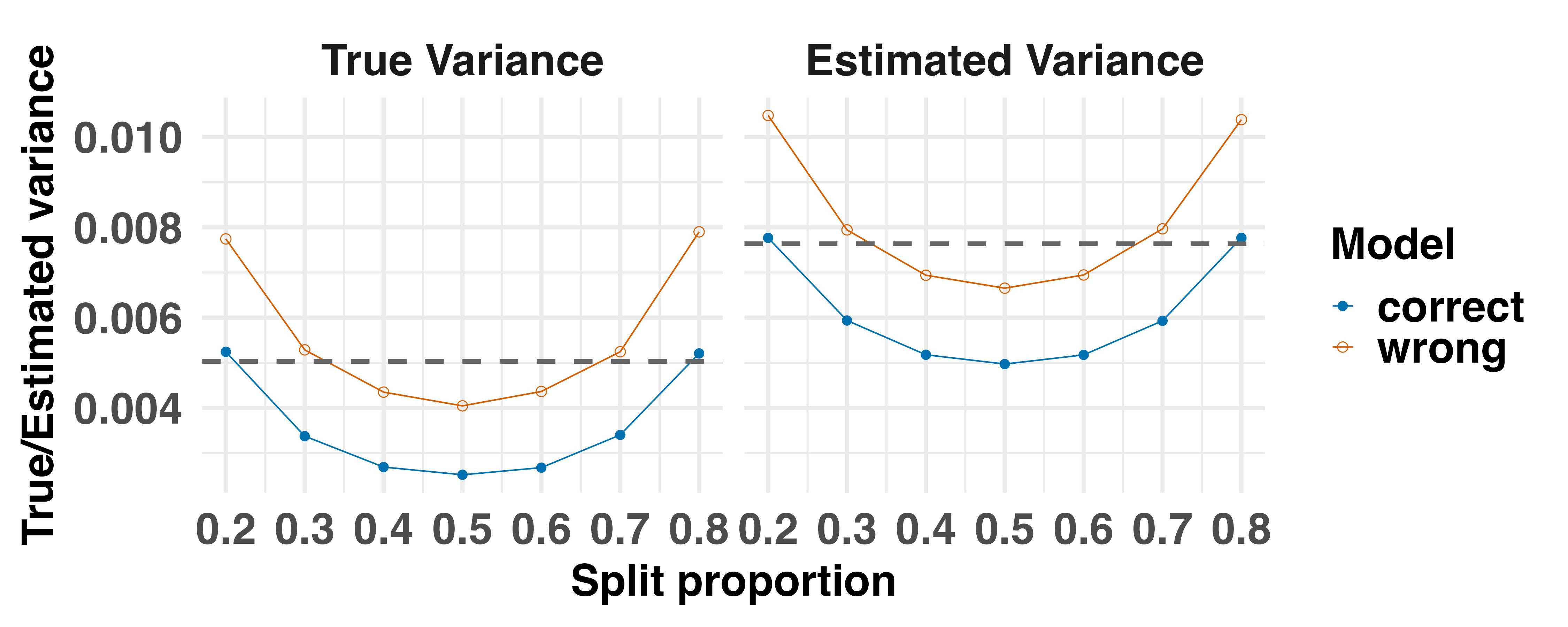}
    \caption{True and estimated variances against different splitting proportions under CREs. The left panel reports the true variance and the right panel reports the estimated variance. The dashed horizontal line reports the corresponding metric for the difference-in-means estimator. }
    \label{fig:split}
\end{figure}

In the third experiment, we validate our optimal sample splitting results in Section \ref{sec:optimal-algorithm} in numerical experiments based on CREs. We generate pretreatment covariates $\bs{x}_i$ from a bivariate standard Gaussian distribution. Then we simulate potential outcomes using the following Poisson regression model:
\begin{align*}
    Y_i(1) \sim \textup{Poisson}\{\exp(\bs{x}_i^\top \bs{\beta}_1)\}, \quad 
    Y_i(0) \sim \textup{Poisson}\{\exp(\bs{x}_i^\top \bs{\beta}_0)\},\quad
    \text{ for } i = 1,\dots,N,
\end{align*}
where $\bs{\beta}_1$ and $\bs{\beta}_0$ are generated from a pair of normalized random vectors whose elements are sampled independently from a Student's $t$-distribution with degree-of-freedom $3$. For each finite population, we set $N = 1,000$, then randomize $ N_1 = 500$ and $N_0 = 500$ units to the treatment arm and the control arm, respectively, under the mechanism described in Example \ref{example:introduce-CR}. To apply the cross-fitting algorithm, we split each realization into two subsets, $\ms_{[1]}$ and $\ms_{[2]}$, with $N_{[1]} = N_{[2]} = 500$. We then vary the proportion of treated units, $N_{[1]1}/N_{[1]} = r$ and $N_{[2]1}/N_{[2]} = 1 - r$, in these two subsets within the set of ratios: $r\in\{0.2, 0.3, 0.4, 0.5, 0.6, 0.7, 0.8\}$. In particular, $r = 0.5$ satisfies Assumption \ref{a:constant-treatment-assignment-probability} and is expected to generate the optimal splitting according to Theorem \ref{thm:constant-probability-optimal}. For a given ratio, we apply conditional cross-fitting (Algorithm \ref{algorithm:conditional-cross-fitting-algorithm}) with two Poisson regressions under no-harm calibration: one with correct model specification and the other with a mis-specified working model that only includes one covariate. To compare the true and estimated variances under different splitting proportions, we generate $50$ populations and implement $1,000$ rounds of randomization under each population to obtain the variance of the estimator as well as the mean of the variance estimator. In Figure \ref{fig:split}, we report the median of these metrics across $50$ populations. 

From Figure \ref{fig:split}, we can verify that, under CREs, Algorithm \ref{algorithm:conditional-cross-fitting-algorithm} achieves optimal variance under the splitting proportion that satisfies Assumption \ref{a:constant-treatment-assignment-probability}. In addition, we also empirically verify that the estimated variance achieves optimal performance when $r = 0.5$. In contrast, splitting proportions that are far away from the optimal choice will lead to larger true and estimated variances. As shown by the results, the estimator with a correct model specification but an extreme splitting proportion ($r = 0.2$ or $0.8$) demonstrates an inferior behavior than that with a wrong model specification but an optimal splitting ($r = 0.5$), as well as the difference-in-means estimator. This result emphasizes the use of the optimal splitting strategy to guarantee a better efficiency in practice. 

\section{Conclusion}
\label{sec:discussion}
In this paper, we propose the conditional cross-fitting method for estimating the ATE in randomized experiments. This method enables unbiased and efficient estimation of the ATE using machine learning methods while maintaining valid inference, without requiring i.i.d. assumptions on the data-generating process. Our framework relies crucially on the sample-splitting algorithm satisfying \Cref{a:conditional-independent-subsamples}.

We have developed feasible algorithms for Bernoulli randomized experiments, completely randomized experiments, and stratified randomized experiments. However, it remains unclear whether feasible algorithms can be constructed under other designs, such as rerandomization \citep{morgan2012rerandomization} and the Gram–Schmidt walk design \citep{Harshaw2024Balancing}.

% Furthermore, our variance estimation relies on the stability of the estimated prediction functions (\Cref{a:negligible-estimateion-error-for-prediction-function}). One potential approach is to adopt \(K\)-fold cross-fitting, where a proportion of \((K-1)/K\) of the data is used to estimate the prediction function; with a slightly larger \(K\), \Cref{a:negligible-estimateion-error-for-prediction-function} is more likely to hold.

Our framework assumes no interference, i.e., an individual's outcome is only affected by their own treatment. When interference is present (e.g., experimental units interact through networks), it remains an open question how to extend the conditional cross-fitting framework to such settings.

\bibliographystyle{agsm}
\bibliography{causal}

@article{lei2021regression,
  title={Regression adjustment in completely randomized experiments with a diverging number of covariates},
  author={Lei, Lihua and Ding, Peng},
  journal={Biometrika},
  volume={108},
  pages={815--828},
  year={2021}
}

@article{Lin2013Agnostic,
	author = {Lin, Winston},
	journal = {The Annals of Applied Statistics},
	pages = {295--318},
	title = {Agnostic notes on regression adjustments to experimental data: {R}eexamining {F}reedman's critique},
	volume = {7},
	year = {2013}}

@book{imbens2015causal,
  title={Causal Inference for Statistics, Social, and Biomedical Sciences: An Introduction},
  author={Imbens, Guido W and Rubin, Donald B},
  year={2015},
  publisher={Cambridge University Press}
}

@article{bloniarz2016lasso,
  title={Lasso adjustments of treatment effect estimates in randomized experiments},
  author={Bloniarz, Adam and Liu, Hanzhong and Zhang, Cun-Hui and Sekhon, Jasjeet S and Yu, Bin},
  journal={Proceedings of the National Academy of Sciences},
  volume={113},
  pages={7383--7390},
  year={2016},
}

@article{negi2021revisiting,
  title={Revisiting regression adjustment in experiments with heterogeneous treatment effects},
  author={Negi, Akanksha and Wooldridge, Jeffrey M},
  journal={Econometric Reviews},
  volume={40},
  pages={504--534},
  year={2021},
}

@article{wager2016high,
  title={High-dimensional regression adjustments in randomized experiments},
  author={Wager, Stefan and Du, Wenfei and Taylor, Jonathan and Tibshirani, Robert J},
  journal={Proceedings of the National Academy of Sciences},
  volume={113},
  pages={12673--12678},
  year={2016}
}

@article{liu2020regression,
  title={Regression-adjusted average treatment effect estimates in stratified randomized experiments},
  author={Liu, Hanzhong and Yang, Yuehan},
  journal={Biometrika},
  volume={107},
  pages={935--948},
  year={2020}
}

@article{morgan2012rerandomization,
	author = {Morgan, K. L. and Rubin, D. B.},
	journal = {Annals of Statistics},
	pages = {1263--1282},
	title = {Rerandomization to improve covariate balance in experiments},
	volume = {40},
	year = {2012}}

@article{guo2023generalized,
  title={The generalized oaxaca-blinder estimator},
  author={Guo, Kevin and Basse, Guillaume},
  journal={Journal of the American Statistical Association},
  volume={118},
  pages={524--536},
  year={2023}
}

@article{chiang2023regression,
  title={Regression adjustment, cross-fitting, and randomized experiments with many controls},
  author={Chiang, Harold D and Matsushita, Yukitoshi and Otsu, Taisuke},
  journal={arXiv preprint arXiv:2302.00469v4},
  year={2025}
}

@article{su2023decorrelation,
  title={A decorrelation method for general regression adjustment in randomized experiments},
  author={Su, Fangzhou and Mou, Wenlong and Ding, Peng and Wainwright, Martin J},
  journal={arXiv preprint arXiv:2311.10076},
  year={2023}
}

@article{freedman2008regression,
  title={On regression adjustments to experimental data},
  author={Freedman, David A},
  journal={Advances in Applied Mathematics},
  volume={40},
  pages={180--193},
  year={2008},
  publisher={Elsevier}
}

@article{lu2023debiased,
  title={Debiased regression adjustment in completely randomized experiments with moderately high-dimensional covariates},
  author={Lu, Xin and Yang, Fan and Wang, Yuhao},
  journal={arXiv preprint arXiv:2309.02073},
  year={2023}
}

@article{chernozhukov2018double,
    author = {Chernozhukov, Victor and Chetverikov, Denis and Demirer, Mert and Duflo, Esther and Hansen, Christian and Newey, Whitney and Robins, James},
    title = "{Double/debiased machine learning for treatment and structural parameters}",
    journal = {The Econometrics Journal},
    volume = {21},
    pages = {C1-C68},
    year = {2018},
}

@article{cohen2020no,
    author = {Cohen, P L and Fogarty, C B},
    title = "{No-harm calibration for generalized Oaxaca–Blinder estimators}",
    journal = {Biometrika},
    volume = {111},
    pages = {331-338},
    year = {2023},
}

@article{lu2024tyranny,
  title={Tyranny-of-the-minority regression adjustment in randomized experiments},
  author={Lu, Xin and Liu, Hanzhong},
  journal={Journal of the American Statistical Association},
  volume = {in press},
  year={2024}
}

@article{liu2KRandomization2024,
author = {Liu, Hanzhong and Ren, Jiyang and Yang, Yuehan},
title = {Randomization-based Joint Central Limit Theorem and Efficient Covariate Adjustment in Randomized Block $2^{K}$ Factorial Experiments},
journal = {Journal of the American Statistical Association},
volume = {119},
pages = {136--150},
year = {2024},

}

@article{fogarty2018regression,
  title={Regression-assisted inference for the average treatment effect in paired experiments},
  author={Fogarty, Colin B},
  journal={Biometrika},
  volume={105},
  pages={994--1000},
  year={2018}
}

@article{ding2021frisch,
  title={The {F}risch--{W}augh--{L}ovell theorem for standard errors},
  author={Ding, Peng},
  journal={Statistics \& Probability Letters},
  volume={168},
  pages={108945},
  year={2021},
}

@article{chang2024exact,
  title={Exact bias correction for linear adjustment of randomized controlled trials},
  author={Chang, Haoge and Middleton, Joel A and Aronow, PM},
  journal={Econometrica},
  volume={92},
  pages={1503--1519},
  year={2024}
}

@article{angelopoulos2023prediction,
  title={Prediction-powered inference},
  author={Angelopoulos, Anastasios N and Bates, Stephen and Fannjiang, Clara and Jordan, Michael I and Zrnic, Tijana},
  journal={Science},
  volume={382},
  pages={669--674},
  year={2023}
}

@article{
zrnic2024cross,
author = {Tijana Zrnic  and Emmanuel J. Candès },
title = {Cross-prediction-powered inference},
journal = {Proceedings of the National Academy of Sciences},
volume = {121},
pages = {e2322083121},
year = {2024},
}

@article{neyman1923,
	author = {Neyman, Jerzy},
	journal = {Statistical Science},
	pages = {465--72},
	title = {On the application of probability theory to agricultural experiments. \textsc{E}ssay on principles (with discussion). \textsc{S}ection 9 (translated). \textsc{R}eprinted},
	year = {1923}}

@article{wu2018loop,
  title={The LOOP estimator: Adjusting for covariates in randomized experiments},
  author={Wu, Edward and Gagnon-Bartsch, Johann A},
  journal={Evaluation Review},
  volume={42},
  pages={458--488},
  year={2018}
}

@article{farrell2021deep,
  title={Deep neural networks for estimation and inference},
  author={Farrell, Max H and Liang, Tengyuan and Misra, Sanjog},
  journal={Econometrica},
  volume={89},
  pages={181--213},
  year={2021}
}

@article{abadie2020sampling,
  title={Sampling-based versus design-based uncertainty in regression analysis},
  author={Abadie, Alberto and Athey, Susan and Imbens, Guido W and Wooldridge, Jeffrey M},
  journal={Econometrica},
  volume={88},
  pages={265--296},
  year={2020}
}

@article{wu2021design,
  title={Design-based covariate adjustments in paired experiments},
  author={Wu, Edward and Gagnon-Bartsch, Johann A},
  journal={Journal of Educational and Behavioral Statistics},
  volume={46},
  pages={109--132},
  year={2021}
}

@article{qu2025randomization,
  title={Randomization-based {Z}-estimation for evaluating average and individual treatment effects},
  author={Qu, Tianyi and Du, Jiangchuan and Li, Xinran},
  journal={Biometrika},
  volume={112},
  pages={asaf002},
  year={2025}
}

@article{pashley2021insights,
  title={Insights on variance estimation for blocked and matched pairs designs},
  author={Pashley, Nicole E and Miratrix, Luke W},
  journal={Journal of Educational and Behavioral Statistics},
  volume={46},
  pages={271--296},
  year={2021}
}

@article{tian2025stratified,
  title={Stratified Permutational Berry--Esseen Bounds and Their Applications to Statistics},
  author={Tian, Pengfei and Yang, Fan and Ding, Peng},
  journal={arXiv preprint arXiv:2503.13986},
  year={2025}
}

@article{Harshaw2024Balancing,
author = {Christopher Harshaw and Fredrik Sävje and Daniel A. Spielman and Peng Zhang},
title = {Balancing Covariates in Randomized Experiments with the Gram–Schmidt Walk Design},
journal = {Journal of the American Statistical Association},
volume = {119},
pages = {2934--2946},
year = {2024},
publisher = {ASA Website},
}

@book{ding2024book,
  title={A first course in causal inference},
  author={Ding, Peng},
  year={2024},
  publisher={Chapman and Hall/CRC}
}

\newpage

\appendix

\centerline{ \Large\bf SUPPLEMENTARY MATERIAL}
\vspace{1cm}

\spacingset{1.5}

\renewcommand{\theassumption}{S\arabic{assumption}}
\setcounter{assumption}{0}
\renewcommand{\thetheorem}{S\arabic{theorem}}
\setcounter{theorem}{0}
\renewcommand{\theproposition}{S\arabic{proposition}}
\setcounter{proposition}{0}
\renewcommand{\thelemma}{S\arabic{lemma}}
\setcounter{lemma}{0}
\renewcommand{\thedefinition}{S\arabic{definition}}
\setcounter{definition}{0}
\renewcommand{\thecondition}{S\arabic{condition}}
\setcounter{condition}{0}

\Cref{sec:additional-results} provides additional results, including a detailed comparison with the decorrelation method proposed by \cite{su2023decorrelation}, the no-harm calibration procedure, and the properties of several linear regression adjustments for SREs and MPEs.

\Cref{sec:proofs} provides proofs of the results in the main text (Theorems \ref{thm:unbiased}--\ref{thm:valid-confidence-interval-2}, Propositions \ref{prop:BT}--\ref{prop:valid-inference-matched-pair}) as well as proofs of the additional results (Propositions \ref{prop:stability-assumption-linear-SR-no-stratum-indicator}--\ref{mp:regression-of-outcome-difference-on-covariate-difference}).

\section{Additional results}
\label{sec:additional-results}

\subsection{Comparison with \cite{su2023decorrelation}}

 \cite{su2023decorrelation} proposed a decorrelation method for BREs. Define the index set of the treatment group $z$: $\ms_z = \{i: Z_i=z\}$. Their decorrelation algorithm partitions each treatment group $\ms_z$, $z\in \{0,1\}$, into overlapping subsets $\ms_{[1]z}$ and $\ms_{[2]z}$ such that $\ms_z = \ms_{[1]z}\cup \ms_{[2]z}$ but $\ms_{[1]z}\cap \ms_{[2]z} \ne \emptyset$. $\ms_{[1]z}$ and $\ms_{[2]z}$ are used in the 
prediction and estimation step, respectively. Let $r_0=1-r_1$. Although $\ms_{[1]z}$ and $\ms_{[2]z}$ overlap, a larger expected size of $\ms_{[1]z}$ corresponds to a smaller expected size of $\ms_{[2]z}$, as suggested by their construction:
$
r_z = \prob(i\in \ms_{[2]z}) + \prob(i\in \ms_{[1]z}) - \prob(i\in \ms_{[2]z})\prob(i\in \ms_{[1]z}), \ z\in \{0,1\}.
$ 
%\cite{su2023decorrelation} showed that $\hat{\tau}_{\dc}$ defined in \Cref{algorithm:decorrelation-algorithm} is unbiased. 
To ensure the decorrelated estimator has the same asymptotic variance as $\hat{\tau}_{\oracleadj}$, two conditions must be satisfied. First, the prediction function must be estimated with sufficient accuracy such that $\|\hat{f}_z - f_z^\ast\|_N = \op(1),$ 
where $\|f - g\|^2_N := N^{-1} \sumi \{f(\bs{x}_i)-g(\bs{x}_i)\}^2$ for two functions $f,g:\mathbb{R}^d\rightarrow \mathbb{R}$. This requires that the expected size of the subsample used in the prediction step, i.e., $N \prob(i\in\ms_{[1]z})$, to be 
sufficiently large. Second, because the decorrelated estimator uses only $\ms_{[2]z}$ in the estimation step while $\hat{\tau}_{\oracleadj}$ uses the full set $\ms_z$, the expected size of $\ms_{[2]z}$ must be close to that of $\ms_z$ to prevent the inflation of the asymptotic variance due to sample loss. That is, it must hold that $\prob(i\in \ms_{[2]z})\rightarrow r_z$, $z=0,1$. These two conditions introduce the first challenge: one must carefully balance $\prob(i \in \ms_{[1]z})$ and $\prob(i \in \ms_{[2]z})$. The optimal trade-off is generally nontrivial and depends on the specific function class to which $f_z^\ast$ belongs. Moreover, discarding part of the sample when estimating the ATE is sometimes criticized for being less efficient in finite samples than using the full sample. Our conditional cross-fitting algorithm solves these challenges.

Another challenge for \cite{su2023decorrelation}'s decorrelation procedure is that this procedure solely utilizes the subset $\{Y_i,\bs{x}_i\}_{i\in \ms_{[1]z}}$ to estimate the prediction function $f^\ast_z$. Consequently, $f^\ast_z$ is limited to prediction functions defined using $\{(Y_i(z),\bs{x}_i)\}_{i=1}^N$, thus excluding many other functions defined on $\{(Y_i(1), Y_i(0),\bs{x}_i)\}_{i=1}^N$. One such example is the no-harm calibration function, which conducts a further linear prediction using the base prediction function: 
\begin{align}
\label{eq:defition-of-cal-fast}
    &f^\ast_z(\bs{x}_i) = \alpha_{z}^\ast + \beta_{z}^\ast g^\ast_{0}(\bs{x}_i) + \gamma_{z}^\ast g^\ast_1(\bs{x}_i),\\
    \nonumber
    &(\alpha_{z}^\ast, \beta_{z}^\ast, \gamma_{z}^\ast) = \argmin_{(\alpha,{\beta}, \gamma)} \frac{1}{N}\sumi \{ Y_i(z) - \alpha - \beta g^\ast_{0}(\bs{x}_i) - \gamma g^\ast_1(\bs{x}_i) \}^2, \quad z=0,1,
\end{align}
where $g^\ast_z$ is the base prediction function defined on $\{Y_i(z),\bs{x}_i\}_{i=1}^N$, typically determined through risk minimization procedures. When \( g_z^\ast \) is a poor predictor, the asymptotic variance of \( \hat{\tau}_{\oracleadj} \) may exceed that of \( \hat{\tau}_{\HT} \). In contrast, the use of the calibration function \( f_z^\ast \) in $\hat{\tau}_{\oracleadj}$ ensures an asymptotic variance no larger than than both that obtained using \( g_z^\ast \) and that of \( \hat{\tau}_{\HT} \).
 Another example is the pooled regression adjustment which imposes common coefficients for both treatment and control groups \citep{negi2021revisiting}. A typical case is the tyranny-of-the-minority (ToM) regression, which uses a common regression coefficient $\bs{\beta}^\ast$ for both treatment and control groups \citep{lu2024tyranny}:
\begin{align*}
    &f^\ast_z(\bs{x}_i) = \alpha_z^\ast + \bs{x}_i^\top \bs{\beta}^\ast,\quad (\alpha_0^\ast,\alpha_1^\ast,\bs{\beta}^\ast) = \argmin_{(\alpha_0,\alpha_1,\bs{\beta})} \sum_{z\in \{0,1\}}\sumi \frac{1}{N_z}\{Y_i(z)-{\alpha}_z - \bs{x}_i^\top {\bs{\beta}}\}^2,
\end{align*}
where $N_z$ is defined in \Cref{example:introduce-CR}. Compared with the interacted regression \citep{Lin2013Agnostic}, which uses separate coefficients $\bs{\beta}_z^\ast$ for each group, \begin{align*}
    &f^\ast_z(\bs{x}_i) = \alpha_z^\ast + \bs{x}_i^\top \bs{\beta}^\ast_z,\quad (\alpha_z^\ast,\bs{\beta}^\ast_z) = \argmin_{(\alpha_z,\bs{\beta}_z)} \sumi \frac{1}{N}\{Y_i(z)-{\alpha}_z - \bs{x}_i^\top {\bs{\beta}_z}\}^2,
\end{align*}
ToM regression estimates fewer regression coefficients and yields better finite-sample performance. Our conditional cross-fitting algorithm resolves this limitation of the decorrelation procedure by enabling the use of such enhanced prediction functions while preserving design-based validity.

\subsection{No-harm calibration procedure}

We illustrate how to adapt the no-harm calibration procedure \citep{cohen2020no} to the cross-fitting algorithm in CREs. Recall the definition of the no-harm calibration function $f_z^\ast$ in \Cref{eq:defition-of-cal-fast}.
We compute $\hat{f}_{[-q]z}$ in \Cref{algorithm:no-harm-calibration} below. Since under BREs, $f^\ast_z(\bs{x}_i)$ defined in \Cref{eq:defition-of-cal-fast} also guarantees the variance reduction of $\hat{\tau}_{\oracleadj}$ compared to $\hat{\tau}_{\HT}$, \Cref{algorithm:no-harm-calibration} also works for BREs.

\begin{algorithm}
\caption{No-harm calibration under CRE}
\label{algorithm:no-harm-calibration}
\begin{algorithmic}[1]
\State \textbf{Input:} Dataset $ \{( Y_i,Z_i,\bs{x}_i)\}_{i=1}^N$. A sample split $(\ms_{[1]},\ms_{[2]})$. A function class $\mathcal{G}_z$ with its loss function $\ell_{i,z}\big(Y_i(z),\bs{x}_i, g\big)$ for $g\in \mathcal{G}_z$. The prediction function
    $
g_z^\ast = \argmin_{g\in \mathcal{G}_z} N^{-1}\sumi \ell_{i,z}\big(Y_i(z),\bs{x}_i, g\big).
$

\For {$q = 1, 2$}
    \State  Estimate the prediction function $\hat{g}_{[-q]z}$, $z\in \{0,1\},$ by 
\[
 \hat{g}_{[-q]z} = \argmin_{g\in \mathcal{G}_z} \sum_{i:i\in \ms_{[-q]},Z_i=z} \frac{1}{N_{[-q]z}}\ell_{i,z}\big(Y_i(z),\bs{x}_i, g\big).
\]
    
    \State Estimate the calibration function by
\begin{align*}
    &\hat{f}_{[-q]z} = \alpha_{[-q]z} + \beta_{[-q]z} \hat{g}_{[-q]1} + \gamma_{[-q]z}\hat{g}_{[-q]0},\\
    &(\alpha_{[-q]z}, \beta_{[-q]z}, \gamma_{[-q]z}) = \argmin_{(\alpha,\beta,\gamma)}\sum_{i:i\in \ms_{[-q]},Z_i=z} \frac{1}{N_{[-q]z}} \{Y_i(z) - \alpha - \beta \hat{g}_{[-q]1}(\bs{x}_i) - \gamma\hat{g}_{[-q]0}(\bs{x}_i)\}^2.
\end{align*}
   
\EndFor

\State \textbf{Return:} The estimated calibration function $\hat{f}_{[-q]z}$.
\end{algorithmic}
\end{algorithm}

In Sections~\ref{sec:linear-regression-adjustments-under-SRE} and \ref{sec:linear-regression-adjustments-under-MPEs}, we verify the stability assumption (\Cref{a:negligible-estimateion-error-for-prediction-function}) and investigate the efficiency gain for the cross-fitting algorithm with linear prediction functions under both SREs and MPEs. We leave the analysis of nonlinear functions for future work. Let $\bs{x}_i \in \mathbb R^{d}$ and we assume that $d$ is fixed.

\subsection{Linear regression adjustments of SREs}
\label{sec:linear-regression-adjustments-under-SRE}

For SREs, we consider three linear regression adjustments:

(1) OLS:
\begin{align*}
    & f^\ast_z(\bs{x}_i) = \bs{x}_i^\top \bs{\beta}_z^\ast + \alpha_{z}^\ast, \quad \textnormal{where } (\alpha_z^\ast,\bs{\beta}_z^\ast) = \argmin_{\alpha,\bs{\beta}} \frac{1}{N}\sumi \{Y_i(z)-\bs{x}_i^\top \bs{\beta} - \alpha\}^2 .
\end{align*}

Two weighted least squares (WLS) using stratification information:

(2) WLS with stratum indicators:
\begin{align*}
    & f^\ast_z(\bs{x}_i) = \bs{x}_i^\top \bs{\beta}_z^\ast + \sum_{k=1}^K \alpha_{\setk z}^\ast I(A_i = k), \quad \textnormal{where}\\
    & (\{\alpha_{\setk z}^\ast\}_{k\in \mk},\bs{\beta}_z^\ast) = \argmin_{\{\alpha_{\setk}\}_{k\in \mk},\bs{\beta}} \frac{1}{N}\sum_{k=1}^K \sum_{i: A_i = k} \{Y_i(z)-\bs{x}_i^\top \bs{\beta}- \alpha_{\setk}\}^2 \omega_{\setk z},
\end{align*}
with $\omega_{\setk z} = \sum_{q \in \{1,2\}} {N_{\setk [q]}^2}/\{N_{\setk [q]z}(N_{\setk}-1)\}.$

(3) ToM regression adjustment \citep{lu2024tyranny}:
\begin{align*}
     f^\ast_z(\bs{x}_i) = & \bs{x}_i^\top \bs{\beta}^\ast + \sum_{k=1}^K \alpha_{\setk z}^\ast I(A_i = k), \quad \textnormal{where}\\
    (\{\alpha_{\setk z}^\ast\}_{k\in \mk,z=0,1},\bs{\beta}^\ast) = & \argmin_{\{\alpha_{\setk z}\}_{k\in \mk,z=0,1},\bs{\beta}} \frac{1}{N} \sum_{k=1}^K \sum_{i: A_i = k} \sum_{z\in\{0,1\}} \{Y_i(z)-\bs{x}_i^\top \bs{\beta}-\alpha_{\setk z}\}^2 \omega_{\setk z}.
\end{align*}

%\begin{itemize}
%    \item Linear regression adjustment without stratum indicators:
%\begin{align*}
%    & f^\ast_z(\bs{x}_i) = \bs{x}_i^\top \bs{\beta}_z^\ast + \alpha_{z}^\ast, \quad \textnormal{where } (\alpha_z^\ast,\bs{\beta}_z^\ast) = \argmin_{\alpha,\bs{\beta}} \frac{1}{N}\sumi \{Y_i(z)-\bs{x}_i^\top \bs{\beta} - \alpha\}^2 .
%\end{align*}
%    \item Linear regression adjustment with stratum indicators:
%\begin{align*}
%    & f^\ast_z(\bs{x}_i) = \bs{x}_i^\top \bs{\beta}_z^\ast + \sum_{k=1}^K \alpha_{\setk z}^\ast I(A_i = k), \quad \textnormal{where}\\
%    & (\{\alpha_{\setk z}^\ast\}_{k\in \mk},\bs{\beta}_z^\ast) = \argmin_{\{\alpha_{\setk}\}_{k\in \mk},\bs{\beta}} \frac{1}{N}\sum_{k=1}^K \sum_{i: A_i = k} \{Y_i(z)-\bs{x}_i^\top \bs{\beta}- \alpha_{\setk}\}^2 \omega_{\setk z}
%\end{align*}
%with $\omega_{\setk z}$ being defined later.
%    \item ToM regression adjustment \citep{lu2024tyranny}:
%\begin{align*}
%     f^\ast_z(\bs{x}_i) = & \bs{x}_i^\top \bs{\beta}^\ast + \sum_{k=1}^K \alpha_{\setk z}^\ast I(A_i = k), \quad \textnormal{where}\\
%    (\{\alpha_{\setk z}^\ast\}_{k\in \mk,z=0,1},\bs{\beta}^\ast) = & \argmin_{\{\alpha_{\setk z}\}_{k\in \mk,z=0,1},\bs{\beta}} \frac{1}{N} \sum_{k=1}^K \sum_{i: A_i = k} \sum_{z\in\{0,1\}} \{Y_i(z)-\bs{x}_i^\top \bs{\beta}-\alpha_{\setk z}\}^2 \omega_{\setk z}.
%\end{align*}
%\end{itemize}

The key results are summarized as follows:

%\begin{itemize}
%    \item 
(1) The stability assumption holds for the three linear regressions described above. However, the number of strata may tend to infinity, such that the parameters to be estimated in the second and third regressions also tend to infinity. To ensure that the stability assumption holds, we must adjust the degrees of freedom loss in the empirical loss function.

%    \item 
(2) WLS with stratum indicators yields a shorter confidence interval compared to the unadjusted method. ToM regression adjustment not only guarantees an efficiency gain but also produces a shorter confidence interval compared to the unadjusted method.

%    \item 

%\end{itemize}

\subsubsection{OLS}

%\liu{seems an odd method? --- to be honest, I am a little lost here.  Make sure the appendix has some correspondence with the main paper.}

Based on the sample-splitting algorithm, for OLS, we have
 \begin{align}
 \label{eq:SR-hat-f-linear-separate-model-no-stratum-indicator}
     & \hat{f}_{[-q]z}(\bs{x}_i) = \bs{x}_i^\top \hat{\bs{\beta}}_{[-q]z} + \hat{\alpha}_{[-q]z},
\end{align}
where
\begin{align*}
     & (\hat{\alpha}_{[-q] z}, \hat{\bs{\beta}}_{[-q]z}) \\
      = & \argmin_{\alpha,\bs{\beta}} \frac{1}{N} \sum_{k=1}^K \sum_{i:A_i=k,Z_i=z,i\in\ms_{[-q]}} (Y_i-\bs{x}_i^\top \bs{\beta} - \alpha)^2 \frac{1}{\prob(Z_i=z,i\in\ms_{[-q]})} \\
      = & \argmin_{\alpha,\bs{\beta}} \frac{1}{N} \sum_{k=1}^K \sum_{i:A_i=k,Z_i=z,i\in\ms_{[-q]}} (Y_i-\bs{x}_i^\top \bs{\beta} - \alpha)^2 \frac{N_{\setk}}{N_{\setk[-q]z}}.
 \end{align*}

\Cref{prop:stability-assumption-linear-SR-no-stratum-indicator} below shows that the stability assumption (\Cref{a:negligible-estimateion-error-for-prediction-function}) holds for $\hat{f}_{[-q]z}$ defined by \eqref{eq:SR-hat-f-linear-separate-model-no-stratum-indicator}. Let $\|\cdot\|_\infty$ denote the $\ell_\infty$-norm of a vector.

\begin{proposition}
\label{prop:stability-assumption-linear-SR-no-stratum-indicator}
    Consider the algorithm in \Cref{example:algorithm-split-by-treatment-sr} and $\hat{f}_{[-q]z}$ defined by \eqref{eq:SR-hat-f-linear-separate-model-no-stratum-indicator}. If (i) $N_{\setk [q]z}/N_{\setk} \in (c,1-c)$, for a constant $c\in (0,0.5)$, $q=1,2$, $z=0,1$, $k\in \mk$, (ii) $\max\{ N^{-1}\sumi \|\bs{x}_i\|^4_{\infty}$, $ N^{-1}\sumi Y_i^4(1), N^{-1}\sumi Y_i^4(0)\} = O(1)$, and (iii) $\bs{\Sigma}_{\tilde{\bs{x}}\tilde{\bs{x}}} := N^{-1} \sumi \tilde{\bs{x}}_i \tilde{\bs{x}}_i^\top$ has an invertible limit, where $\tilde{\bs{x}}_i = (1,\bs{x}_i^\top)^\top$, then $\|\hat{f}_{[-q]z}-f^\ast_z\|_N^2 = \op(1)$. 
\end{proposition}

\subsubsection{WLS with stratum indicators}
WLS with stratum indicators includes the stratum indicators to account for stratum-level heterogeneity, as well as weights $\omega_{\setk z}$ to allow
for variation in data importance across treatment groups and strata.

We obtain $\hat{f}_{[-q]z}$ through:
 \begin{align}
 \label{eq:SR-hat-f-linear-separate-model}
     & \hat{f}_{[-q]z}(\bs{x}_i) = \bs{x}_i^\top \hat{\bs{\beta}}_{[-q]z} + \sum_{k=1}^K\hat{\alpha}_{\setk [-q] z} I(A_i=k),
\end{align}
where
\begin{align*}
      (\{\hat{\alpha}_{\setk [-q] z}\}_{k\in \mk}, \hat{\bs{\beta}}_{[-q]z})  &= \argmin_{\{\alpha_{\setk}\}_{k\in \mk},\bs{\beta}} \frac{1}{N} \sum_{k=1}^K \sum_{i:Z_i=z,i\in \ms_{[-q]},A_i = k} (Y_i-\bs{x}_i^\top \bs{\beta}-  \alpha_{\setk})^2 \omega_{\setk [-q] z}, \\
    \omega_{\setk [-q] z} &= \omega_{\setk z} \frac{N_{\setk}-1}{N_{\setk [-q]z}-1}.
\end{align*}

Note that the weight $\omega_{\setk [-q] z}$ in the empirical loss function is not obtained by naive inverse probability weighting, i.e., $$\omega_{\setk [-q] z} \ne \frac{\omega_{\setk z}}{\prob(Z_i=z,i\in \ms_{[-q]},A_i=k)} = \omega_{\setk z} \frac{N_{\setk}}{N_{\setk [-q]z}}.$$  This adjustment to the weights accounts for the degrees of freedom loss, as the number of parameters to be estimated in the linear regression model grows with the number of strata. A similar adjustment was also made in \cite{lu2024tyranny}.

\Cref{prop:stability-assumption-linear-SR} below shows that the stability assumption (\Cref{a:negligible-estimateion-error-for-prediction-function}) holds for $\hat{f}_{[-q]z}$ defined by \eqref{eq:SR-hat-f-linear-separate-model}.

\begin{proposition}
\label{prop:stability-assumption-linear-SR}
    Consider the algorithm in \Cref{example:algorithm-split-by-treatment-sr} and $\hat{f}_{[-q]z}$ defined by \eqref{eq:SR-hat-f-linear-separate-model}. If (i) $N_{\setk [q]z} \geq 2$, $N_{\setk [q]z}/N_{\setk},\omega_{\setk z} \in (c,1-c)$, for a constant $c\in (0,0.5)$, $q=1,2$, $z=0,1$, $k\in \mk$, (ii) $\max\{ N^{-1}\sumi \|\bs{x}_i\|^4_{\infty}, N^{-1}\sumi Y_i^4(1), N^{-1}\sumi Y_i^4(0)\} = O(1)$, and (iii) for $z=0,1$, $\bs{\Sigma}_{z,\bs{x}\bs{x}} := N^{-1} \sum_{k=1}^K\sum_{i:A_i=k} \omega_{\setk z} (\bs{x}_i-\bar{\bs{x}}_{\setk}) (\bs{x}_i-\bar{\bs{x}}_{\setk})^\top$ have invertible limits and $\bs{\beta}_z^\ast$ have finite limits, then $\|\hat{f}_{[-q]z}-f^\ast_z\|_N^2 = \op(1)$. 
\end{proposition}

Let $\tilde{\sigma}^2_{\SR}$ be the probability limit of $N\hat{V}_{\oraclecf}$. We define $\tilde{\sigma}^2_{\textnormal{SR-unadj}}$ as the unadjusted analogy of $\tilde{\sigma}^2_{\SR}$ by replacing $\varepsilon_i(z)$ with $Y_i(z)$ in the definition of $\tilde{\sigma}^2_{\SR}$. Intuitively, $\tilde{\sigma}^2_{\textnormal{SR-unadj}}$ is the probability limit of $N\hat{V}_{\oraclecf}$ if we consider $f^\ast_z\equiv 0$.
 
\begin{proposition}
\label{eq:separate-model-using-optimal-weights}
 Consider the algorithm in \Cref{example:algorithm-split-by-treatment-sr} and $\hat{f}_{[-q]z}$ defined by \eqref{eq:SR-hat-f-linear-separate-model}. Under the assumptions of \Cref{prop:stability-assumption-linear-SR}, we have $\tilde{\sigma}_{\SR}^2 \leq \tilde{\sigma}^2_{\textnormal{SR-unadj}}$.
\end{proposition}

 \Cref{eq:separate-model-using-optimal-weights} shows that $\tilde{\sigma}_{\SR}^2 \leq \tilde{\sigma}^2_{\textnormal{SR-unadj}}$. Consequently, the cross-fitted variance estimator, as well as the resulting confidence interval, is no longer than that of the unadjusted method. The intuition is as follows: in an intermediate step of proving \Cref{prop:valid-inference-sr}, we show that
\begin{align*}
\tilde{\sigma}^2_{\SR} = & \lim\limits_{N \rightarrow \infty} \sum_{q \in \{1,2\}}\sum_{k=1}^K \frac{N_{\setk [q]}^2}{N}\sum_{i:A_i=k} \sum_{z\in \{0,1\}} \frac{ \{\varepsilon_i(z) - {\bar{\varepsilon}_{\setk}}(z)\}^2}{N_{\setk [q]z}(N_{\setk}-1)} \\
=& \lim\limits_{N \rightarrow \infty} \frac{1}{N}\sum_{k=1}^K \sum_{i:A_i=k} \sum_{z\in\{0,1\}} \{\varepsilon_i(z) - {\bar{\varepsilon}_{\setk}}(z)\}^2 \omega_{\setk z}.
\end{align*}
 Therefore, $f^\ast_z$ minimizes the estimated variance.
\subsubsection{ToM regression adjustment}

%Let $\sigma_{\SR}^2$ and $\sigma_{\textnormal{SR-unadj}}^2$ denote the asymptotic variances of the cross-fitted ATE estimator and the unadjusted ATE estimator, respectively. 

Define $\sigma^2_{\SR}$ as the probability limit of $N\Var(\hat{\tau}_{\oraclecf})$ and $\sigma^2_{\textnormal{SR-unadj}}$ as the unadjusted analogy of $\sigma^2_{\SR}$ by setting $f^\ast_z\equiv 0$ in the definition of $\sigma^2_{\SR}$. \Cref{eq:separate-model-using-optimal-weights} demonstrates that the estimated variance is no greater than that of the unadjusted method. However, this may not hold true for the sampling variance, i.e., $\sigma_{\SR}^2$ may be greater than $\sigma_{\textnormal{SR-unadj}}^2$. To address this issue, we consider using the ToM regression \citep{lu2024tyranny}.

We obtain $\hat{f}_{[-q]z}$ through:
 \begin{align}
 \label{eq:SR-hat-f-linear-tom}
     & \hat{f}_{[-q]z}(\bs{x}_i) = \bs{x}_i^\top \hat{\bs{\beta}}_{[-q]} + \sum_{k=1}^K\hat{\alpha}_{\setk [-q] z} I(A_i=k),
\end{align}
where
\begin{align*}
%\label{eq:SR-hat-f-linear-tom}
     & (\{\hat{\alpha}_{\setk [-q] z}\}_{k\in \mk,z=0,1}, \hat{\bs{\beta}}_{[-q]}) \\
     \nonumber
     & = \argmin_{\{\alpha_{\setk z}\}_{k\in \mk, z=0,1},\bs{\beta}} \frac{1}{N}\sum_{z\in\{0,1\}}\sum_{k=1}^K \sum_{i:A_i=k,Z_i=z,i\in \ms_{[-q]}} (Y_i-\bs{x}_i^\top \bs{\beta}- \alpha_{\setk z})^2 \omega_{\setk [-q] z}. 
 \end{align*}
%\begin{align*}
%    \omega_{\setk [-q] z} = \omega_{\setk z} ({N_{\setk}-1})/({N_{\setk [-q]z}-1}).
%\end{align*}

\begin{proposition}
\label{prop:stability-assumption-linear-SR-tom}
   Consider the algorithm in \Cref{example:algorithm-split-by-treatment-sr} and $\hat{f}_{[-q]z}$ defined by \eqref{eq:SR-hat-f-linear-tom}. If (i) $N_{\setk [q]z} \geq 2$, $N_{\setk [q]z}/N_{\setk},\omega_{\setk z} \in (c,1-c)$, for a constant $c\in (0,0.5)$, $q=1,2$, $z=0,1$, $k\in \mk$, (ii) $\max\{N^{-1}\sumi \|\bs{x}_i\|^4_\infty, N^{-1}\sumi Y_i^4(1),N^{-1}\sumi Y_i^4(0)\} = O(1)$, and (iii) $\bs{\Sigma}_{\bs{x}\bs{x}} := N^{-1} \sum_{z\in\{0,1\}}$ $\sum_{k=1}^K\sum_{i:A_i=k} \omega_{\setk z} \bs{x}_i \bs{x}_i^\top$ has an invertible limit and $\bs{\beta}^\ast$ has a finite limit, then we have (i) $\|\hat{f}_{[-q]z}-f^\ast_z\|_N^2 = \op(1)$, and (ii) $\tilde{\sigma}_{\SR}^2 \leq \tilde{\sigma}^2_{\textnormal{SR-unadj}}$ and $\sigma_{\SR}^2 \leq \sigma^2_{\textnormal{SR-unadj}}$. 
\end{proposition}

\Cref{prop:stability-assumption-linear-SR-tom} shows that (i) the stability assumption holds for $\hat{f}_{[-q]z}$ defined by \eqref{eq:SR-hat-f-linear-tom}, and (ii) ToM regression yields a smaller sampling variance and a shorter confidence interval.

\subsection{Linear regression adjustments of MPEs}
\label{sec:linear-regression-adjustments-under-MPEs}
For MPEs, we consider two linear prediction functions, where the latter incorporates pairing information:

%\begin{itemize}
%    \item
(1) OLS of outcome on covariates:
\begin{align*}
    & f^\ast_z(\bs{x}_i) = \bs{x}_i^\top \bs{\beta}_z^\ast + \alpha_z^\ast,\quad \textnormal{where }  (\alpha_z^\ast,\bs{\beta}_z^\ast) = \argmin_{\alpha, \bs{\beta}} \frac{1}{N}\sumi \{Y_i(z)-\bs{x}_i^\top \bs{\beta}-\alpha\}^2.
\end{align*}

(2) OLS of paired outcome difference on covariates \citep{fogarty2018regression}.
%\end{itemize}

We verify that the stability assumption holds for both regression adjustments. Moreover, echoing the results of \cite{fogarty2018regression}, the second regression not only guarantees an efficiency gain but also produces a shorter confidence interval compared to the unadjusted method.

\subsubsection{OLS of outcome on covariates}

We obtain $\hat{f}_{[-q]z}(\bs{x}_i)$ through:
\begin{align}
\label{eq:matched-pair-hat-f-[-q]-z}
    & \hat{f}_{[-q]z}(\bs{x}_i) = \bs{x}_i^\top \hat{\bs{\beta}}_{[-q]z} + \hat{\alpha}_{[-q]z},\\
  \nonumber  
    & (\hat{\alpha}_{[-q]z}, \hat{\bs{\beta}}_{[-q]z}) = \argmin_{(\alpha,\bs{\bs{\beta}})} \frac{1}{N_{[-q]}/2}\sum_{i:Z_i=z,i\in \ms_{[-q]}} (Y_i-\bs{x}_i^\top \bs{\beta}-\alpha)^2. 
\end{align}

\Cref{prop:stability-assumption-linear-MP} below shows that the stability assumption (\Cref{a:negligible-estimateion-error-for-prediction-function}) holds for $\hat{f}_{[-q]z}$ defined by \eqref{eq:matched-pair-hat-f-[-q]-z}.
\begin{proposition}
\label{prop:stability-assumption-linear-MP}
   Consider the algorithm in \Cref{example:split-by-stratum-sr} and $\hat{f}_{[-q]z}$ defined by \eqref{eq:matched-pair-hat-f-[-q]-z}. If (i) $N_{ [q]}/N \in (c,1-c)$, for a constant $c\in (0,0.5)$, $q=1,2$, (ii) $\max\{N^{-1}\sumi \|\bs{x}_i\|^4_\infty,  N^{-1}\sumi Y_i^4(1), $ $ N^{-1}\sumi Y_i^4(0)\} = O(1)$, and (iii) $\bs{\Sigma}_{\bs{x}\bs{x}}^{\MP} := N^{-1}\sumi (\bs{x}_i-\bar{\bs{x}})(\bs{x}_i-\bar{\bs{x}})^\top$ has an invertible limit and $\bs{\beta}^\ast_z$, $z=0,1$, have finite limits, then we have $\|\hat{f}_{[-q]z}-f^\ast_z\|_N^2 = \op(1)$.
\end{proposition}

\subsubsection{OLS of paired outcome difference on covariates}

Let $2k-1$ and ${2k}$ indicate the two units in the $k$th pair, $k=1,\ldots, N/2$. Let $E_k = I(Z_{2k-1}=1)$.  We consider the OLS regression of the observed difference of outcomes on the observed difference of the covariates \citep{fogarty2018regression,imbens2015causal}:
\begin{align}
\label{eq:hat-f-[-q]-MP-2}
    & \hat{f}_{[-q]1}(\bs{x}_i) = \hat{f}_{[-q]0}(\bs{x}_i) = \bs{x}_i^\top \hat{\bs{\beta}}_{[-q]},\\
    \nonumber
    & \hat{\bs{\beta}}_{[-q]} = \argmin_{\bs{\beta}} \frac{1}{N_{[-q]}/2} \sum_{k\in \mk_{[-q]}} \Big[\{Y_{2k-1}-Y_{2k}-(\bs{x}_{2k-1}-\bs{x}_{2k})^\top \bs{\beta}\}^2 E_k + \\
    & \qquad \qquad \qquad \{Y_{2k}-Y_{2k-1}-(\bs{x}_{2k}-\bs{x}_{2k-1})^\top \bs{\beta}\}^2 (1-E_k)\Big],
\end{align}
which corresponds to the following prediction function:
\begin{align*}
    & f^\ast_1(\bs{x}_i) = f^\ast_0(\bs{x}_i) = \bs{x}_i^\top \bs{\beta}^\ast,\\
       & \bs{\beta}^\ast = \argmin_{\bs{\beta}} \frac{1}{N} \sum_{k=1}^{N/2} \Big[\{Y_{2k-1}(1)-Y_{2k}(0)-(\bs{x}_{2k-1}-\bs{x}_{2k})^\top \bs{\beta}\}^2  + \\
    &  \qquad \qquad \qquad \{Y_{2k}(1)-Y_{2k-1}(0)-(\bs{x}_{2k}-\bs{x}_{2k-1})^\top \bs{\beta}\}^2 \Big].
\end{align*}

Similar to $\tilde{\sigma}_{\SR}^2$, $\sigma_{\SR}^2$, $\tilde{\sigma}^2_{\textnormal{SR-unadj}}$,  $\sigma^2_{\textnormal{SR-unadj}}$, define $\tilde{\sigma}_{\MP}^2$, $\sigma_{\MP}^2$, $\tilde{\sigma}^2_{\textnormal{MP-unadj}}$,  $\sigma^2_{\textnormal{MP-unadj}}$ as the probability limits of $N\hat{V}_{\oraclecf}$, $N\Var(\hat{\tau}_{\oraclecf})$, and their unadjusted analogies in MPEs. \Cref{mp:regression-of-outcome-difference-on-covariate-difference} below shows that the above regression reduces the confidence interval length and the sampling variance.

\begin{proposition}
\label{mp:regression-of-outcome-difference-on-covariate-difference}
 Consider the algorithm in \Cref{example:split-by-stratum-sr} and $\hat{f}_{[-q]z}$ defined by \eqref{eq:hat-f-[-q]-MP-2}. If (i) $N_{ [q]}/N \in (c,1-c)$, for a constant $c\in (0,0.5)$, $q=1,2$, (ii) $\max\{N^{-1}\sumi \|\bs{x}_i\|^4_\infty, N^{-1}\sumi Y_i^4(1),  $ $ N^{-1}\sumi Y_i^4(0)\} = O(1)$, and (ii) $\bs{\Sigma}_{\bs{x}\bs{x}}^{\MP} := (2/N)\sum_{k=1}^{N/2} (\bs{x}_{2k}-\bs{x}_{2k-1})(\bs{x}_{2k}-\bs{x}_{2k-1})^\top$ has an invertible limit and $\bs{\beta}^\ast$ has a finite limit, then we have (i) $\|\hat{f}_{[-q]z}-f^\ast_z\|_N^2 = \op(1)$, and (ii) $\tilde{\sigma}_{\MP}^2 \leq \tilde{\sigma}^2_{\textnormal{MP-unadj}}$ and $\sigma_{\MP}^2 \leq \sigma^2_{\textnormal{MP-unadj}}$. 
\end{proposition}

\section{Proofs}
\label{sec:proofs}

\subsection{Additional notation}
\label{sec:notation}

Let $R_i = I(i\in \ms_{[1]})$ and $\bs{R} = (R_i)_{i=1}^N$. For any $k$-dimensional vector $\bs{v}=(v_1,\ldots,v_k)^\top$, let $||\bs{v}||_1 = \sum_{i=1}^k |v_i|$, $||\bs{v}||_2 = ( \sum_{i=1}^k v_i^2)^{1/2}$, and $||\bs{v}||_\infty = \max_{i=1,\ldots,k} |v_i|$ denote its $\ell_1$, $\ell_2$, and $\ell_\infty$ norms, respectively. For a finite-population $\{\bs{a}_i\}_{i=1}^N$, $z\in\{0,1\}$, $k \in \mathcal{K}$, and $q \in \{1,2\}$, we define $\bar{\bs{a}} = N^{-1}\sumi \bs{a}_i$, $\bar{\bs{a}}_{\setk} = N_{\setk}^{-1}\sum_{i:A_i=k} \bs{a}_i$, $\bar{\bs{a}}_{z} = N_{z}^{-1}\sum_{i:Z_i=z} \bs{a}_i$, $\bar{\bs{a}}_{[q]} = N_{[q]}^{-1}\sum_{i:i\in\ms_{[q]}} \bs{a}_i$, $\bar{\bs{a}}_{\setk z} = N_{\setk z}^{-1}\sum_{i:A_i=k,Z_i=z} \bs{a}_i$, and $\bar{\bs{a}}_{\setk [q] z} = N_{\setk 
[q]z}^{-1}\sum_{i:A_i=k,Z_i=z, i\in\ms_{[q]}} \bs{a}_i$. For a vector $\bs{v} = (v_1,\ldots,v_N)^\top$, we define the following subvectors: $\bs{v}_{\setk} = (v_i)_{i:A_i=k}$, $\bs{v}_{[q]} = (v_i)_{i:i\in\ms_{[q]}}$, and $\bs{v}_{\setk [q]} = (v_i)_{i:A_i=k,i\in\ms_{[q]}}$. For two finite populations $\{\bs{a}_i\}_{i=1}^N$, $\{\bs{b}_i\}_{i=1}^N$, define
\begin{align*}
    &\bs{S}_{\bs{a}\bs{b}} = \frac{1}{N-1}\sumi (\bs{a}_i-\bar{\bs{a}})(\bs{b}_i-\bar{\bs{b}})^\top,\\
    & \bs{S}_{z,\bs{a}\bs{b}} = \frac{1}{N_{z}-1}\sum_{i:Z_i=z} (\bs{a}_i-\bar{\bs{a}}_{z})(\bs{b}_i-\bar{\bs{b}}_{z})^\top, \\
    &\bs{S}_{\setk,\bs{a}\bs{b}} = \frac{1}{N_{\setk}-1}\sum_{i:A_i=k} (\bs{a}_i-\bar{\bs{a}}_{\setk})(\bs{b}_i-\bar{\bs{b}}_{\setk})^\top, \\
    & \bs{S}_{[q],\bs{a}\bs{b}} = \frac{1}{N_{[q]}-1}\sum_{i:i\in \ms_{[q]}} (\bs{a}_i-\bar{\bs{a}}_{[q]})(\bs{b}_i-\bar{\bs{b}}_{[q]})^\top, \\
    &\bs{S}_{\setk z,\bs{a}\bs{b}} = \frac{1}{N_{\setk z}-1}\sum_{i:A_i=k,Z_i=z} (\bs{a}_i-\bar{\bs{a}}_{\setk z})(\bs{b}_i-\bar{\bs{b}}_{\setk z})^\top, \\
    &\bs{S}_{\setk [q] z,\bs{a}\bs{b}} = \frac{1}{N_{\setk [q] z}-1}\sum_{i:A_i=k,i\in \ms_{[q]},Z_i=z} (\bs{a}_i-\bar{\bs{a}}_{\setk [q] z})(\bs{b}_i-\bar{\bs{b}}_{\setk [q] z})^\top.
\end{align*}

\subsection{Useful lemmas}

\begin{lemma}
\label{lem:variance-of-CR}
    If $\prob_{\bs{Z}} = \CR(N,N_1)$ with $N_0 = N - N_1$, then for $\{a_i\}_{i=1}^N$ and $\{b_i\}_{i=1}^N$, we have
    \begin{align*}
      (i)\quad  \Var\Big(\sumi \frac{Z_i a_i}{N_1} - \sumi\frac{(1-Z_i) b_i}{N_0}\Big) = \frac{S_{aa}}{N_1} + \frac{S_{bb}}{N_0} - \frac{S_{(a-b)(a-b)}}{N},
    \end{align*}
    \begin{align*}
       (ii)\quad \Var\Big(\sumi Z_i a_i\Big) \leq  \frac{N_0 N_1 }{(N-1)N } \sumi a_i^2.
    \end{align*}
\end{lemma}

\begin{proof}
 \Cref{lem:variance-of-CR}(i) follows from Theorem 6.2 in \cite{imbens2015causal} or Theorem 4.1 in \cite{ding2024book}. \Cref{lem:variance-of-CR}(ii) follows by setting $b_i = 0$ in the first statement and $S_{aa} \leq (N-1)^{-1} \sumi a_i^2$.
\end{proof}

\begin{lemma}
    \label{lem:unbiasedness-of-variance-estimator}
    If $\prob_{\bs{Z}} = \CR(N,N_1)$, we have $\ope (S_{z,aa}) = S_{aa}.$
%    \begin{align*}
%       \ope S_{z,aa} = S_{aa}.
%    \end{align*}
\end{lemma}

\begin{proof}
 The conclusion follows from \citet[][Appendix A]{imbens2015causal} or the proof of Theorem 4.1 in \cite{ding2024book}.
\end{proof}

%$\bs{S}_{\bs{a}\bs{b}} = \frac{1}{N-1}\sumi (\bs{a}_i-\bar{\bs{a}})(\bs{b}_i-\bar{\bs{b}})^\top$, $\bs{S}_{\setk,\bs{a}\bs{b}} = \frac{1}{N_{\setk}-1}\sum_{i:A_i=k} (\bs{a}_i-\bar{\bs{a}}_{\setk})(\bs{b}_i-\bar{\bs{b}}_{\setk})^\top$, $\bs{S}_{\setk z,\bs{a}\bs{b}} = \frac{1}{N_{\setk z}-1}\sum_{i:A_i=k,Z_i=z} (\bs{a}_i-\bar{\bs{a}}_{\setk z})(\bs{b}_i-\bar{\bs{b}}_{\setk z})^\top$, $\bs{S}_{\setk [q] z,\bs{a}\bs{b}} = \frac{1}{N_{\setk [q] z}-1}\sum_{i:A_i=k,i\in \ms_{[q]},Z_i=z} (\bs{a}_i-\bar{\bs{a}}_{\setk [q] z})(\bs{b}_i-\bar{\bs{b}}_{\setk [q] z})^\top$ 

\subsection{Proof of \Cref{thm:unbiased}}
\label{sec:proof-of-thm-1}

\begin{proof}
    
Recall that, for $q \in \{1,2\}$ and $z\in \{0,1\}$, $\hat{\tau}_{\cfadj, [q]}  = \hat{\mu}_{\cfadj, [q]}(1)- \hat{\mu}_{\cfadj, [q]}(0)$, where
        \begin{align*}
     \hat{\mu}_{\cfadj, [q]}(z) & = \frac{1}{N_{[q]}} \sum_{i:Z_i=z,i\in \ms_{[q]}} \frac{Y_i-\hat{f}_{[-q]z}(\bs{x}_i)}{\prob(Z_i=z\mid \ms_{[1]},\ms_{[2]})} + \frac{1}{N_{[q]}} \sum_{i\in \ms_{[q]}} \hat{f}_{[-q]z}(\bs{x}_i). 
\end{align*}
Since we use $\{( Y_i,Z_i,\bs{x}_i)\}_{i\in\ms_{[-q]}}$ to estimate the prediction function $\hat{f}_{[-q]z}(\bs{x}_i)$, conditional on $(\ms_{[1]},\ms_{[2]},\bs{Z}_{[-q]})$, $\hat{f}_{[-q]z}(\bs{x}_i)$ is considered fixed. Therefore,
\begin{align*}
    &\ope ( \hat{\mu}_{\cfadj, [q]}(z)  \mid \ms_{[1]},\ms_{[2]},\bs{Z}_{[-q]}) \\
     = &   \frac{1}{N_{[q]}} \sum_{i\in \ms_{[q]}} \frac{\big\{Y_i(z) -\hat{f}_{[-q]z}(\bs{x}_i)\big\} \prob(Z_i=z\mid \ms_{[1]},\ms_{[2]},\bs{Z}_{[-q]})}{\prob(Z_i=z\mid \ms_{[1]},\ms_{[2]})}  +  \frac{1}{N_{[q]}} \sum_{i\in \ms_{[q]}} \hat{f}_{[-q]z}(\bs{x}_i).
\end{align*}

By \Cref{a:conditional-independent-subsamples}, for $i \in \ms_{[q]}$, we have $\prob(Z_i=z\mid \ms_{[1]},\ms_{[2]},\bs{Z}_{[-q]}) = \prob(Z_i=z\mid \ms_{[1]},\ms_{[2]})$. Therefore, 
\begin{align*}
   \ope ( \hat{\mu}_{\cfadj, [q]}(z)  \mid \ms_{[1]},\ms_{[2]},\bs{Z}_{[-q]}) 
   =&   \frac{1}{N_{[q]}} \sum_{i\in \ms_{[q]}} \big\{Y_i(z)-\hat{f}_{[-q]z}(\bs{x}_i)\big\} + \frac{1}{N_{[q]}}  \sum_{i\in \ms_{[q]}} \hat{f}_{[-q]z}(\bs{x}_i) \\
   = & \frac{1}{N_{[q]}} \sum_{i\in \ms_{[q]}} Y_i(z).
\end{align*}
As a consequence, 
\begin{align*}
    &\ope (\hat{\mu}_{\cfadj, [q]}(z)  \mid \ms_{[1]},\ms_{[2]}) =  N_{[q]}^{-1} \sum_{i\in \ms_{[q]}} Y_i(z), \quad \ope ( \hat{\tau}_{\cfadj,[q]} \mid \ms_{[1]},\ms_{[2]}) = N_{[q]}^{-1} \sum_{i\in \ms_{[q]}} \{ Y_i(1) - Y_i(0) \}.
\end{align*}
%$$\ope (\hat{\mu}_{\cfadj, [q]}(z)  \mid \ms_{[1]},\ms_{[2]}) =  N_{[q]}^{-1} \sum_{i\in \ms_{[q]}} Y_i(z), \quad \ope ( \hat{\tau}_{\cfadj,[q]} \mid \ms_{[1]},\ms_{[2]}) = N_{[q]}^{-1} \sum_{i\in \ms_{[q]}} \{ Y_i(1) - Y_i(0) \}.$$
It follows that 
\begin{align*}
    \ope ( \hat{\tau}_{\cfadj} \mid \ms_{[1]},\ms_{[2]}) =& \sum_{q \in \{1,2\}} \frac{N_{[q]}}{N} \ope ( \hat{\tau}_{\cfadj,[q]} \mid \ms_{[1]},\ms_{[2]}) 
  =   \sum_{q \in \{1,2\}} \frac{N_{[q]}}{N} \frac{1}{N_{[q]}} \sum_{i\in \ms_{[q]}} \{Y_i(1) - Y_i(0)\} \\
  = & \frac{1}{N} \sum_{q \in \{1,2\}}  \sum_{i\in \ms_{[q]}} \{Y_i(1) - Y_i(0)\}
  =  \bar{\tau}. 
\end{align*}

\end{proof}

\subsection{Proof of \Cref{prop:BT}--\ref{prop:sr-2}}
\label{sec:verify-conditional-independence}
Recall that $R_i = I(i\in \ms_{[1]})$ and $\bs{R} = (R_i)_{i=1}^N$. Therefore, it suffices to show that $\bs{Z}_{[1]}\indep \bs{Z}_{[2]} \mid \bs{R}$.

\begin{proof}[Proof of \Cref{prop:BT}]
It follows from the fact that $\bs{R}$ and $\bs{Z}$ are independent Bernoulli random vectors.
\end{proof}

\begin{proof}[Proof of \Cref{prop:CR}]
Define $\Omega =\{(\bs{r},\bs{z}):$ $\bs{r} = (r_i)_{i=1}^N \in \{0,1\}^N$, $\bs{z} = (z_i)_{i=1}^N \in \{0,1\}^N$, $\|\bs{z}\|_1 = N_1$, $\|\bs{r}\|_1 = N_{[1]}$, $|\{i: z_i = 1, R_i=1\}| = N_{[1]1}\}$. For $(\bs{r},\bs{z}) \in \Omega$, by the construction of the split-by-treatment algorithm in \Cref{example:algorithm-split-by-treatment}, we have
\[
\prob(\bs{R} = \bs{r} \mid \bs{Z} = \bs{z}) = \frac{N_{[1]1}! N_{[2]1}! }{N_1!} \frac{N_{[1]0}N_{[2]0}}{N_0 !}.
\]
Therefore, for $(\bs{r},\bs{z}) \in \Omega$,
\begin{align*}
    &\prob(\bs{R} = \bs{r},\bs{Z} = \bs{z}) = \prob(\bs{R} = \bs{r} \mid \bs{Z} = \bs{z}) \prob(\bs{Z} = \bs{z}) \\
    =& \frac{N_{[1]1}! N_{[2]1}! }{N_1!} \frac{N_{[1]0}N_{[2]0}}{N_0 !} \frac{N_1 ! N_0 !}{N!} = \frac{N_{[1]1}! N_{[2]1}! N_{[1]0}!N_{[2]0}!}{N!}.
\end{align*} 
Otherwise, $\prob(\bs{R} = \bs{r},\bs{Z} = \bs{z}) = 0$.

For $(\bs{r},\bs{z}) \in \Omega$, we have
\begin{align*}
    &\prob(\bs{Z} = \bs{z}\mid \bs{R} = \bs{r}) = \frac{\prob(\bs{R} = \bs{r}, \bs{Z} = \bs{z})}{\sum_{\bs{z'}\in \{0,1\}^N}\prob(\bs{R} = \bs{r}, \bs{Z} = \bs{z'})}.
\end{align*}
There are $(N_{[1]}! N_{[2]}!)/ (N_{[1]1}! N_{[2]1}! N_{[1]0}! N_{[2]0}!)$ choices of $\bs{z'}$ such that $(\bs{r},\bs{z'})\in \Omega$. It follows that
\begin{align*}
 \prob(\bs{Z} = \bs{z}\mid \bs{R} = \bs{r})  = \frac{N_{[1]1}! N_{[2]1}! N_{[1]0}! N_{[2]0}!}{N_{[1]}! N_{[2]}!}.
\end{align*}

Denote $\tilde{\Omega} = \{(\bs{z}_{[1]},\bs{z}_{[2]},\bs{r}):\bs{z}_{[q]} \in \{0,1\}^{N_{[q]}}, \ \|\bs{z}_{[q]}\|_1 = N_{[q]1},  \ q=1,2, \ \bs{r} = (r_i)_{i=1}^N\in \{0,1\}^N, \ \|\bs{r}\|_1 = N_{[1]}\}$. For $(\bs{z}_{[1]},\bs{z}_{[2]},\bs{r}) \in \tilde{\Omega}$, we have
\begin{align*}
    &\prob(\bs{Z}_{[1]} = \bs{z}_{[1]},\bs{Z}_{[2]} = \bs{z}_{[2]}\mid \bs{R}=\bs{r}) = \prob(\bs{Z} = \bs{z}\mid \bs{R} = \bs{r}) = \frac{N_{[1]1}! N_{[2]1}! N_{[1]0}! N_{[2]0}!}{N_{[1]}! N_{[2]}!}.
\end{align*}
Since 
\begin{align*}
   \prob(\bs{Z}_{[1]} =& \bs{z}_{[1]}\mid \bs{R}=\bs{r}) = \sum_{\bs{z}_{[2]}\in \{0,1\}^{N_{[2]}}}\prob(\bs{Z}_{[1]} = \bs{z}_{[1]},\bs{Z}_{[2]} = \bs{z}_{[2]}\mid \bs{R}=\bs{r}),
\end{align*}
and there are $N_{[2]}!/(N_{[2]1}!N_{[2]0}!)$ positive terms in the summand, we have
\begin{align*}
    \prob(\bs{Z}_{[1]} = \bs{z}_{[1]}\mid \bs{R}=\bs{r}) = \frac{N_{[1]1}! N_{[2]1}! N_{[1]0}! N_{[2]0}!}{N_{[1]}! N_{[2]}!} \frac{N_{[2]}!}{N_{[2]1}!N_{[2]0}!} =  \frac{N_{[1]1}!  N_{[1]0}! }{N_{[1]}!}.
\end{align*}

Therefore, $\prob_{\bs{Z}_{[1]}\mid \ms_{[1]},\ms_{[2]}} = \prob_{\bs{Z}_{[1]}\mid \bs{R}} = \CR(N_{[1]},N_{[1]1})$.
Similarly, we have $\prob_{\bs{Z}_{[2]}\mid \ms_{[1]},\ms_{[2]}} = \CR(N_{[2]},N_{[2]1})$. As a consequence, we have
\begin{align*}
    & \prob(\bs{Z}_{[1]} = \bs{z}_{[1]},\bs{Z}_{[2]} = \bs{z}_{[2]}\mid \bs{R}=\bs{r}) = \prob(\bs{Z}_{[1]} = \bs{z}_{[1]}\mid \bs{R}=\bs{r}) 
 \prob(\bs{Z}_{[2]} = \bs{z}_{[2]}\mid \bs{R}=\bs{r}). 
\end{align*}
The conclusion follows.
\end{proof}

 \begin{proof}[Proof of \Cref{prop:sr-1}]
       By \Cref{prop:CR}, we have $\bs{Z}_{\setk [1]} \indep \bs{Z}_{\setk [2]} \mid \bs{R}_{\setk}$ and $\prob_{\bs{Z}_{\setk [q]} \mid \bs{R}_{\setk}} = \CR(N_{\setk [q]}, N_{\setk [q] 1})$. Since $(\bs{R}_{\setk},\bs{Z}_{\setk})$, $k \in \mk$, are independent by our algorithm, by the property of stratified randomized experiments, the conclusion follows.   
 \end{proof}

\begin{proof}[Proof of \Cref{prop:sr-2}]
   Let $D_k = I(k\in \mk_{[1]})$ and $\bs{D} = (D_k)_{k=1}^K$. 
We see that $\bs{Z}_{[1]}= (\bs{Z}_{\setk})_{D_k=1}$, $\bs{Z}_{[2]}= (\bs{Z}_{\setk})_{D_k=0}$. The conclusion follows from the fact that $\bs{D}$ and $\bs{Z}$ are independent and $\bs{Z}_{\setk}$, $k\in \mk$, are independent.  
\end{proof}

%Next, we prove \Cref{prop:assumption-2-holds}.

\subsection{Proof of \Cref{prop:assumption-2-holds}}

\begin{proof}%[Proof of \Cref{prop:assumption-2-holds}]
(i) For $\prob_{\bs{Z}_{[q]}\mid \ms_{[1]},\ms_{[2]}} = \BT(N_{[q]}, r_1)$, we have $\Var(I(Z_i=z)\mid \ms_{[1]},\ms_{[2]}) = r_1r_0$, $\prob(Z_i=z\mid \ms_{[1]},\ms_{[2]}) = r_z$ for $z\in \{0,1\}$. Therefore,
\begin{align*}
    &\Var\Big(\frac{1}{N}\sum_{i\in \ms_{[q]}} \frac{I(Z_i=z)}{\prob(Z_i=z\mid \ms_{[1]},\ms_{[2]})}a_i \mid \ms_{[1]},\ms_{[2]}\Big) = \frac{ r_{1-z}}{N^2 r_z} \sum_{i\in \ms_{[q]}} a_i^2.
\end{align*}
Therefore, \Cref{a:conditional-experimental-design-bounded-variance} is satisfied with $C_z = r_{1-z}/r_z$ for $z \in \{0,1\}$.

(ii) For $\prob_{\bs{Z}_{[q]}\mid \ms_{[1]},\ms_{[2]}} = \CR(N_{[q]},N_{[q]1})$, we have $\prob(Z_i=z\mid \ms_{[1]},\ms_{[2]}) = N_{[q]z}/N_{[q]}$, for $z=0,1$, $i\in \ms_{[q]}$. By \Cref{lem:variance-of-CR}(ii), we have
\begin{align*}
    &\Var\Big(\frac{1}{N}\sum_{i\in \ms_{[q]}} \frac{I(Z_i=1)}{\prob(Z_i=1\mid \ms_{[1]},\ms_{[2]})}a_i \mid \ms_{[1]},\ms_{[2]}\Big)
    \leq \frac{N_{[q]}N_{[q]0}}{N^2N_{[q]1}(N_{[q]}-1)} \sum_{i\in \ms_{[q]}} a_i^2,\\
    &\Var\Big(\frac{1}{N}\sum_{i\in \ms_{[q]}} \frac{I(Z_i=0)}{\prob(Z_i=0\mid \ms_{[1]},\ms_{[2]})}a_i \mid \ms_{[1]},\ms_{[2]}\Big) \leq \frac{N_{[q]}N_{[q]1}}{N^2N_{[q]0}(N_{[q]}-1)} \sum_{i\in \ms_{[q]}} a_i^2.
\end{align*}

Therefore, when $c < N_{[q]z}/N < 1-c$ for $z=0,1$, $q=0,1$, we have $N_{[q]0}/N_{[q]1}$, $N_{[q]1}/N_{[q]0}$ are bounded by $(1-c)/c$ and \Cref{a:conditional-experimental-design-bounded-variance} is satisfied with $C_z =  2(1-c)/c$ for $z \in \{0,1\}$.

(iii) For $\prob_{\bs{Z}_{[q]}\mid \ms_{[1]},\ms_{[2]}}=\SR(N_{[q]},\bs{A}_{[q]},(N_{\setk [q] 1})_{k\in \mk})$, we have
\begin{align*}
    \frac{1}{N}\sum_{i\in \ms_{[q]}} \frac{I(Z_i=z)}{\prob(Z_i=z\mid \ms_{[1]},\ms_{[2]})}a_i  = \frac{1}{N}\sum_{k=1}^K\sum_{i:i\in \ms_{[q]},A_i=k} \frac{I(Z_i=z) N_{\setk[q]}}{N_{\setk[q]z} }a_i,
\end{align*}
 $\prob_{\bs{Z}_{\setk [q]}\mid \ms_{[1]},\ms_{[2]}} = \CR(N_{\setk[q]},N_{\setk[q]1})$, and $\bs{Z}_{\setk [q]}$, $k\in \mk$, are independent conditional on $(\ms_{[1]},\ms_{[2]})$. Applying \Cref{lem:variance-of-CR}(ii) to the subset $\{i:i\in \ms_{[q]},A_i=k\}$, $k\in \mk$,
we have
\begin{align*}
    & \Var\Big(\frac{1}{N}\sum_{i\in \ms_{[q]}} \frac{I(Z_i=z)}{\prob(Z_i=z\mid \ms_{[1]},\ms_{[2]})}a_i \mid \ms_{[1]},\ms_{[2]}\Big) \\
   \leq & 
    \sum_{k=1}^K \frac{N_{\setk[q]}N_{\setk[q],1-z}}{N^2N_{\setk [q]z}(N_{\setk [q]}-1)} \sum_{i\in \ms_{[q]}, A_i = k} a_i^2\\
   \leq & \frac{2(1-c)}{c} \frac{1}{N^2} \sum_{i\in \ms_{[q]}} a_i^2\quad \Big(\text{using}\quad  \frac{N_{\setk [q]z}}{N_{\setk}} \in (c,1-c)\Big).
\end{align*}
Therefore, \Cref{a:conditional-experimental-design-bounded-variance} is satisfied with $C_z = 2(1-c)/c$ for $z\in \{0,1\}$.

(iv) For $\prob_{\bs{Z}_{[q]}\mid \ms_{[1]},\ms_{[2]}}=\SR(N_{[q]},\bs{A}_{[q]},(\bs{N}_{\setk 1})_{k\in \mk_{[q]}})$, we have
\begin{align*}
    \frac{1}{N}\sum_{i\in \ms_{[q]}} \frac{I(Z_i=z)}{\prob(Z_i=z\mid \ms_{[1]},\ms_{[2]})}a_i  = \frac{1}{N}\sum_{k\in \mk_{[q]}}\sum_{i:A_i=k} \frac{I(Z_i=z) N_{\setk}}{N_{\setk z} }a_i,
\end{align*}
 $\prob_{\bs{Z}_{\setk}\mid \ms_{[1]},\ms_{[2]}} = \CR(N_{\setk},N_{\setk1})$, and $\bs{Z}_{\setk}$, $k\in \mk_{[q]}$, are independent conditional on $(\ms_{[1]},\ms_{[2]})$. Applying \Cref{lem:variance-of-CR}(ii) to $\{i:A_i=k\}$, $k\in \mk_{[q]}$, we have
\begin{align*}
    & \Var\Big(\frac{1}{N}\sum_{i\in \ms_{[q]}} \frac{I(Z_i=z)}{\prob(Z_i=z\mid \ms_{[1]},\ms_{[2]})}a_i \mid \ms_{[1]},\ms_{[2]}\Big) \\
    \leq & \sum_{k\in \mk_{[q]}}\frac{N_{\setk }N_{\setk ,1-z}}{N^2N_{\setk  z}(N_{\setk  }-1)} \sum_{i: A_i = k} a_i^2 \leq \frac{2(1-c)}{c} \frac{1}{N^2} \sum_{i\in \ms_{[q]}} a_i^2,
\end{align*}
where the last inequality uses $N_{\setk z}/N_{\setk} \in (c,1-c)$, $z\in \{0,1\}$. Therefore, \Cref{a:conditional-experimental-design-bounded-variance} holds with $C_z = 2(1-c)/c$ for $z\in \{0,1\}$.
\end{proof}

\subsection{Proof of \Cref{thm:valid-confidence-interval}}

 \Cref{lem:property-of-op-1} below states that if $X_n = \op(1)$, then $|X_n|$ can be controlled by a diminishing deterministic sequence $\varepsilon_n$, with high probability.
\begin{lemma}
\label{lem:property-of-op-1}
    If $X_n = \op(1)$, as $n\rightarrow \infty$, then there exists $\varepsilon_n \rightarrow 0$ such that
\begin{align*}
    \prob(|X_n| \geq \varepsilon_n) \rightarrow 0.
\end{align*}
\end{lemma}
\begin{proof}%[Proof of \Cref{lem:property-of-op-1}]
   \( X_n = \op(1) \) means that
   \[
   \forall \varepsilon > 0, \ \prob(|X_n| \geq \varepsilon) \rightarrow 0 \text{ as } n \rightarrow \infty.
   \]
Take \(\varepsilon_k = 1/k \) (so that \(\varepsilon_k \to 0\) as \(k \to \infty\)).
   Given \( X_n = \op(1) \), for each fixed \(\varepsilon_k = 1/k \), we have
   \[
   \prob(|X_n| \geq \varepsilon_k) \rightarrow 0 \text{ as } n \to \infty.
   \]
   Then, there exists an integer \(N_k\) such that for all \(n \geq N_k\),
   \[
   \prob(|X_n| \geq \varepsilon_k) < \frac{1}{k}.
   \]

   Define a new sequence \(\varepsilon_n\) as follows:
   \[
   \varepsilon_n = \varepsilon_k = \frac{1}{k} \quad \text{for } N_k \leq n < N_{k+1}.
   \]
   By construction, we have \(\varepsilon_n \rightarrow 0\) as \(n \to \infty\), and
   \[
   \prob(|X_n| \geq \varepsilon_n) = \prob(|X_n| \geq \varepsilon_k) < \frac{1}{k} \rightarrow 0 \text{ as } n \to \infty.
   \]
Therefore, there exists a sequence \(\varepsilon_n \rightarrow 0\) such that \(\prob(|X_n| \geq \varepsilon_n) \rightarrow 0\). This completes the proof.
\end{proof}

\begin{proof}[Proof of \Cref{thm:valid-confidence-interval}]
 We decompose $\hat{\tau}_{\oraclecf} - \hat{\tau}_{\cfadj} = \sum_{q\in\{1,2\} } (T_{[q]1} - T_{[q]0})$,
%  \begin{align*}
%      \hat{\tau}_{\oraclecf} - \hat{\tau}_{\cfadj} = \sum_{q\in\{1,2\} } (T_{[q]1} - T_{[q]0}),
%  \end{align*}  
  where
  \begin{align*}
      T_{[q]z} = \frac{1}{N} \sum_{i\in \ms_{[q]}}\Big\{1-\frac{I(Z_i=z)}{\prob(Z_i=z\mid \ms_{[1]},\ms_{[2]})}\Big\} \{f^\ast_z (\bs{x}_i)-\hat{f}_{[-q]z}(\bs{x}_i)\}, \quad z \in \{0,1\}.
  \end{align*}
By \Cref{a:negligible-estimateion-error-for-prediction-function} and \Cref{lem:property-of-op-1}, there exists $C_N\rightarrow 0$, such that
\begin{align*}
    \prob(\|f^\ast_z-\hat{f}_{[-q]z}\|_N^2 > C_N) = o(1).
\end{align*}
Denote the event
\begin{align*}
   \mathcal{A} := \Big\{\frac{1}{N} \sum_{i\in \ms_{[q]}} \{f^\ast_z (\bs{x}_i)-\hat{f}_{[-q]z}(\bs{x}_i)\}^2 \leq C_N \Big\}.
\end{align*}
We have
\begin{align*}
    \prob(\mathcal{A}^c) \leq \prob(\|f^\ast_z-\hat{f}_{[-q]z}\|_N^2 > C_N) = o(1).
\end{align*}

\Cref{a:conditional-independent-subsamples} implies that $\prob_{\bs{Z}_{[q]}\mid \ms_{[1]},\ms_{[2]},\bs{Z}_{[-q]},\mathcal{A}} = \prob_{\bs{Z}_{[q]}\mid \ms_{[1]},\ms_{[2]}}$. Then, by \Cref{a:conditional-experimental-design-bounded-variance} with $a_i=f^\ast_z (\bs{x}_i)-\hat{f}_{[-q]z}(\bs{x}_i)$, we have  
\begin{align*}
   &\Var(T_{[q]z}\mid \bs{Z}_{[-q]},\ms_{[1]},\ms_{[2]},\mathcal{A}) \leq \frac{C_z}{N^2} \sum_{i\in \ms_{[q]}} \{f^\ast_z (\bs{x}_i)-\hat{f}_{[-q]z}(\bs{x}_i)\}^2 
   \leq \frac{C_z C_N}{N} .
\end{align*}

Therefore, for any fixed constant $C^\prime > 0$, we have
\begin{align*}
    &\prob(N^{1/2}|T_{[q]z}| > C^{\prime}) \leq \prob(N^{1/2}|T_{[q]z}| > C^{\prime}\mid \mathcal{A}) + \prob( \mathcal{A}^c)\\
    \leq  & \ope \Big\{\prob(N^{1/2}|T_{[q]z}| > C^{\prime} \mid \bs{Z}_{[-q]},\ms_{[1]},\ms_{[2]},\mathcal{A} )~\Big |~ \mathcal{A} \Big\} + \prob(\mathcal{A}^c)\\
    \leq & \ope \Big\{N\Var(T_{[q]z}\mid \bs{Z}_{[-q]},\ms_{[1]},\ms_{[2]},\mathcal{A})/C^{\prime 2}~\Big |~ \mathcal{A} \Big\} + o(1) \\
    \leq & \frac{C_zC_N}{C^{\prime 2}} + o(1) = o(1),
\end{align*}
where the last equality is because $C_N \rightarrow 0$.  

As a consequence, $N^{1/2}T_{[q]z} = \op(1)$, for $q\in \{1,2\}$, $z\in \{0,1\}$. Therefore, $N^{1/2} (\hat{\tau}_{\oraclecf} - \hat{\tau}_{\cfadj}) =  \op(1)$. By \Cref{a:nondegenerate-variance} and $(\hat{\tau}_{\oraclecf}-\bar{\tau})/\Var(\hat{\tau}_{\oraclecf})^{1/2} \rightsquigarrow \mathcal{N}(0,1)$, we have $(\hat{\tau}_{\cfadj}-\bar{\tau})/\Var(\hat{\tau}_{\oraclecf})^{1/2} \rightsquigarrow \mathcal{N}(0,1)$.
\end{proof}

\subsection{Proof of \Cref{thm:constant-probability-optimal}}
%\begin{proof}%[Proof of \Cref{prop:cf-oracle}]
%    Under \Cref{a:constant-treatment-assignment-probability}, we have $\hat{\tau}_{\oraclecf} = \hat{\tau}_{\oracleadj}$. The conclusion follows from \Cref{prop:oraclecf-and-cf}.
%\end{proof}

%\section{Proof of \Cref{thm:constant-probability-optimal}}
%\label{sec:proof-of-thm-2}
\begin{proof}
By the law of total variance, we have
    \begin{align*}
    \Var(\hat{\tau}_{\oraclecf}) & = \ope\Big\{\Var(\hat{\tau}_{\oraclecf}\mid \bs{Z})\Big\} + \Var\{\ope (\hat{\tau}_{\oraclecf}\mid \bs{Z})\}\\
    & \geq  \Var\{\ope (\hat{\tau}_{\oraclecf}\mid \bs{Z})\}.
\end{align*}

Next, we will show that 
$
\Var\{\ope (\hat{\tau}_{\oraclecf}\mid \bs{Z})\} = \Var(\hat{\tau}_{\oracleadj}),
$
from which the first conclusion follows.

We decompose $\hat{\tau}_{\oraclecf} = T_1-T_0$, where, for $z \in \{0,1\}$,
\begin{align*}
    T_z &= \frac{1}{N} \sum_{i:Z_i=z} \Big\{\frac{(Y_i-f^\ast_z(\bs{x}_i))R_i}{\prob(Z_i=z\mid \bs{R})} + \frac{(Y_i-f^\ast_z(\bs{x}_i))(1-R_i)}{\prob(Z_i=z\mid \bs{R})}\Big\} + \frac{1}{N} \sumi f^\ast_z(\bs{x}_i).
\end{align*}
\Cref{a:a-class-of-design} suggests that, for $z \in \{0,1\}$ and $r \in \{0,1\}$,
\begin{align*}
         &\prob(Z_i=z\mid \bs{R}) = \prob(Z_i=z\mid R_i),\quad \prob(R_i = r \mid \bs{Z}) = \prob(R_i = r \mid Z_i).
\end{align*}
Therefore, 
\begin{align*}
      T_z & = \frac{1}{N} \sum_{i:Z_i=z} \Big\{\frac{(Y_i-f^\ast_z(\bs{x}_i))R_i}{\prob(Z_i=z\mid R_i)} + \frac{(Y_i-f^\ast_z(\bs{x}_i))(1-R_i)}{\prob(Z_i=z\mid R_i)}\Big\} + \frac{1}{N} \sumi f^\ast_z(\bs{x}_i)\\
      & = \frac{1}{N} \sum_{i:Z_i=z} \Big\{\frac{(Y_i-f^\ast_z(\bs{x}_i))R_i}{\prob(Z_i=z\mid R_i=1)} + \frac{(Y_i-f^\ast_z(\bs{x}_i))(1-R_i)}{\prob(Z_i=z\mid R_i=0)}\Big\} + \frac{1}{N} \sumi f^\ast_z(\bs{x}_i).
\end{align*}

For $r\in \{0,1\}$, we have
 \begin{align*}
      & \ope \Big\{\frac{I(R_i=r)}{\prob(Z_i=z\mid R_i=r)}\mid \bs{Z}\Big\} = \frac{\prob(R_i=r\mid\bs{Z})}{\prob(Z_i=z\mid R_i=r)} = \frac{\prob(R_i=r\mid Z_i)}{\prob(Z_i=z\mid R_i=r)}.
 \end{align*}
In light of the above, we have
\begin{align*}
   & \ope(T_z\mid \bs{Z}) \\
    =& \frac{1}{N} \sum_{i:Z_i=z} \{Y_i-f^\ast_z(\bs{x}_i)\} \Big\{\frac{\prob(R_i=0\mid Z_i=z)}{\prob(Z_i=z\mid R_i=0)} + \frac{\prob(R_i=1\mid Z_i=z)}{\prob(Z_i=z\mid R_i=1)} \Big\} + \frac{1}{N} \sumi f^\ast_z(\bs{x}_i) \\
    =&\frac{1}{N} \sum_{i:Z_i=z} \{Y_i-f^\ast_z(\bs{x}_i)\} \frac{1}{\prob(Z_i=z)} + \frac{1}{N} \sumi f^\ast_z(\bs{x}_i),
\end{align*}
where the last equality holds because
\begin{align*}
    \frac{\prob(R_i=0\mid Z_i=z)}{\prob(Z_i=z\mid R_i=0)} =  \frac{\prob(R_i=0)}{\prob(Z_i=z)},\quad \frac{\prob(R_i=1\mid Z_i=z)}{\prob(Z_i=z\mid R_i=1)} = \frac{\prob(R_i=1)}{\prob(Z_i=z)}.
\end{align*}
As a consequence, we have
\begin{align*}
    \ope(\hat{\tau}_{\oraclecf}\mid \bs{Z}) = \hat{\tau}_{\oracleadj}.
\end{align*}
Therefore, 
\begin{align*}
    \Var(\hat{\tau}_{\oraclecf}) \geq \Var\{\ope (\hat{\tau}_{\oraclecf}\mid \bs{Z})\} = \Var(\hat{\tau}_{\oracleadj}).
\end{align*}
When \Cref{a:constant-treatment-assignment-probability} holds, $\hat{\tau}_{\oracleadj} = \hat{\tau}_{\oraclecf}$ and the equality is achieved.

% \begin{align*}
%     \sum_{q \in \{1,2\}}\ope \Big\{\frac{I(i\in \ms_{[q]},Z_i=z)}{\prob(Z_i=z\mid \ms_{[1]},\ms_{[2]})}\mid Z_i\Big\} = \sum_{q \in \{1,2\}} \frac{I(Z_i=z)\prob(i\in \ms_{[q]}\mid Z_i)}{\prob(Z_i=z\mid i\in \ms_{[q]})}= \frac{I(Z_i=z)}{\prob(Z_i=z)}
% \end{align*}
\end{proof}

\subsection{Proof of \Cref{thm:valid-confidence-interval-2}}
\label{sec:proof-for-confidence-intervals}

Let $\Delta_i = \hat{\varepsilon}_i - \varepsilon_i = \sum_{q \in \{1,2\}} \sum_{z\in \{0,1\}}\{f_z^\ast(\bs{x}_i)-\hat{f}_{[-q]z}(\bs{x}_i)\}I(Z_i = z,i\in\ms_{[q]})$ and $\Delta_i(z) = \sum_{q \in \{1,2\}}\{f_z^\ast(\bs{x}_i)-\hat{f}_{[-q]z}(\bs{x}_i)\}I(i\in\ms_{[q]})$. Let $$\hat{V}_{\oraclecf} = \sum_{q\in \{1,2\}} (N_{[q]}/N)^2\hat{V}_{\oraclecf, [q]},$$
where
\begin{align*}
  \hat{V}_{\oraclecf, [q]} =  \hat V\Big(( \varepsilon_i)_{i\in \ms_{[q]}},\bs{Z}_{[q]};\prob_{\bs{Z}_{[q]}\mid \ms_{[1]},\ms_{[2]}} \Big).
\end{align*}
Here, $\hat V$ is a generic variance estimator and should be constructed case by case.

%We now verify conditions \eqref{eq:conservative-variance} and \eqref{eq:clt-for-oracle-cf}. Let $$\hat{V}_{\oraclecf} = \sum_{q\in \{1,2\}} (N_{[q]}/N)^2\hat{V}_{\oraclecf, [q]}.$$ For condition \eqref{eq:conservative-variance}, we can decompose
%\begin{align*}
%     & N \{\hat{V}_{\cfadj} -\Var(\hat{\tau}_{\oraclecf})\} =  \underset{\text{Term I}}{\underbrace{N\{\hat{V}_{\cfadj} - \hat{V}_{\oraclecf}\}}} + \underset{\text{Term II}}{\underbrace{N\big\{\hat{V}_{\oraclecf} - \Var(\hat{\tau}_{\oraclecf})\big\}}}.
%\end{align*}
%Term I is the error when replacing the oracle prediction function with the estimated one. Term II is the estimation error of $\hat{V}_{\oraclecf}$. 
% Our routine to establish \eqref{eq:conservative-variance} is to show that
%  \[
%  \text{Term I} = \op(1),\quad \plim_{N\rightarrow \infty}\text{Term II} \geq 0.
%  \]
%For condition \eqref{eq:clt-for-oracle-cf}, we will use the established CLT from the literature.

\begin{proof}[Proof of \Cref{thm:valid-confidence-interval-2}]
Let $\Delta = \plim_{N \rightarrow \infty} N \{\hat{V}_{\cfadj} -\Var(\hat{\tau}_{\oraclecf})\} \geq 0$. Then, 
\[
N\hat{V}_{\cfadj} - \{N\Var(\hat{\tau}_{\oraclecf}) + \Delta\} = \op(1).
\]
On the other hand, we have
\begin{align*}
    N^{1/2}(\hat{\tau}_{\cfadj} - \bar{\tau}) &= N^{1/2}(\hat{\tau}_{\oraclecf} - \bar{\tau}) +  N^{1/2}(\hat{\tau}_{\cfadj}-\hat{\tau}_{\oraclecf}) \\
    &= N^{1/2}(\hat{\tau}_{\oraclecf} - \bar{\tau}) + \op(1),
\end{align*}
where the last equality holds because by \Cref{thm:valid-confidence-interval}, we have
\[
N^{1/2}(\hat{\tau}_{\oraclecf} - \hat{\tau}_{\cfadj}) = \op(1).
\]
Hence, if Condition \eqref{eq:clt-for-oracle-cf} in \Cref{thm:valid-confidence-interval-2} holds, i.e., $(\hat{\tau}_{\oraclecf}-\bar{\tau})/\Var(\hat{\tau}_{\oraclecf})^{1/2} \rightsquigarrow \mathcal{N}(0,1)$, we have $(\hat{\tau}_{\cfadj}-\bar{\tau})/\Var(\hat{\tau}_{\oraclecf})^{1/2} \rightsquigarrow \mathcal{N}(0,1)$. Moreover,
\begin{align*}
    \frac{\hat{\tau}_{\cfadj} - \bar{\tau}}{ (\hat{V}_{\cfadj})^{1/2} } =& \frac{N^{1/2}(\hat{\tau}_{\cfadj} - \bar{\tau})}{(N\hat{V}_{\cfadj})^{1/2}} 
    = \frac{ N^{1/2}(\hat{\tau}_{\oraclecf} - \bar{\tau}) + \op(1)} { \{N\Var(\hat{\tau}_{\oraclecf}) + \Delta\}^{1/2} + \op(1)} \\
    =& \frac{N^{1/2}(\hat{\tau}_{\oraclecf} - \bar{\tau})}{\{N\Var(\hat{\tau}_{\oraclecf}) + \Delta\}^{1/2}} + \op(1) \quad \textnormal{(by \Cref{a:nondegenerate-variance})}.
\end{align*}

Let $\sigma^2_{\oraclecf}$ denote the limit of $N\Var(\hat{\tau}_{\oraclecf})$. Using $(\hat{\tau}_{\oraclecf}-\bar{\tau})/\Var(\hat{\tau}_{\oraclecf})^{1/2} \rightsquigarrow \mathcal{N}(0,1)$, we have
\begin{align*}
    \frac{\hat{\tau}_{\cfadj} - \bar{\tau}}{\hat{V}_{\cfadj}} \rightsquigarrow \mathcal{N}\Big(0,\frac{\sigma^2_{\oraclecf}}{\sigma^2_{\oraclecf} + \Delta}\Big).
\end{align*}  
Therefore,
$$
\liminf_{n \rightarrow \infty} \prob( \bar{\tau} \in [\hat{\tau}_{\cfadj} - z_{1-\alpha/2}\hat{V}_{\cfadj}^{1/2}, \hat{\tau}_{\cfadj} + z_{1-\alpha/2}\hat{V}_{\cfadj}^{1/2}] )
\geq 1 - \alpha.
$$
This completes the proof.
\end{proof}

\subsection{Proof of \Cref{prop:valid-inference-bt}}

\begin{proof}

\noindent \textbf{Part I: We verify Condition \eqref{eq:conservative-variance} by proving that}
\begin{align*}
    N\{\hat{V}_{\cfadj} - \Var(\hat{\tau}_{\oraclecf})\} -\frac{1}{N-1}\sumi (\tau_{\varepsilon,i} - \bar{\tau}_{\varepsilon})^2 =  \op(1).
\end{align*}

\textbf{Step I:} We prove that
\begin{equation}
\label{eqn:ora-cf}
    N(\hat{V}_{\oraclecf} - \hat{V}_{\cfadj}) = \op(1).
\end{equation}

Recall the definition of $\hat{D}_{\varepsilon,i}$ and $\hat{\bar{D}}_{\varepsilon,[q]}$ in the main text. Analogously, we define
\begin{align*}
      &{D}_{\varepsilon,i} =  \frac{Z_i \varepsilon_i}{r_1} - \frac{(1-Z_i) {\varepsilon}_i}{r_0},\quad  {\bar{D}}_{\varepsilon,[q]} = \frac{1}{N_{[q]}}\sum_{i\in \ms_{[q]}} {D}_{\varepsilon,i},\\
      &{D}_{\Delta,i} =  \frac{Z_i \Delta_i}{r_1} - \frac{(1-Z_i) {\Delta}_i}{r_0},\quad  {\bar{D}}_{\Delta,[q]} = \frac{1}{N_{[q]}}\sum_{i\in \ms_{[q]}} {D}_{\Delta,i}.
\end{align*}
We have $N(\hat{V}_{\oraclecf} - \hat{V}_{\cfadj}) = \sum_{q \in \{1,2\}}  T_{[q]}$ with 
%\begin{align*}
%    N(\hat{V}_{\oraclecf} - \hat{V}_{\cfadj}) = \sum_{q \in \{1,2\}}  T_{[q]}, 
%\end{align*}
\begin{align*}
  T_{[q]} &= \frac{1}{N} \sum_{i\in \ms_{[q]}} \Big\{(D_{\varepsilon,i} - \bar{D}_{\varepsilon,[q]})^2  - (\hat{D}_{\varepsilon,i} - \hat{\bar{D}}_{\varepsilon,[q]})^2 \Big\}\\
  & =  - \frac{1}{N} \sum_{i\in \ms_{[q]}} \Big\{({D}_{\Delta,i}  - {\bar{D}}_{\Delta,[q]})^2 + 2({D}_{\Delta,i} - {\bar{D}}_{\Delta,[q]})({D}_{\varepsilon,i} - {\bar{D}}_{\varepsilon,[q]})\Big\},
  % &= \frac{1}{N} \sum_{i\in \ms_{[q]}} \Big\{\hat{D}_{\varepsilon,i}^2 - D_{\varepsilon,i}^2 \Big\} - \frac{N_{[q]}}{N} \Big(\hat{\bar{D}}_{\varepsilon,[q]}^2 - \bar{D}_{\varepsilon,[q]}^2\Big)
\end{align*}
where the last equality follows from
$
D_{\varepsilon,i} -\hat{D}_{\varepsilon,i} = - D_{\Delta,i},\ \bar{D}_{\varepsilon,i}-\hat{\bar{D}}_{\varepsilon,i} = - \bar{D}_{\Delta,i}.
$

We bound the two terms separately. For the first term, we have
\begin{align*}
    \frac{1}{N} \sum_{i\in \ms_{[q]}} ({D}_{\Delta,i} - {\bar{D}}_{\Delta,[q]})^2 \leq & \frac{1}{N} \sum_{i\in \ms_{[q]}} {D}_{\Delta,i}^2 =    \frac{1}{N}\sum_{i\in \ms_{[q]}} \Big\{\frac{Z_i \Delta_i^2(1)}{r_1^2} +  \frac{(1-Z_i) \Delta_i^2(0)}{(1-r_1)^2}\Big\}\\
      \leq  &\frac{1}{N}\sum_{i\in \ms_{[q]}} \Big\{\frac{ \Delta_i^2(1)}{r_1^2} +  \frac{\Delta_i^2(0)}{(1-r_1)^2}\Big\} \\
      = &   O(\|f^\ast_0-\hat{f}_{[-q]0}\|_N^2+\|f^\ast_1-\hat{f}_{[-q]1}\|_N^2)
      =  \op(1),
\end{align*}
and
\begin{align*}
    \frac{1}{N} \sum_{i\in \ms_{[q]}} ({D}_{\varepsilon,i} - {\bar{D}}_{\varepsilon,[q]})^2 \leq  \frac{1}{N} \sum_{i\in \ms_{[q]}} {D}_{\varepsilon,i}^2 = O\Big(\frac{1}{N}\sumi \{\varepsilon_i^2(1)+\varepsilon_i^2(0)\}\Big) = O(1).
\end{align*}

By the Cauchy--Schwarz inequality, we have
\begin{align*}
   &\Big| \frac{1}{N} \sum_{i\in \ms_{[q]}} ({D}_{\Delta,i} - {\bar{D}}_{\Delta,[q]})({D}_{\varepsilon,i} - {\bar{D}}_{\varepsilon,[q]}) \Big| \\
   \leq & \Big\{\frac{1}{N} \sum_{i\in \ms_{[q]}} ({D}_{\Delta,i} - {\bar{D}}_{\Delta,[q]})^2\Big\}^{1/2} \Big\{\frac{1}{N} \sum_{i\in \ms_{[q]}} ({D}_{\varepsilon,i} - {\bar{D}}_{\varepsilon,[q]})^2\Big\}^{1/2}
   = \op(1)O(1) = \op(1).
\end{align*}
As a consequence, we have $T_{[q]} = \op(1)$ for $q \in \{1,2\}$. Therefore, $N(\hat{V}_{\oraclecf} - \hat{V}_{\cfadj}) = \op(1)$.

\textbf{Step II:} We prove that
$$N\{\hat{V}_{\oraclecf}-\Var(\hat{\tau}_{\oraclecf})\} - \frac{1}{N-1}\sumi (\tau_{\varepsilon,i}-\bar{\tau}_{\varepsilon})^2 = \op(1).$$
By the definition of $\hat{V}_{\oraclecf}$, we have
\begin{align*}
  N \hat{V}_{\oraclecf} =  \frac{1}{N} \sum_{q \in \{1,2\}}\sum_{i\in \ms_{[q]}} (D_{\varepsilon,i} - \bar{D}_{\varepsilon,[q]})^2 =   \frac{1}{N} \sumi D_{\varepsilon,i}^2 - \sum_{q \in \{1,2\}} \frac{N_{[q]}}{N} \bar{D}_{\varepsilon,[q]}^2.
\end{align*}
Let $R_{i,q} = I(i\in\ms_{[q]})$ and $\pi_q=\pi$ if $q=1$ and $\pi_q = 1-\pi$ if $q=2$. We have
\begin{align*}
  \bar{D}_{\varepsilon,[q]} =  \frac{1}{N_{[q]}} \sumi \Big\{\frac{\varepsilon_i(1) Z_iR_{i,q}}{r_1}- \frac{\varepsilon_i(0) (1-Z_i)R_{i,q}}{r_0}\Big\}.
\end{align*}
Since $(Z_i R_{i,q})_{i=1}^N$, $((1-Z_i) R_{i,q})_{i=1}^N$, and $(R_{i,q})_{i=1}^N$ follow from $\BT(N,r_1 \pi_q )$, $\BT(N,r_0 \pi_q )$, and $\BT(N, \pi_q )$, respectively, and by Chebyshev's inequality and \Cref{a:regularity-condition-bernoulli-trial}, we have
\[
 \frac{1}{N} \sum_{i:Z_i=z} \frac{\varepsilon_i(z) R_i}{r_1} =  \frac{1}{N} \sumi \pi_q \varepsilon_i(z) + \Op(N^{-1/2}),\quad  \frac{N_{[q]}}{N} = \pi_q + \Op(N^{-1/2}).
\]
Therefore, 
\[
\bar{D}_{\varepsilon,[q]} - \bar{\tau}_{\varepsilon} = \Op(N^{-1/2}),\quad \sum_{q \in \{1,2\}} \frac{N_{[q]}}{N} \bar{D}_{\varepsilon,[q]}^2 = \bar{\tau}_{\varepsilon}^2 + \Op(N^{-1/2}).
\]

It remains to derive the limit of $N^{-1} \sumi D_{\varepsilon,i}^2$. Simple calculation yields
\begin{align*}
    \frac{1}{N} \sumi D_{\varepsilon,i}^2 = \frac{1}{N}\sumi \Big[\frac{ Z_i \varepsilon_i^2(1)}{r_1^2} +  \frac{(1-Z_i) \varepsilon_i^2(0)}{r_0^2}\Big].
\end{align*}
By Chebyshev's inequality and \Cref{a:regularity-condition-bernoulli-trial}, we have
\begin{align*}
    & \frac{1}{N} \sumi D_{\varepsilon,i}^2 - \frac{1}{N}\sumi \Big\{\frac{ \varepsilon_i^2(1)}{r_1} +  \frac{\varepsilon_i^2(0)}{r_0}\Big\}
    = \Op\Big(\frac{1}{N}\Big[\sumi \{\varepsilon_i^4(1)+ \varepsilon_i^4(0)\} \Big]^{1/2}\Big)
    = \Op\Big(\frac{1}{\sqrt{N}}\Big).
\end{align*}
Therefore, 
\begin{align}
\label{eq:hat-V-oracle-cf-BT}
    N\hat{V}_{\oraclecf} - \frac{1}{N}\sumi \Big\{\frac{\varepsilon_i^2(1)}{r_1} +  \frac{\varepsilon_i^2(0)}{r_0}-\bar{\tau}_{\varepsilon}^2\Big\}  = \op(1).
\end{align}

%  = \frac{1}{N r_1r_0} \sumi (r_1\varepsilon_i(0)+r_0\varepsilon_i(1))^2
On the other hand, we have
\begin{align}
\nonumber
 N\Var(\hat{\tau}_{\oraclecf}) =& N\Var\Big(\frac{1}{N}\sumi\frac{(Z_i-r_1)\{r_1\varepsilon_i(0)+r_0\varepsilon_i(1)\}}{r_1r_0}\Big)\\
\nonumber
    =& \frac{1}{N r_1r_0} \sumi \{r_1\varepsilon_i(0)+r_0\varepsilon_i(1)\}^2\\
\label{eq:Var-oracle-cf-BT}
    =& \frac{1}{N}\sumi \Big\{\frac{ \varepsilon_i^2(1)}{r_1} +  \frac{\varepsilon_i^2(0)}{r_0}\Big\} - \frac{1}{N}\sumi \tau_{\varepsilon, i}^2, 
\end{align}
where for the last equality, we use the equality $(r_1 a + r_0 b)^2 = r_1 a^2 + r_0 b^2 - r_1r_0(a-b)^2$ with $a = \varepsilon_i(0)$ and $b = \varepsilon_i(1)$. Comparing \eqref{eq:hat-V-oracle-cf-BT} and \eqref{eq:Var-oracle-cf-BT}, and using the fact that $N^{-1} \sumi \tau_{\varepsilon, i}^2 - \bar{\tau}_{\varepsilon}^2 = N^{-1} \sumi (\tau_{\varepsilon, i}-\bar{\tau}_{\varepsilon})^2$, we have
   \begin{align*}
       N\{\hat{V}_{\oraclecf} - \Var(\hat{\tau}_{\oraclecf})\} -\frac{1}{N-1}\sumi (\tau_{\varepsilon,i} - \bar{\tau}_{\varepsilon})^2 =  \op(1).
   \end{align*}
Combining this with \eqref{eqn:ora-cf}, we have
\begin{align*}
    N\{\hat{V}_{\cfadj} - \Var(\hat{\tau}_{\oraclecf})\} -\frac{1}{N-1}\sumi (\tau_{\varepsilon,i} - \bar{\tau}_{\varepsilon})^2 =  \op(1).
\end{align*}

\textbf{Part II: We verify Condition \eqref{eq:clt-for-oracle-cf}}.

 Since the sample-splitting algorithm for BREs satisfies Assumptions~\ref{a:a-class-of-design} and \ref{a:constant-treatment-assignment-probability}, we have $\hat{\tau}_{\oraclecf} = \hat{\tau}_{\oracleadj}$. Note that
\[
 N^{-3/2}\sumi\ope |{D}_{\varepsilon,i}|^3 = O\Big( N^{-3/2}\sumi \big\{\varepsilon_i^{3}(1)+\varepsilon_i^{3}(0)\big\}\Big) = o(1).
\]
By Lyapunov's central limit theorem (CLT), we have $(\hat{\tau}_{\oraclecf}-\bar{\tau})/\Var(\hat{\tau}_{\oraclecf})^{1/2} \rightsquigarrow  \mathcal{N}(0,1).$
\end{proof}

\subsection{Proof of \Cref{prop:valid-inference-cr}}
\begin{proof}

\noindent \textbf{Part I: We verify Condition \eqref{eq:conservative-variance} by proving that}
\[
N\{\hat{V}_{\cfadj} - \Var(\hat{\tau}_{\oraclecf}) \} -\frac{1}{N-1}\sumi (\tau_{i,\varepsilon} -  \bar{\tau}_{\varepsilon})^2 = \op(1).
\]

\textbf{Step I:} We prove that
\begin{equation}
\label{eqn:cfadj-ora}
    N(\hat{V}_{\cfadj} - \hat{V}_{\oraclecf}) = \op(1).
\end{equation}

Recall 
\begin{align*}
   &\hat{V}_{\cfadj[q]} =  \frac{S_{[q]1,\hat{\varepsilon}\hat{\varepsilon}}}{N_{[q]1}} + \frac{S_{[q]0,\hat{\varepsilon}\hat{\varepsilon}}}{N_{[q]0}}.
\end{align*}
Therefore, 
\[
    N(\hat{V}_{\cfadj} - \hat{V}_{\oraclecf}) = \sum_{q \in \{1,2\}} \frac{N_{[q]}^2}{N} (\hat{V}_{\cfadj,[q]} - \hat{V}_{\oraclecf, [q]}) = \sum_{q \in \{1,2\}} \sum_{z\in\{0,1\}} M_{[q]z},
\]
where
    \begin{align*}
       M_{[q]z} =& \frac{N_{[q]}^2}{N_{[q]z}N} (S_{[q]z,\hat{\varepsilon}\hat{\varepsilon}} - S_{[q]z,{\varepsilon}{\varepsilon}})
       = \frac{N_{[q]}^2}{N_{[q]z}N} (S_{[q]z,\Delta\Delta} + 2S_{[q]z,\Delta \varepsilon}).
    \end{align*}

    Using the facts that
\begin{align*}
        & S_{[q]z,\Delta\Delta} \leq \frac{1}{N-1}\sum_{i\in \ms_{[q]},Z_i=z}  \Delta_i^2 = O(\|f^\ast_z-\hat{f}_{[-q]z}\|_N^2) = \op(1),\\
        & S_{[q]z,\varepsilon\varepsilon} \leq \frac{1}{N-1}\sum_{i\in \ms_{[q]},Z_i=z} \varepsilon_i^2 \leq \frac{1}{N-1}\sumi \varepsilon_i^2(z) = O(1),
\end{align*}
and $| S_{[q]z,\Delta \varepsilon} | \leq S_{[q]z,\Delta\Delta}^{1/2} S_{[q]z,\varepsilon\varepsilon}^{1/2} = \op(1)$, we have $M_{[q]z} = \op(1)$. Therefore, 
 $$N(\hat{V}_{\cfadj} - \hat{V}_{\oraclecf}) = \op(1).$$

% We note that
% \begin{align*}
%     \bar{\Delta}_{[q]z}-\bar{\Delta}_{[q]}(z) = \frac{1}{N_{[q]}}\sum_{i\in \ms_{[q]}} \Big(\frac{N_{[q]}}{N_{[q]z}} Z_i - 1\Big)\Delta_i(z)
% \end{align*}
% is the term $T_{[q]z}$ in the proof of \Cref{prop:oraclecf-and-cf} multiplied by $N/N_{[q]}$. Therefore 
% \[
% \bar{\Delta}_{[q]z}-\bar{\Delta}_{[q]}(z) = \op(N^{-1/2}).
% \]

% On the other hand, we have
%  \begin{align*}
%      & N_{[q]}^{-1} \sum_{i\in \ms_{[q]},Z_i=z}  (\Delta_i-\bar{\Delta}_{[q]}(z))^2 \leq N_{[q]}^{-1} \sum_{i\in \ms_{[q]}}  (\Delta_i(z)-\bar{\Delta}_{[q]}(z))^2 \\
%      \leq & N_{[q]}^{-1} \sum_{i\in \ms_{[q]}}  \Delta_i(z)^2 
%      = O(\|f^\ast_z - \hat{f}_{[-q]z}\|_N^2) = \op(1).
%  \end{align*}
% Putting together, we have
% \[
%  \frac{1}{N}\sum_{i\in \ms_{[q]},Z_i=z} (\Delta_i-\bar{\Delta}_{[q]z})^2 = \op(1)
% \]
% For the second term
%     \item 
% \begin{align*}
%     \sum_{i\in \ms_{[q]},Z_i=z}  (\varepsilon_i - \bar{\varepsilon}_{[q]z})(\Delta_i-\bar{\Delta}_{[q]z}) = \sum_{i\in \ms_{[q]},Z_i=z}  (\varepsilon_i \Delta_i -\bar{\varepsilon}_{[q]z}\bar{\Delta}_{[q]z})
% \end{align*}
% \item Therefore $T_z = \op(1)$
% \item 
% \begin{align*}
%     N(\hat{V}_{\oraclecf} - \Var(\hat{\tau}_{\oracleadj})) = T_1 + T_0
% \end{align*}

\textbf{Step II:} We prove that 
\begin{align}
    \label{eqn:Eora-Vora}
    N\{\ope (\hat{V}_{\oraclecf}) - \Var(\hat{\tau}_{\oraclecf}) \} =\frac{1}{N-1}\sumi (\tau_{i,\varepsilon} -  \bar{\tau}_{\varepsilon})^2.
\end{align}

 Using $\prob_{\bs{Z}_{[q]}\mid \ms_{[1]},\ms_{[2]}} = \CR(N_{[q]}, N_{[q]1})$ and \Cref{lem:variance-of-CR}(i), we have
\begin{align*}
    \Var(\hat{\tau}_{\oraclecf,[q]}\mid \ms_{[1]},\ms_{[2]})  = & \sum_{z\in\{0,1\}} \frac{S_{[q],\varepsilon(z)\varepsilon(z)}}{N_{[q]z}} - \frac{S_{[q],\tau_\varepsilon \tau_\varepsilon}}{N_{[q]}}.
\end{align*}
Let $R_{i,q} = I(i\in \ms_{[q]})$ and $\bs{R}_q = (R_{i,q})_{i=1}^N$, we have $\prob_{\bs{R}_q} = \CR(N,N_{[q]})$. Therefore, by \Cref{lem:unbiasedness-of-variance-estimator} with $a_i = \varepsilon_i(z)$ or $\tau_{\varepsilon,i}$, we have 
\begin{align}
\label{eq:Var-oracle-cf-CR}
  &\Var(\hat{\tau}_{\oraclecf[q]})  = \ope \{\Var(\hat{\tau}_{\oraclecf[q]}\mid \ms_{[1]},\ms_{[2]}) \}
  =  \sum_{z\in \{0,1\}} \frac{S_{\varepsilon(z) \varepsilon(z)}}{N_{[q]z}} - \frac{S_{\tau_\varepsilon\tau_\varepsilon}}{N_{[q]}}. 
\end{align}

By the definition of $\hat{V}_{\oraclecf,[q]}$, we have
\begin{align*}
   &\hat{V}_{\oraclecf,[q]} = \sum_{z\in\{0,1\}}  \frac{S_{[q]z,\varepsilon\varepsilon}}{N_{[q]z}}. 
\end{align*}
Since $(I(Z_i=z, i\in\ms_{[q]}))_{i=1}^N$ follows $\CR(N,N_{[q]z})$, then applying \Cref{lem:unbiasedness-of-variance-estimator} with $a_i = \varepsilon_i(z)$ and $\bs{Z} = (I(Z_i=z, i\in\ms_{[q]}))_{i=1}^N$, we have
\begin{align*}
   &\ope \bigg( \frac{S_{[q]z,\varepsilon\varepsilon}}{N_{[q]z}} \bigg) = \frac{S_{\varepsilon(z)\varepsilon(z)}}{N_{[q]z}}.
\end{align*}
Therefore,
\begin{align}
\label{eq:hat-V-oracle-cf-CR}
    \ope (\hat{V}_{\oraclecf,[q]}) =  \sum_{z\in\{0,1\}}  \frac{S_{\varepsilon(z)\varepsilon(z)}}{N_{[q]z}}.
\end{align}
Combining \eqref{eq:Var-oracle-cf-CR} and \eqref{eq:hat-V-oracle-cf-CR}, we have
\begin{align*}
    \ope (\hat{V}_{\oraclecf,[q]}) - \ope \{ \Var(\hat{\tau}_{\oraclecf[q]}\mid \ms_{[1]},\ms_{[2]}) \} = \frac{1}{(N-1)N_{[q]}} \sumi (\tau_{\varepsilon,i} -  \bar{\tau}_{\varepsilon})^2.
\end{align*}
Recall 
$$\Var(\hat{\tau}_{\oraclecf}) = \sum_{q \in \{1,2\}} \frac{N_{[q]}^2}{N^2} \ope \{ \Var(\hat{\tau}_{\oraclecf,[q]}\mid \ms_{[1]},\ms_{[2]}) \}.$$
Therefore,
\begin{align*}
    N\big\{\ope (\hat{V}_{\oraclecf}) - \Var(\hat{\tau}_{\oraclecf})\big\}  = & \sum_{q \in \{1,2\}} \frac{N_{[q]}^2}{N} \frac{1}{(N-1)N_{[q]}}\sumi (\tau_{\varepsilon,i} -  \bar{\tau}_{\varepsilon})^2 \\
    = &  \frac{1}{N-1}\sumi (\tau_{\varepsilon,i} -  \bar{\tau}_{\varepsilon})^2.
\end{align*}

\textbf{Step III:} We prove that
\begin{align}
    \label{eqn:ora-Eora}
N \{ \hat{V}_{\oraclecf} - \ope (\hat{V}_{\oraclecf}) \} = \op(1).
\end{align}

It suffice to prove that $N\{\hat{V}_{\oraclecf,[q]} - \ope( \hat{V}_{\oraclecf,[q]})\} = \op(1)$. We have
\begin{align*}
 \frac{S_{[q]z,\varepsilon\varepsilon}}{N_{[q]z}} = \sum_{i\in \ms_{[q]},Z_i=z} \frac{\{{\varepsilon}_i - {\bar{\varepsilon}}(z)\}^2}{N_{[q]z}(N_{[q]z}-1)} - \frac{\{\bar{\varepsilon}(z) - {\bar{\varepsilon}}_{[q]z}\}^2}{N_{[q]z}-1}.
\end{align*}
Therefore,
\begin{align*}
    \Var\Big(\frac{S_{[q]z,\varepsilon\varepsilon}}{N_{[q]z}}\Big) \leq 2\Var\Big(\sum_{i\in \ms_{[q]},Z_i=z} \frac{\{{\varepsilon}_i - {\bar{\varepsilon}}(z)\}^2}{N_{[q]z}(N_{[q]z}-1)}\Big) + 2\Var\Big(\frac{\{\bar{\varepsilon}(z) - {\bar{\varepsilon}}_{[q]z}\}^2}{N_{[q]z}-1}\Big).
\end{align*}

Using the fact that $(I(i\in \ms_{[q]},Z_i=z))_{i=1}^N$ follows $\CR(N,N_{[q]z})$ and \Cref{lem:variance-of-CR}(ii), we have
\begin{align*}
    \Var\Big(\sum_{i\in \ms_{[q]},Z_i=z} \frac{\{{\varepsilon}_i - {\bar{\varepsilon}}(z)\}^2}{N_{[q]z}(N_{[q]z}-1)}\Big) \leq \frac{N_{[q]1}N_{[q]0}}{N(N-1)} \frac{1}{N_{[q]z}^2(N_{[q]z}-1)^2} \sumi  \{{\varepsilon}_i(z) - {\bar{\varepsilon}}(z)\}^4 = O(N^{-3}),
\end{align*}
and
\begin{align*}
    \Var\Big(\frac{\{\bar{\varepsilon}(z) - {\bar{\varepsilon}}_{[q]z}\}^2}{N_{[q]z}-1}\Big) \leq & \ope \bigg[ \frac{\{\bar{\varepsilon}(z) - {\bar{\varepsilon}}_{[q]z}\}^4}{(N_{[q]z}-1)^2} \bigg] \leq \max_{i\in [N]} |\varepsilon_i(z)|^2 \frac{\Var({\bar{\varepsilon}}_{[q]z})}{(N_{[q]z}-1)^2}\\
    =& \max_{i\in [N]} |\varepsilon_i(z)|^2 \frac{\Var({\bar{\varepsilon}}_{[q]z})}{(N_{[q]z}-1)^2} = \max_{i\in [N]} |\varepsilon_i(z)|^2 \frac{(N-N_{[q]z}) S_{\varepsilon(z)\varepsilon(z)}}{N N_{[q]z}(N_{[q]z}-1)^2} \\
    =& O\Big( N^{-4}\max_{i\in [N]} |\varepsilon_i(z)|^2\sumi {\varepsilon}_i^2(z) \Big) = O( N^{-5/2}),
\end{align*}
where the last equality is due to
\begin{align*}
    \max_{i\in [N]} |\varepsilon_i(z)|^2 \leq \Big(\sumi |\varepsilon_i(z)|^4 \Big)^{1/2} = O(N^{1/2}).
\end{align*}

Therefore, we have
\begin{align*}
    \Var\Big(\hat{V}_{\oraclecf,[q]}\Big) = o(N^{-2}), \quad N\{\hat{V}_{\oraclecf,[q]} - \ope (\hat{V}_{\oraclecf,[q]}) \}= \op(1).
\end{align*}
%and 
%\begin{equation*}
%    N\{\hat{V}_{\oraclecf,[q]} - \ope (\hat{V}_{\oraclecf,[q]}) \}= \op(1).
%\end{equation*}

Combining \eqref{eqn:cfadj-ora}, \eqref{eqn:Eora-Vora}, and \eqref{eqn:ora-Eora}, we have
\[
N\{\hat{V}_{\cfadj} - \Var(\hat{\tau}_{\oraclecf}) \} -\frac{1}{N-1}\sumi (\tau_{i,\varepsilon} -  \bar{\tau}_{\varepsilon})^2 = \op(1).
\]

\textbf{Part II: We verify Condition \eqref{eq:clt-for-oracle-cf}.} See the proof of \Cref{prop:valid-inference-sr} below, as CRE is a special case of SRE with $K=1$.
\end{proof}

\subsection{Proof of \Cref{prop:valid-inference-sr}}

The following lemma is useful for proving \Cref{prop:valid-inference-sr}.

\begin{lemma}
\label{lem:property-of-stratified-randomized-experiments}
If $\prob_{\bs{Z}} = \SR(N,\bs{A},(N_{\setk 1})_{k=1}^K)$, $N_{\setk z}/N_{\setk} \in (c,1-c)$, for a constant $c\in (0,0.5)$, $k\in \mk$, $z\in \{0,1\}$, and $N^{-1}\sumi a_i^4 = O(1)$ and $N^{-1}\sumi b_i^4 = O(1)$ for two finite populations $\{a_i\}_{i=1}^N$ and $\{b_i\}_{i=1}^N$, then
\begin{align*}
(i)&\quad \Var\Big(\frac{1}{N}\sum_{k=1}^K\sum_{i:A_i=k,Z_i=z}  a_ib_i\Big) = O(N^{-1}),\\
(ii)&\quad \Var\Big\{\sum_{k=1}^K \frac{N_{\setk z}}{N} (\bar{a}_{\setk z} - \bar{a}_{\setk})(\bar{b}_{\setk z} - \bar{b}_{\setk})\Big\} = O(N^{-1/2}),\\
(iii)&\quad \Var\Big\{\frac{1}{N}\sum_{k=1}^K \sum_{i: A_i=k, Z_i=z}  (a_i - \bar{a}_{\setk z})(b_i - \bar{b}_{\setk z})\Big\} = O( N^{-1/2}).
\end{align*}
\end{lemma}
\begin{proof}%[Proof of \Cref{lem:property-of-stratified-randomized-experiments}]
\noindent \textbf{(i)} By \Cref{lem:variance-of-CR}(ii) with $a_i \equiv a_i b_i$, we have
\begin{align*}
& \Var\Big(\frac{1}{N_{\setk z}}\sum_{i: A_i=k, Z_i=z}  a_i b_i \Big)
\leq  \frac{(N_{\setk} -N_{\setk z}) \sum_{i:A_i=k} a_i^2 b_i^2}{N_{\setk}(N_{\setk}-1)N_{\setk z}}.
\end{align*}
Therefore, 
\begin{align*}
 \Var\Big(\frac{1}{N}\sum_{k=1}^K \sum_{i: A_i=k, Z_i=z}  a_i b_i\Big) 
  \leq & \sum_{k=1}^K  \frac{N_{\setk z}^2}{N^2} \frac{(N_{\setk} -N_{\setk z}) \sum_{i:A_i=k} a_i^2 b_i^2}{N_{\setk}(N_{\setk}-1)N_{\setk z}}\\
   \leq  & \sum_{k=1}^K  \frac{N_{\setk z}^2}{N^2} \frac{(N_{\setk} -N_{\setk z}) \sum_{i:A_i=k} (a_i^4 + b_i^4)}{N_{\setk}(N_{\setk}-1)N_{\setk z}}\\
   = & O\Big(N^{-2} \sum_{k=1}^K \sum_{i: A_i=k} a_i^4 + b_i^4\Big)  \quad \Big(\text{Using}~\frac{N_{\setk z}}{N_{\setk}}  \in (c,1-c) \Big) \\
   =& O\Big(N^{-2}\sumi (a_i^4 + b_i^4)\Big) =O(N^{-1}).
\end{align*}

\noindent \textbf{(ii)}
We have
\begin{align*}
    & \Var\Big\{\sum_{k=1}^K\frac{N_{\setk}}{N} (\bar{a}_{\setk} - \bar{a}_{\setk z})(\bar{b}_{\setk} - \bar{b}_{\setk z})\Big\} \\
    \leq & \sum_{k=1}^K\frac{N_{\setk}^2}{N^2} \ope \big\{(\bar{a}_{\setk} - \bar{a}_{\setk z})^2(\bar{b}_{\setk} - \bar{b}_{\setk z})^2\big\}\\
    \leq & \sum_{k=1}^K\frac{N_{\setk}^2}{N^2} \{\ope (\bar{a}_{\setk} - \bar{a}_{\setk z})^4 + \ope(\bar{b}_{\setk} - \bar{b}_{\setk z})^4\big\}.
\end{align*}
We bound the two terms separately. For the first term, we have
\begin{align*}
   &  \sum_{k=1}^K\frac{N_{\setk}^2}{N^2} \ope (\bar{a}_{\setk} - \bar{a}_{\setk z})^4 
   \leq  4\max_{i\in [N]} |a_i|^2\sum_{k=1}^K\frac{N_{\setk}^2}{N^2} \ope (\bar{a}_{\setk} - \bar{a}_{\setk z})^2 \\
   = & 4\max_{i\in [N]} |a_i|^2\sum_{k=1}^K\frac{N_{\setk}^2}{N^2} \frac{(N_{\setk} -N_{\setk z}) \sum_{i:A_i=k} (a_i - \bar{a}_{\setk})^2}{N_{\setk}(N_{\setk}-1)N_{\setk z}}\\
   =&O\Big(\big(\max_{i\in [N]} a_i^2\big) N^{-2}\sumi a_i^2 \Big).
\end{align*}
Since $\sumi a_i^4 = O(N)$, then we have $\max_{i\in [N]} a_i^2\leq  (\max_{i\in [N]} a_i^4)^{1/2} \leq (\sumi a_i^4)^{1/2} = O(N^{1/2})$. Therefore,
\[
\sum_{k=1}^K\frac{N_{\setk}^2}{N^2} \ope (\bar{a}_{\setk} - \bar{a}_{\setk z})^4=O(N^{-1/2}).
\]
Similarly, 
$$\sum_{k=1}^K\frac{N_{\setk}^2}{N^2} \ope (\bar{b}_{\setk} - \bar{b}_{\setk z})^4=O(N^{-1/2}).$$

\noindent \textbf{(iii)}
Using $\Var(a+b) \leq 2 \Var(a) + 2\Var(b)$, we have
\begin{align*}
    & \Var\Big\{\frac{1}{N}\sum_{k=1}^K \sum_{i: A_i=k, Z_i=z}  (a_i - \bar{a}_{\setk z})(b_i - \bar{b}_{\setk z})\Big\}   \\
   \leq & 2\Var\Big\{\frac{1}{N}\sum_{k=1}^K \sum_{i: A_i=k, Z_i=z}  (a_i - \bar{a}_{\setk})(b_i - \bar{b}_{\setk})\Big\} + \\
    & 2\Var\Big\{\sum_{k=1}^K \frac{N_{\setk z}}{N}  (\bar{a}_{\setk z} - \bar{a}_{\setk})(\bar{b}_{\setk z} - \bar{b}_{\setk})\Big\}\\
    = & O(N^{-1}) + O(N^{-1/2}) = O(N^{-1/2}),  
\end{align*}
where the second to last equality is due to \Cref{lem:property-of-stratified-randomized-experiments}(i) and (ii).
\end{proof}

\begin{proof}[Proof of \Cref{prop:valid-inference-sr}]
\textbf{Part I: We verify Condition \eqref{eq:conservative-variance} by proving that}
\[
N\{\hat{V}_{\cfadj} - \Var(\hat{\tau}_{\oraclecf}) \} - \sum_{k=1}^K \frac{N_{\setk}}{N} \frac{1}{N_{\setk}-1} \sum_{i:A_i=k} (\tau_{\varepsilon,i} -  \bar{\tau}_{\varepsilon,\setk})^2 = \op(1).
\]

\textbf{Step I:} We prove that
    \begin{equation}
\label{eqn:cfadj-ora-sre}
    N(\hat{V}_{\cfadj} - \hat{V}_{\oraclecf}) = \op(1).
\end{equation}

    We decompose $N(\hat{V}_{\cfadj}-\hat{V}_{\oraclecf}) = \sum_{q \in \{1,2\}} \sum_{z\in \{0,1\}} M_{[q]z}$, where
    \begin{align*}
      M_{[q]z} = \sum_{k=1}^K\frac{N_{\setk [q]}^2}{N} \sum_{i:A_i = k, Z_i = z, i\in \ms_{[q]}} \frac{(\hat{\varepsilon}_i - \hat{\bar{\varepsilon}}_{\setk [q] z})^2-({\varepsilon}_i - {\bar{\varepsilon}}_{\setk [q] z})^2}{N_{\setk [q] z}(N_{\setk [q] z}-1)}. 
    \end{align*}
    Next, we show that $M_{[q]z} = \op(1)$. Note that
    \begin{align*}
       M_{[q]z} =  \sum_{k=1}^K\frac{N_{\setk [q]}^2}{N} \sum_{i:A_i = k, Z_i = z, i\in \ms_{[q]}} \frac{(\Delta_i - {\bar{\Delta}}_{\setk [q] z})^2 +2(\Delta_i - {\bar{\Delta}}_{\setk [q] z})({\varepsilon}_i - {\bar{\varepsilon}}_{\setk [q] z})}{N_{\setk [q] z}(N_{\setk [q] z}-1)}.
    \end{align*}
    By Assumptions~\ref{a:negligible-estimateion-error-for-prediction-function} and \ref{a:constant-probability-stratified-randomized-experiments}, we have
    \begin{align*}
        & \sum_{k=1}^K\frac{N_{\setk [q]}^2}{N} \sum_{i:A_i = k, Z_i = z, i\in \ms_{[q]}} \frac{(\Delta_i - {\bar{\Delta}}_{\setk [q] z})^2}{N_{\setk [q] z}(N_{\setk [q] z}-1)} \\
        \leq & \sum_{k=1}^K\frac{N_{\setk [q]}^2}{N} \sum_{i:A_i = k} \frac{\Delta_i^2(z)}{N_{\setk [q] z}(N_{\setk [q] z}-1)} 
        = O(\|f^\ast_z-\hat{f}_{[-q]z}\|_N^2) = \op(1),\\
        & \sum_{k=1}^K\frac{N_{\setk [q]}^2}{N} \sum_{i:A_i = k, Z_i = z, i\in \ms_{[q]}} \frac{({\varepsilon}_i - {\bar{\varepsilon}}_{\setk [q] z})^2}{N_{\setk [q] z}(N_{\setk [q] z}-1)} \\
        \leq & \sum_{k=1}^K\frac{N_{\setk [q]}^2}{N} \sum_{i:A_i = k} \frac{\varepsilon_i^2(z)}{N_{\setk [q] z}(N_{\setk [q] z}-1)} 
        =  O\Big(N^{-1}\sumi\varepsilon_i^2(z)\Big) = O(1).
    \end{align*}
    Again, using the Cauchy--Schwarz inequality, we have $M_{[q]z} = \op(1)$. Therefore,
\begin{equation*}
    N(\hat{V}_{\cfadj} - \hat{V}_{\oraclecf}) = \op(1).
\end{equation*}

\textbf{Step II:} We prove that 
\begin{align}
    \label{eqn:Eora-Vora-sre}
    N\{\ope (\hat{V}_{\oraclecf}) - \Var(\hat{\tau}_{\oraclecf}) \} =\sum_{k=1}^K \frac{N_{\setk}}{N} \frac{1}{N_{\setk}-1} \sum_{i:A_i=k} (\tau_{\varepsilon,i} -  \bar{\tau}_{\varepsilon,\setk})^2.
\end{align}
It is straightforward to verify that
\begin{align*}
\hat{\tau}_{\oraclecf} = \sum_{k=1}^K \frac{N_{\setk}}{N} \hat{\tau}_{\oraclecf,\setk},\quad \hat{V}_{\oraclecf} = \sum_{k=1}^K \frac{N_{\setk}^2}{N^2} \hat{V}_{\oraclecf,\setk},
\end{align*}
where $\hat{\tau}_{\oraclecf,\setk}$ is the oracle regression-adjusted estimator for the ATE within stratum $k$ and $\hat{V}_{\oraclecf,\setk}$ denotes the corresponding oracle cross-fitted variance estimator. From the conclusion of \textbf{Step II} in the proof of \Cref{prop:valid-inference-cr}, we have
\begin{align*}
    N_{\setk}\{\ope (\hat{V}_{\oraclecf,\setk}) - \Var(\hat{\tau}_{\oraclecf,\setk}) \} - \frac{1}{N_{\setk}-1} \sum_{i:A_i=k} (\tau_{\varepsilon,i} -  \bar{\tau}_{\varepsilon,\setk})^2 = \op(1).
\end{align*}
Using the independence of $(\bs{R}_{\setk},\bs{Z}_{\setk})$, $k\in \mk$, we have 
\[
\Var(\hat{\tau}_{\oraclecf}) = \sum_{k=1}^K \frac{N_{\setk}^2}{N^2} \Var(\hat{\tau}_{\oraclecf,\setk}).
\]
It follows that
\begin{align*}
     N\{\ope (\hat{V}_{\oraclecf}) - \Var(\hat{\tau}_{\oraclecf}) \} 
    =& \sum_{k=1}^K \frac{N_{\setk}^2}{N}\{\ope (\hat{V}_{\oraclecf,\setk}) - \Var(\hat{\tau}_{\oraclecf,\setk})\} \\
    =& \sum_{k=1}^K \frac{N_{\setk}}{N} \frac{1}{N_{\setk}-1} \sum_{i:A_i=k} (\tau_{\varepsilon,i} -  \bar{\tau}_{\varepsilon,\setk})^2.
\end{align*}

% Using the results from CRE, we have
% \begin{align*}
%     \ope\Var(\hat{\tau}_{\oraclecf[q]}\mid \ms_{[1]},\ms_{[2]}) = & \sum_{k=1}^K \frac{N_{\setk [q]}^2}{N_{[q]}^2}\sum_{i:A_i=k} \sum_{z\in \{0,1\}} \frac{ \{\varepsilon_i(z) - {\bar{\varepsilon}_{\setk}}(z)\}^2}{N_{\setk [q]z}(N_{\setk}-1)} \\
%     & - \sum_{k=1}^K \frac{N_{\setk [q]}^2}{N_{[q]}^2} \frac{1}{(N_{\setk}-1)N_{\setk [q]}} \sumi (\tau_{\varepsilon,i} -  \bar{\tau}_{\varepsilon,\setk})^2
% \end{align*}
%     and 
% \begin{align*}
%     \ope \hat{V}_{\oraclecf,[q]} =  \sum_{k=1}^K \frac{N_{\setk [q]}^2}{N_{[q]}^2}\sum_{i:A_i=k} \sum_{z\in \{0,1\}} \frac{ \{\varepsilon_i(z) - {\bar{\varepsilon}_{\setk}}(z)\}^2}{N_{\setk [q]z}(N_{\setk}-1)}.
% \end{align*}
% As a consequence, we have
% \begin{align*}
%     & N\{\ope \hat{V}_{\oraclecf} - \Var(\hat{\tau}_{\oraclecf})\} \nonumber \\
%     =& \sum_{q \in \{1,2\}}\frac{N_{[q]}^2}{N}\big\{\ope \hat{V}_{\oraclecf,[q]} - \ope\Var(\hat{\tau}_{\oraclecf[q]}\mid \ms_{[1]},\ms_{[2]})\Big\} \nonumber \\
%     =& \sum_{k=1}^K \frac{N_{\setk}}{N} \frac{1}{N_{\setk}-1} \sum_{i:A_i=k} (\tau_{\varepsilon,i} -  \bar{\tau}_{\varepsilon,\setk})^2.
% \end{align*}

\textbf{Step III:} We prove that
\begin{align}
    \label{eqn:ora-Eora-sre}
N\{\hat{V}_{\oraclecf} - \ope (\hat{V}_{\oraclecf}) \} = \op(1).
\end{align}

As $(I(Z_i=z,i\in \ms_{[q]}))_{i=1}^N$ follows $\SR(N_{[q]},\bs{A}_{[q]},(N_{\setk [q] z})_{k\in \mk})$, applying \Cref{lem:property-of-stratified-randomized-experiments} with $a_i = b_i = \sqrt{\frac{N_{\setk [q]}^2}{N_{\setk [q] z}(N_{\setk [q] z}-1)}}\varepsilon_i(z)$, we have
\begin{align*}
    \Var\Big\{\sum_{k=1}^K\frac{N_{\setk [q]}^2}{N} \sum_{i:A_i = k, Z_i = z, i\in \ms_{[q]}} \frac{({\varepsilon}_i - {\bar{\varepsilon}}_{\setk [q] z})^2}{N_{\setk [q] z}(N_{\setk [q] z}-1)} \Big\} = o(1).
\end{align*}
Therefore,  $N\{\hat{V}_{\oraclecf,[q]}-\ope (\hat{V}_{\oraclecf,[q]}) \}= \op(1)$ and
$ N\{ \hat{V}_{\oraclecf}-\ope (\hat{V}_{\oraclecf}) \} = \op(1).$

Combining  \eqref{eqn:cfadj-ora-sre}--\eqref{eqn:ora-Eora-sre}, we have
\[
N\{\hat{V}_{\cfadj} - \Var(\hat{\tau}_{\oraclecf}) \} - \sum_{k=1}^K \frac{N_{\setk}}{N} \frac{1}{N_{\setk}-1} \sum_{i:A_i=k} (\tau_{\varepsilon,i} -  \bar{\tau}_{\varepsilon,\setk})^2 = \op(1).
\]

\textbf{Part II: We verify Condition \eqref{eq:clt-for-oracle-cf}.}

By definition, we have
\[
\hat{\tau}_{\oraclecf} - \bar{\tau} = \sum_{k=1}^K \sum_{q \in \{1,2\}}\frac{N_{\setk q}}{N_{\setk}}\Big(\sum_{i:Z_i=1,i\in\ms_{[q]}} \frac{\varepsilon_i}{N_{\setk [q]1}}-\sum_{i:Z_i=0,i\in\ms_{[q]}} \frac{\varepsilon_i}{N_{\setk [q]0}}\Big) - \bar{\tau}_{\varepsilon}.
\]
Define 
$$
\tilde{Z}_i = \begin{cases}
    1,\quad \text{if} \quad  Z_i=1,i\in\ms_{[1]}, \\
    2,\quad \text{if} \quad  Z_i=1,i\in\ms_{[2]}, \\
    3,\quad  \text{if} \quad Z_i=0,i\in\ms_{[1]}, \\
    4, \quad \text{if} \quad Z_i=0,i\in\ms_{[2]}.  \\
\end{cases}
$$ 

We can verify that $\tilde{\bs{Z}}=(\tilde{Z}_i)_{i=1}^N$ follows a randomized block $2^2$ experimental design with parameters $(N_{\setk [1]1}, N_{\setk [2]1}, N_{\setk [1]0}, N_{\setk [2]0})_{k=1}^K$, i,e.,
\[
\prob(\tilde{\bs{Z}} = \tilde{\bs{z}}) = \prod_{k=1}^K \Big(\frac{N_{\setk [1]1}! N_{\setk [2]1}! N_{\setk [1]0}! N_{\setk [2]0}!}{N_{\setk}!} \Big),
\]
if $\tilde{\bs{z}}= (\tilde{z}_1,\ldots,\tilde{z}_N)$ satisfies, for all $k\in \mk$, $\sumi I(\tilde{z}_i=q,A_i=k) = N_{\setk [1]1},  N_{\setk [2]1},  N_{\setk [1]0}$, $ N_{\setk [2]0}$ for $q=1,2,3,4$, respectively; otherwise $\prob(\tilde{\bs{Z}} = \tilde{\bs{z}}) = 0$.

 By Theorem 3 of \citet{liu2KRandomization2024} with $Z_i = \tilde{Z}_i$ and $$(Y_i(1), Y_i(2),Y_i(3),Y_i(4)) = \Big(\frac{N_{\{k\}[1]}}{N_{\{k\}}}\varepsilon_i(1),\frac{N_{\{k\}[2]}}{N_{\{k\}}}\varepsilon_i(1),\frac{N_{\{k\}[1]}}{N_{\{k\}}}\varepsilon_i(0),\frac{N_{\{k\}[2]}}{N_{\{k\}}}\varepsilon_i(0)\Big),$$ for $A_i=k$, the CLT holds for the joint distribution of $(\frac{N_{[q]}}{N}\hat{\mu}_{\oraclecf,[q]}(z))_{q=1,2;z=0,1}$, and hence also holds for $$\hat{\tau}_{\oraclecf} = \frac{N_{[1]}}{N}\hat{\mu}_{\oraclecf,[1]}(1) + \frac{N_{[2]}}{N}\hat{\mu}_{\oraclecf,[2]}(1) - \frac{N_{[1]}}{N}\hat{\mu}_{\oraclecf,[1]}(0) - \frac{N_{[2]}}{N}\hat{\mu}_{\oraclecf,[2]}(0).$$
\end{proof}

\subsection{Proof of \Cref{prop:valid-inference-matched-pair}}

Let $2k-1$ and ${2k}$ denote the two units in the $k$th pair, $k=1,\ldots, N/2$.  \Cref{lem:variance-of-matched-pair-and-split-by-strutum} below is useful for proving \Cref{prop:valid-inference-matched-pair}

\begin{lemma}
    \label{lem:variance-of-matched-pair-and-split-by-strutum}
    Suppose that $\bs{Z}$ follows $\MP(N,\bs{A})$ and by \Cref{example:algorithm-split-by-treatment-sr} we split $\mk=\mk_{[1]}
    \cup \mk_{[2]}$ with $N_{[q]}/N \in (c,1-c)$, for $c\in (0,1)$ and $q=1,2$. For $z=0,1$, let $E_{k,z} = I(Z_{2k-1}=z)$, $k=1,\ldots,N/2$, be i.i.d. Bernoulli random variables with probability $1/2$. For a finite population $\{c_k\}_{k=1}^{N/2}$ with $N^{-1}\sum_{k=1}^{N/2} c_k^2 = O(1)$, we have, for $z=0,1$, and $q=1,2$,
 \begin{align*}
    (i)\quad  \Var\Big(\frac{1}{N}\sum_{k\in \mk_{[q]}} E_{k,z} c_k \Big) = O(N^{-1});
 \end{align*} 
 for two finite populations $\{a_i\}_{i=1}^N$ and $\{b_i\}_{i=1}^N$ with $N^{-1}\sumi a_i^4 = O(1)$ and $N^{-1}\sumi b_i^4 = O(1)$, we have for $z=0,1$, and $q=1,2$,
\begin{align*}
(ii)&\quad \Var\Big\{  (\bar{a}_{[q]z} - \bar{a})(\bar{b}_{[q] z} - \bar{b})\Big\} = O(N^{-1/2});\\
(iii)&\quad \Var\Big\{\frac{1}{N}\sum_{k\in \mk_{[q]}} \sum_{i: A_i=k, Z_i=z}  (a_i - \bar{a}_{[q] z})(b_i - \bar{b}_{[q] z})\Big\} = O( N^{-1/2}).
\end{align*}
\end{lemma}
\begin{proof}%[Proof of \Cref{lem:variance-of-matched-pair-and-split-by-strutum}]
\noindent \textbf{(i)} By the law of total variance, we have
\begin{align*}
        &\Var\Big(\frac{1}{N}\sum_{k\in \mk_{[q]}} E_{k,z} c_k \Big) 
        = \ope\Big\{\Var\Big(\frac{1}{N}\sum_{k\in \mk_{[q]}} E_{k,z} c_k \mid \mk_{[q]}\Big )\Big\} +  \Var\Big\{\ope\Big(\frac{1}{N}\sum_{k\in \mk_{[q]}} E_{k,z} c_k \mid \mk_{[q]}\Big )\Big\}. 
    \end{align*}

Note that $\big(I(k\in \mk_{[q]})\big)_{k=1}^K$ follows $\CR(N/2,N_{[q]}/2)$, independent of $(E_{k,z})_{k=1}^K$. Therefore, for the above two terms, we have
    \begin{align*}
        & \ope\Big\{\Var\Big(\frac{1}{N}\sum_{k\in \mk_{[q]}} E_{k,z} c_k \mid \mk_{[q]}\Big )\Big\} \\
        = & \ope\Big(\frac{1}{4 N^2}\sum_{k\in \mk_{[q]}}  c_k^2 \Big) 
        =  \frac{N_{[q]}}{4 N^3}\sum_{k=1}^{N/2}  c_k^2 = O\Big(\frac{1}{N^2}\sum_{k=1}^{N/2}  c_k^2 \Big) = O(N^{-1}),
    \end{align*}
    and
   \begin{align*}
       \Var\Big\{\ope\Big(\frac{1}{N}\sum_{k\in \mk_{[q]}} E_{k,z} c_k \mid \mk_{[q]}\Big )\Big\}
       =         \Var\Big(\frac{1}{2N}\sum_{k\in \mk_{[q]}} c_k \Big)
       =    O\Big(\frac{1}{N^2}\sum_{k=1}^{N/2} c_k^2 \Big) = O(N^{-1}).
   \end{align*}
   Therefore, 
   \[
  \Var\Big(\frac{1}{N}\sum_{k\in \mk_{[q]}} E_{k,z} c_k \Big) = O(N^{-1}).
   \]

\textbf{(ii)} First, we have
   \begin{align*}
       \Var\Big\{  (\bar{a}_{[q]z} - \bar{a})(\bar{b}_{[q] z} - \bar{b})\Big\} \leq 2\max_{i\in [N]} b_i^2 \ope (\bar{a}_{[q]z} - \bar{a})^2 = O( N^{1/2}) \Var (\bar{a}_{[q]z}).
   \end{align*}
Applying \Cref{lem:variance-of-matched-pair-and-split-by-strutum} (i), we have   
\[
\Var (\bar{a}_{[q]z}) = O\Big(\frac{1}{N^2}\sumi a_i^2\Big) = O(N^{-1}).
\]
As a consequence, we have
\[
\Var\Big\{  (\bar{a}_{[q]z} - \bar{a})(\bar{b}_{[q] z} - \bar{b})\Big\} = O(N^{-1/2}).
\]

\textbf{(iii)} Simple calculation yields
\begin{align*}
& \frac{1}{N}\sum_{k\in \mk_{[q]}} \sum_{i: A_i=k, Z_i=z}  (a_i - \bar{a}_{[q] z})(b_i - \bar{b}_{[q] z}) \\
= &\frac{1}{N}\sum_{k\in \mk_{[q]}} \sum_{i: A_i=k, Z_i=z}  (a_i - \bar{a})(b_i - \bar{b}) +  \frac{N_{[q]}/2}{N} (\bar{a} - \bar{a}_{[q] z})(\bar{b} - \bar{b}_{[q] z}).
\end{align*}
Using $\Var(a+b) \leq 2\Var(a) + 2\Var(b)$, it suffices to show that the variances of the two terms are $O(N^{-1/2})$. The variance of second term is $O(N^{-1/2})$ by (ii). We now calculate the variance of the first term.

By the definition of $E_{k,z}$, we have
\begin{align*}
    & \frac{1}{N}\sum_{k\in \mk_{[q]}}\sum_{i:A_i=k,Z_i=z}  (a_i-\bar{a})(b_i - \bar{b})  \\
    = &  \frac{1}{N}\sum_{k\in \mk_{[q]}} \{ E_{k,z} (a_{2k-1} -\bar{a}) (b_{2k-1} - \bar{b}) + E_{k,1-z} (a_{2k} -\bar{a})(b_{2k} - \bar{b}) \}.
\end{align*}

 Applying (i) with $c_k = (a_{2k-1}-\bar{a})(b_{2k-1}-\bar{b})$ and  $c_k = (a_{2k}-\bar{a})(b_{2k}-\bar{b})$ and using that $N^{-1}\sum_{k=1}^{N/2} c_k^2 = O(N^{-1}\sumi a_i^4 + N^{-1}\sumi b_i^4) = O(1)$, we have
   \[
   \Var\Big(\frac{1}{N}\sum_{k\in \mk_{[q]}}\sum_{i:A_i=k,Z_i=z}  (a_i -\bar{a}) (b_i - \bar{b})\Big) = O(N^{-1}).
   \]
   This completes the proof.
\end{proof}

\begin{proof}[Proof of \Cref{prop:valid-inference-matched-pair}]

Let $E_k = I(Z_{2k-1}=1)$, $\hat{B}_{\varepsilon, \setk} =  E_{k}(\hat{\varepsilon}_{2k-1}-\hat{\varepsilon}_{2k}) + (1-E_{k})(\hat{\varepsilon}_{2k}-\hat{\varepsilon}_{2k-1})$, and $\hat{\bar{B}}_{\varepsilon, [q]} = (2 / N_{[q]}) \sum_{k\in \mk_{[q]}}\hat{B}_{\varepsilon, \setk}$. Define ${B}_{\varepsilon, \setk}$ and ${\bar{B}}_{\varepsilon, [q]}$ (resp. ${B}_{\Delta, \setk}$ and ${\bar{B}}_{\Delta, [q]}$) the same as $\hat{B}_{\varepsilon, \setk}$ and $\hat{\bar{B}}_{\varepsilon, [q]}$ with $\hat{\varepsilon}_i$ replaced by $\varepsilon_i$ (resp. $\Delta_i$). 

\textbf{Part I: We verify Condition \eqref{eq:conservative-variance} by proving that}
\begin{align*}
    N\{\hat{V}_{\oraclecf} - \Var(\hat{\tau}_{\oraclecf})\} = \frac{4}{N}\sum_{k=1}^{N/2} (\bar{\tau}_{\varepsilon,\setk}-\bar{\tau}_{\varepsilon})^2 + \op(1).
\end{align*}

\textbf{Step I:} We prove that
     \begin{equation}
\label{eqn:cfadj-ora-pair}
    N(\hat{V}_{\cfadj} - \hat{V}_{\oraclecf}) = \op(1).
\end{equation}   

By definition, we have 
    \begin{align*}
        & \hat{V}_{\cfadj, [q]} = \frac{4}{(N_{[q]}-2)N_{[q]}}\sum_{k\in \mk_{[q]}}(\hat{B}_{\varepsilon, \setk}-\hat{\bar{B}}_{\varepsilon, [q]})^2,\\
        & \hat{V}_{\oraclecf, [q]} = \frac{4}{(N_{[q]}-2)N_{[q]}}\sum_{k\in \mk_{[q]}}({B}_{\varepsilon, \setk}-{\bar{B}}_{\varepsilon, [q]})^2.
    \end{align*}
    Therefore, $N(\hat{V}_{\cfadj}- \hat{V}_{\oraclecf}) = \sum_{q \in \{1,2\}} T_{[q]},$
    where
    \begin{align*}
        T_{[q]} = \frac{4 N_{[q]}}{(N_{[q]}-2)N}\sum_{k\in \mk_{[q]}}\Big\{({B}_{\Delta, \setk}- \bar{B}_{\Delta, [q]})^2 + 2 ({B}_{\varepsilon, \setk}-{\bar{B}}_{\varepsilon, [q]})({B}_{\Delta, \setk}-\bar{B}_{\Delta, [q]})\Big\}.
    \end{align*}

Note that
    \begin{align*}
         \frac{1}{N}\sum_{k\in \mk_{[q]}} ({B}_{\Delta, \setk} -\bar{B}_{\Delta, [q]})^2 
        \leq & \frac{1}{N}\sum_{k\in \mk_{[q]}} {B}_{\Delta, \setk}^2 = O\Big\{\frac{1}{N}\sum_{k\in \mk_{[q]}} \sum_{i:A_i=k} \{\Delta_i^2(1) + \Delta_i^2(0)\}\Big\} \\
        = & O\Big(\sum_{z\in\{0,1\}}\|f^\ast_z - \hat{f}_{[-q]z}\|_N^2\Big) = \op(1),\\
         \frac{1}{N}\sum_{k\in \mk_{[q]}} ({B}_{\varepsilon, \setk}-{\bar{B}}_{\varepsilon, [q]})^2 
        = & O\Big(\frac{1}{N}\sumi \varepsilon_i^2(1) + \varepsilon_i^2(0)\Big) = O(1).
    \end{align*}
    By the Cauchy--Schwarz inequality, we have $T_{[q]} = \op(1)$. Therefore,
    \begin{equation*}
    N(\hat{V}_{\cfadj} - \hat{V}_{\oraclecf}) = \op(1).
\end{equation*}

\textbf{Step II:} We prove that
\begin{align}
    \label{eqn:ora-Vora-pair}
    N\{\hat{V}_{\oraclecf} - \Var(\hat{\tau}_{\oraclecf})\} = \frac{4}{N}\sum_{k=1}^{N/2} (\bar{\tau}_{\varepsilon,\setk}-\bar{\tau}_{\varepsilon})^2 + \op(1).
\end{align}

First, we have
\begin{align*}
    \hat{V}_{\oraclecf} - \Var(\hat{\tau}_{\oraclecf}) = \sum_{q=1}^2 \frac{N_{[q]}^2}{N^2} \Big[\hat{V}_{\oraclecf,[q]} - \ope\Big\{\Var(\hat{\tau}_{\oraclecf,[q]}\mid \ms_{[1]},\ms_{[2]})\Big\}\Big].
\end{align*}
We will deal with $\hat{V}_{\oraclecf,[q]}$ and $\ope\{\Var(\hat{\tau}_{\oraclecf,[q]}\mid \ms_{[1]},\ms_{[2]})\}$, separately.

By definition, we have $|\mk_{[q]}| = N_{[q]}/2$. Simple calculation yields 
    \begin{align*}
        \hat{V}_{\oraclecf,[q]} = & \frac{4}{(N_{[q]}-2)N_{[q]}}\sum_{k\in \mk_{[q]}}({B}_{\varepsilon, \setk}-{\bar{B}}_{\varepsilon, [q]})^2 \\
       =& \frac{4}{(N_{[q]}-2)N_{[q]}}\sum_{k\in \mk_{[q]}}({B}_{\varepsilon, \setk} - \bar{\tau}_{\varepsilon})^2-\frac{2}{N_{[q]}-2} ({\bar{B}}_{\varepsilon, [q]} - \bar{\tau}_{\varepsilon})^2, \\
       {\bar{B}}_{\varepsilon, [q]} =& \frac{1}{N_{[q]}/2} \Big(\sum_{k\in\mk_{[q]}}  E_{k}\{\varepsilon_{2k-1}(1)-\varepsilon_{2k}(0)\} + (1-E_{k})\{\varepsilon_{2k}(1)-\varepsilon_{2k-1}(0)\} \Big).
    \end{align*}
%\begin{align*}
%    {\bar{B}}_{\varepsilon, [q]} = \frac{1}{N_{[q]}/2} \Big(\sum_{k\in\mk_{[q]}}  E_{k}\{\varepsilon_{2k-1}(1)-\varepsilon_{2k}(0)\} + (1-E_{k})\{\varepsilon_{2k}(1)-\varepsilon_{2k-1}(0)\} \Big).
%\end{align*}
Applying \Cref{lem:variance-of-matched-pair-and-split-by-strutum}(i) with $c_k = \varepsilon_{2k-1}(1)-\varepsilon_{2k}(0)$ and $c_k = \varepsilon_{2k}(1)-\varepsilon_{2k-1}(0)$,  we have $\Var({\bar{B}}_{\varepsilon, [q]}) = O(N^{-2}\sumi \{\varepsilon_i^2(1)+\varepsilon_i^2(0)\})$. Using Chebyshev's inequality and $\ope ({\bar{B}}_{\varepsilon, [q]}) = \bar{\tau}_{\varepsilon}$, we have ${\bar{B}}_{\varepsilon, [q]} -\bar{\tau}_{\varepsilon} = \Op(N^{-1/2})$.  Moreover, 
\begin{align*}
    &\frac{4}{(N_{[q]}-2)N_{[q]}}\sum_{k\in \mk_{[q]}}({B}_{\varepsilon, \setk}-\bar{\tau}_{\varepsilon})^2\\
    =& \frac{4}{(N_{[q]}-2)N_{[q]}}\sum_{k\in \mk_{[q]}} \Big[ \{\varepsilon_{2k-1}(1)-\varepsilon_{2k}(0)-\bar{\tau}_{\varepsilon} \}^2 E_k + \{\varepsilon_{2k}(1) - \varepsilon_{2k-1}(0)-\bar{\tau}_{\varepsilon}\}^2 (1-E_k)\Big]. 
\end{align*}

Applying \Cref{lem:variance-of-matched-pair-and-split-by-strutum}(i) with $c_k = \{\varepsilon_{2k-1}(1)-\varepsilon_{2k}(0)-\bar{\tau}_{\varepsilon} \}^2$ and $c_k = \{\varepsilon_{2k}(1) - \varepsilon_{2k-1}(0)-\bar{\tau}_{\varepsilon}\}^2$ and using that $N^{-1}\sum_{k=1}^{N/2} c_k^2 = O(N^{-1}\sumi \varepsilon_i^4(1)+\varepsilon_i^4(0))$, we have
\begin{align*}
&\frac{4}{(N_{[q]}-2)N_{[q]}}\sum_{k\in \mk_{[q]}}({B}_{\varepsilon, \setk}-\bar{\tau}_{\varepsilon})^2 \\
    =& \frac{2}{(N_{[q]}-2)N}\sum_{k=1}^{N/2} \Big[ \{\varepsilon_{2k-1}(1)-\varepsilon_{2k}(0)-\bar{\tau}_{\varepsilon} \}^2 + \{\varepsilon_{2k}(1) - \varepsilon_{2k-1}(0)-\bar{\tau}_{\varepsilon}\}^2\Big] + \Op(N^{-3/2})\\
    =&\frac{1}{(N_{[q]}-2)N}\sum_{k=1}^{N/2}\Big\{4(\bar{\tau}_{\varepsilon,\setk}-\bar{\tau}_{\varepsilon})^2+ \{\varepsilon_{2k-1}(0)+\varepsilon_{2k-1}(1)-\varepsilon_{2k}(0)-\varepsilon_{2k}(1)\}^2\Big\} \\
    &+ \Op(N^{-3/2}),
\end{align*}
where for the last equality, we use $2{(a_i^2+b_i^2)} = {(a_i+b_i)}^2 + {(a_i-b_i)^2}$ with $a_i = \varepsilon_{2k-1}(1)-\varepsilon_{2k}(0)-\bar{\tau}_{\varepsilon}$ and $b_i=\varepsilon_{2k}(1) - \varepsilon_{2k-1}(0)-\bar{\tau}_{\varepsilon}$.

As a consequence, we have
\begin{align*}
    &\hat{V}_{\oraclecf,[q]} \\
    =& \frac{1}{(N_{[q]}-2)N}\sum_{k=1}^{N/2}\Big\{4(\bar{\tau}_{\varepsilon,\setk}-\bar{\tau}_{\varepsilon})^2+ \{\varepsilon_{2k-1}(0)+\varepsilon_{2k-1}(1)-\varepsilon_{2k}(0)-\varepsilon_{2k}(1)\}^2\Big\} \\
    &+ \Op(N^{-3/2}).
\end{align*}

On the other hand, we have
\begin{align*}
    \Var(\hat{\tau}_{\oraclecf,[q]}\mid \ms_{[1]},\ms_{[2]}) = & \frac{1}{(N_{[q]}/2)^2}\sum_{k\in \mk_{[q]}} \Var(B_{\varepsilon,\setk})  \\
    =& \frac{1}{N_{[q]}^2}\sum_{k\in \mk_{[q]}} \{\varepsilon_{2k-1}(1)- \varepsilon_{2k}(0)- \varepsilon_{2k}(1)+ \varepsilon_{2k-1}(0)\}^2.
\end{align*}
Therefore,
\begin{align*}
    \ope\Big\{\Var(\hat{\tau}_{\oraclecf,[q]}\mid \ms_{[1]},\ms_{[2]})\Big\} = \frac{1}{N_{[q]}N}\sum_{k=1}^{N/2} \{\varepsilon_{2k-1}(1)- \varepsilon_{2k}(0) - \varepsilon_{2k}(1) + \varepsilon_{2k-1}(0)\}^2.
\end{align*}

As a consequence, we have
\[
\hat{V}_{\oraclecf,[q]} - \ope\Big\{\Var(\hat{\tau}_{\oraclecf,[q]}\mid \ms_{[1]},\ms_{[2]})\Big\} = \frac{4}{N_{[q]}N}\sum_{k=1}^{N/2}(\bar{\tau}_{\varepsilon,\setk}-\bar{\tau}_{\varepsilon})^2 + \op(N^{-1}),
\]
\begin{align*}
    N\{\hat{V}_{\oraclecf} - \Var(\hat{\tau}_{\oraclecf})\} = \frac{4}{N}\sum_{k=1}^{N/2} (\bar{\tau}_{\varepsilon,\setk}-\bar{\tau}_{\varepsilon})^2 + \op(1).
\end{align*}

Combining \eqref{eqn:ora-Vora-pair} and \eqref{eqn:cfadj-ora-pair}, we have 
\[
N\{\hat{V}_{\cfadj} - \Var(\hat{\tau}_{\oraclecf}) \} -  \frac{4}{N}  \sum_{k=1}^{N/2} (\bar{\tau}_{\varepsilon,\setk}-\bar{\tau}_{\varepsilon})^2 = \op(1).
\]
%\liu{Here, the coefficient is $4/N$ (or $\sum_{q \in \{0,1\}} \frac{N_{[q]}^2}{N^2} \frac{4}{N_{[q]} - 2} $), but in the main text, it is $2/(N-2)$. Please double-check the definition of $\Delta_{\MP}$ in the main text.}

\textbf{Part II: We verify Condition \eqref{eq:clt-for-oracle-cf}.}

 By definition, we have
\begin{align*}
    \hat{\tau}_{\oraclecf} - \bar{\tau} &= \frac{1}{N/2}\sum_{k=1}^{N/2} \Big\{(\varepsilon_{2k-1}-\varepsilon_{2k})E_{k} + (\varepsilon_{2k}-\varepsilon_{2k-1})(1-E_{k})\Big\} - \bar{\tau}_{\varepsilon}\\
    &= \frac{1}{N/2}\sum_{k=1}^{N/2} \Big[\{\varepsilon_{2k-1}(1)-\varepsilon_{2k}(1)-\varepsilon_{2k}(0) + \varepsilon_{2k-1}(0)\}\big(E_{k}-\frac{1}{2}\big)\Big].
\end{align*}

The terms in the above summand are independent since $E_k$ are i.i.d, following the Bernoulli distribution with probability $1/2$. Moreover,
\begin{align*}
    & \frac{1}{N^{3/2}}\sum_{k=1}^{N/2}\{\varepsilon_{2k-1}(1)-\varepsilon_{2k}(1)-\varepsilon_{2k}(0) + \varepsilon_{2k-1}(0)\}^3 \\
    = & O\Big(\frac{1}{N^{3/2}}\sumi |\varepsilon_i(1)|^3+|\varepsilon_i(0)|^3 \Big) = o(1).
\end{align*}
Therefore, $(\hat{\tau}_{\oraclecf}-\bar{\tau})/\Var(\hat{\tau}_{\oraclecf})^{1/2}\rightsquigarrow \mathcal{N}(0,1) $ follows from Lyapunov's CLT.

\end{proof}

\subsection{Proof of \Cref{prop:stability-assumption-linear-SR-no-stratum-indicator}}

\begin{proof}
Note that
\begin{align*}
    \|\hat{f}_{[-q]z}-f^\ast_z\|_N^2 = & \frac{1}{N} \sumi \big\{\bs{x}_i^\top (\bs{\beta}_z^\ast-\hat{\bs{\beta}}_{[-q]z}) + (\alpha_{z}^\ast -\hat{\alpha}_{[-q] z})\big\}^2 \\
    = &
  \begin{pmatrix}
      \alpha_{z}^\ast-\hat{\alpha}_{[-q] z}\\
     \bs{\beta}_z^\ast - \hat{\bs{\beta}}_{[-q]z}
  \end{pmatrix}^\top \bs{\Sigma}_{\tilde{\bs{x}}\tilde{\bs{x}}} \begin{pmatrix}
      \alpha_{z}^\ast-\hat{\alpha}_{[-q] z}\\
     \bs{\beta}_z^\ast - \hat{\bs{\beta}}_{[-q]z}
  \end{pmatrix}. 
\end{align*}

Since $\bs{\Sigma}_{\tilde{\bs{x}}\tilde{\bs{x}}}$ has an invertible limit, then it suffices to prove that $\alpha_{z}^\ast-\hat{\alpha}_{[-q] z} = \op(1)$ and $\bs{\beta}_z^\ast - \hat{\bs{\beta}}_{[-q]z} = \op(1) $. Note that
\begin{align*}
  \begin{pmatrix}
      \alpha_{z}^\ast\\
     \bs{\beta}_z^\ast
  \end{pmatrix}  =  \bs{\Sigma}_{\tilde{\bs{x}}\tilde{\bs{x}}}^{-1} \bs{\Sigma}_{\tilde{\bs{x}}Y(z)}, \quad 
  \begin{pmatrix}
      \hat{\alpha}_{[-q]z}\\
     \hat{\bs{\beta}}_{[-q]z}
  \end{pmatrix}  = \hat{\bs{\Sigma}}_{[-q]z,\tilde{\bs{x}}\tilde{\bs{x}}}^{-1} \hat{\bs{\Sigma}}_{[-q]z,\tilde{\bs{x}}Y(z)}, 
\end{align*}
where
\begin{align*}
\bs{\Sigma}_{\tilde{\bs{x}}Y(z)} =& \frac{1}{N}\sumi \tilde{\bs{x}}_i Y_i(z),\\
  \hat{\bs{\Sigma}}_{[-q]z,\tilde{\bs{x}}\tilde{\bs{x}}} =& \frac{1}{N}\sum_{k=1}^K \sum_{i:A_i=k,Z_i=z,i\in\ms_{[-q]}} \tilde{\bs{x}}_i \tilde{\bs{x}}_i^\top \frac{N_{\setk}}{N_{\setk[-q]z}},\\
  \hat{\bs{\Sigma}}_{[-q]z,\tilde{\bs{x}}Y(z)}  = & \frac{1}{N}\sum_{k=1}^K \sum_{i:A_i=k,Z_i=z,i\in\ms_{[-q]}} \tilde{\bs{x}}_i Y_i \frac{N_{\setk}}{N_{\setk[-q]z}}.
\end{align*}

Simple calculation yields
$ \ope (\hat{\bs{\Sigma}}_{[-q]z,\tilde{\bs{x}}\tilde{\bs{x}}}) = \bs{\Sigma}_{\tilde{\bs{x}}\tilde{\bs{x}}} $ and $\ope(\hat{\bs{\Sigma}}_{[-q]z,\tilde{\bs{x}}Y(z)}) = \bs{\Sigma}_{\tilde{\bs{x}}Y(z)}$. Note that $(I(Z_i=z,i\in \ms_{[-q]}))_{i=1}^N$ follows $\SR(N,\bs{A},(N_{\setk[-q]z})_{k=1}^K)$. Therefore, applying \Cref{lem:property-of-stratified-randomized-experiments} with $Z_i = I(Z_i=z,i\in \ms_{[-q]})$, $ a_i =  ( N_{\setk}/ N_{\setk[-q]z} )^{1/2}  x_{il}$, $l=1,\ldots, d$, and
$$ 
b_i = \sqrt{\frac{N_{\setk}}{N_{\setk[-q]z}}} x_{im}, \ m=1,\ldots, d, \ \textnormal{or }\sqrt{\frac{N_{\setk}}{N_{\setk[-q]z}}}Y_i(z),$$ 
by Chebyshev's inequality and $$\max \Big\{ N^{-1}\sumi \|\bs{x}_i\|^4_{\infty}, \ N^{-1}\sumi Y_i^4(1), N^{-1}\sumi Y_i^4(0) \Big\} = O(1),$$ we have $\hat{\bs{\Sigma}}_{[-q]z,\tilde{\bs{x}}\tilde{\bs{x}}} = \bs{\Sigma}_{\tilde{\bs{x}}\tilde{\bs{x}}} + \Op(N^{-1/2})$ and $\hat{\bs{\Sigma}}_{[-q]z,\tilde{\bs{x}}Y(z)} = \bs{\Sigma}_{\tilde{\bs{x}}Y(z)} + \Op(N^{-1/2})$.
%\[
%\hat{\bs{\Sigma}}_{[-q]z,\tilde{\bs{x}}\tilde{\bs{x}}} = \bs{\Sigma}_{\tilde{\bs{x}}\tilde{\bs{x}}} + \Op(N^{-1/2}),\quad \hat{\bs{\Sigma}}_{[-q]z,\tilde{\bs{x}}Y(z)} = \bs{\Sigma}_{\tilde{\bs{x}}Y(z)} + \Op(N^{-1/2}).
%\]
Again using that $\bs{\Sigma}_{\tilde{\bs{x}}\tilde{\bs{x}}}$ has an invertible limit, we have
\begin{align*}
     \begin{pmatrix}
      \hat{\alpha}_{[-q]z}\\
     \hat{\bs{\beta}}_{[-q]z}
  \end{pmatrix} =& \hat{\bs{\Sigma}}_{[-q]z,\tilde{\bs{x}}\tilde{\bs{x}}}^{-1} \hat{\bs{\Sigma}}_{[-q]z,\tilde{\bs{x}}Y(z)}
  = \Big(\bs{\Sigma}_{\tilde{\bs{x}}\tilde{\bs{x}}} + \Op(N^{-1/2})\Big)^{-1} \Big(\bs{\Sigma}_{\tilde{\bs{x}}Y(z)} + \Op(N^{-1/2})\Big) \\
  = & \bs{\Sigma}_{\tilde{\bs{x}}\tilde{\bs{x}}}^{-1} \bs{\Sigma}_{\tilde{\bs{x}}Y(z)} +\op(1) = \begin{pmatrix}
      \alpha_{z}^\ast\\
     \bs{\beta}_z^\ast
  \end{pmatrix} +\op(1).
\end{align*}
The conclusion follows.
\end{proof}

\subsection{Proof of \Cref{prop:stability-assumption-linear-SR}}

\begin{proof}
Note that
\begin{align*}
    \|\hat{f}_{[-q]z}-f^\ast_z\|_N^2 =& \frac{1}{N} \sumi \big\{\bs{x}_i^\top (\bs{\beta}_z^\ast-\hat{\bs{\beta}}_{[-q]z}) + \sum_{k=1}^K (\alpha_{\setk z}^\ast -\hat{\alpha}_{\setk [-q] z}) I(A_i = k)\big\}^2 \\
    \leq & \frac{2}{N} \sumi \big\{(\bs{x}_i-\bar{\bs{x}}_{\setk})^\top (\bs{\beta}_z^\ast-\hat{\bs{\beta}}_{[-q]z})\big\}^2  \\
    & + \sum_{k=1}^K \frac{2 N_{\setk}}{N} \big\{\bar{\bs{x}}_{\setk}^\top (\bs{\beta}_z^\ast-\hat{\bs{\beta}}_{[-q]z}) + \alpha_{\setk z}^\ast -\hat{\alpha}_{\setk [-q] z}\big\}^2 . 
\end{align*}

On the other hand, we have
\begin{align*}
    &\alpha_{\setk z}^\ast = \bar{Y}_{\setk}(z) - \bar{\bs{x}}_{\setk}^\top \bs{\beta}^\ast_z ,\quad \hat{\alpha}_{\setk [-q] z} = \bar{Y}_{\setk [-q] z} - \bar{\bs{x}}_{\setk[-q] z}^\top \hat{\bs{\beta}}_{[-q]z}.
\end{align*}
Therefore,
\begin{align*}
     & \|\hat{f}_{[-q]z}-f^\ast_z\|_N^2 \\
     \leq & \frac{2}{N} \sumi \big\{(\bs{x}_i-\bar{\bs{x}}_{\setk})^\top (\bs{\beta}_z^\ast-\hat{\bs{\beta}}_{[-q]z})\big\}^2 + \sum_{k=1}^K \frac{4N_{\setk}}{N} \{\bar{Y}_{\setk}(z) - \bar{Y}_{\setk [-q] z}\}^2  \\
    & + \sum_{k=1}^K \frac{4N_{\setk}}{N} \Big\{\big(\bar{\bs{x}}_{\setk[-q] z}-\bar{\bs{x}}_{\setk}\big)^\top \hat{\bs{\beta}}_{[-q]z})\Big\}^2 .
\end{align*}

Let 
\begin{align*}
    & \bs{\Sigma}_{z,\bs{x}Y(z)} = \frac{1}{N} \sum_{k=1}^K\sum_{i:A_i=k} \omega_{\setk z} (\bs{x}_i-\bar{\bs{x}}_{\setk}) \{Y_i(z) - \bar{Y}_{\setk}(z)\}, \\
    & \hat{\bs{\Sigma}}_{[-q]z,\bs{x}\bs{x}} = \frac{1}{N}\sum_{k=1}^K\sum_{i:A_i=k,Z_i=z,i\in \ms_{[-q]}} \omega_{\setk [-q] z} (\bs{x}_i- \bar{\bs{x}}_{\setk[-q]z})(\bs{x}_i- \bar{\bs{x}}_{\setk[-q]z})^\top ,\\
    &\hat{\bs{\Sigma}}_{[-q]z,\bs{x}Y(z)} = \frac{1}{N} \sum_{k=1}^K\sum_{i:A_i=k,Z_i=z,i\in \ms_{[-q]}} \omega_{\setk [-q] z}  (\bs{x}_i- \bar{\bs{x}}_{\setk[-q]z})(Y_i - \bar{Y}_{\setk[-q]z})^\top.
\end{align*}
By Frisch–Waugh–Lovell (FWL) theorem \citep{ding2021frisch}, we have
\[
\bs{\beta}_z^\ast = \bs{\Sigma}_{z,\bs{x}\bs{x}}^{-1} \bs{\Sigma}_{z,\bs{x}Y(z)},\quad \hat{\bs{\beta}}_{[-q]z} = \hat{\bs{\Sigma}}_{[-q]z,\bs{x}\bs{x}}^{-1}\hat{\bs{\Sigma}}_{[-q]z,\bs{x}Y(z)}.
\]

Applying \Cref{lem:variance-of-CR} to $(I(Z_i=z,i\in\ms_{[-q]}))_{i:A_i=k}$, which follows $\CR(N_{\setk},N_{\setk [-q] z})$, we have
\begin{align*}
    & \ope \bigg\{ \sum_{i:A_i=k,Z_i=z,i\in \ms_{[-q]}} \omega_{\setk [-q] z} (\bs{x}_i- \bar{\bs{x}}_{\setk[-q]z})(\bs{x}_i- \bar{\bs{x}}_{\setk[-q]z})^\top \bigg\} \\
    = & \sum_{i:A_i=k} \omega_{\setk z} (\bs{x}_i-\bar{\bs{x}}_{\setk}) (\bs{x}_i-\bar{\bs{x}}_{\setk})^\top,\\
   & \ope \bigg\{ \sum_{i:A_i=k,Z_i=z,i\in \ms_{[-q]}} \omega_{\setk [-q] z}  (\bs{x}_i- \bar{\bs{x}}_{\setk[-q]z})(Y_i - \bar{Y}_{\setk[-q]z})^\top \bigg\} \\
   = & \sum_{i:A_i=k} \omega_{\setk z} (\bs{x}_i-\bar{\bs{x}}_{\setk}) \{Y_i(z) - \bar{Y}_{\setk}(z)\}.
\end{align*}

Note that $(I(Z_i=z,i\in \ms_{[-q]}))_{i=1}^N$ follows $\SR(N,\bs{A},(N_{\setk [-q]z})_{k\in \mk})$. Applying \Cref{lem:property-of-stratified-randomized-experiments}(ii) to each element of $\hat{\bs{\Sigma}}_{[-q]z,\bs{x}\bs{x}}$ and $\hat{\bs{\Sigma}}_{[-q]z,\bs{x}Y(z)}$, by Chebyshev's inequality and $\max\{ N^{-1}\sumi \|\bs{x}_i\|^4_{\infty}$, $ N^{-1}\sumi Y_i^4(1)$, $N^{-1}\sumi Y_i^4(0)\} = O(1)$, we have $\hat{\bs{\Sigma}}_{[-q]z,\bs{x}\bs{x}} - \bs{\Sigma}_{z,\bs{x}\bs{x}} = \op(1)$ and $\hat{\bs{\Sigma}}_{[-q]z,\bs{x}Y(z)} - \bs{\Sigma}_{z,\bs{x}Y(z)} = \op(1)$.
%\[
%\hat{\bs{\Sigma}}_{[-q]z,\bs{x}\bs{x}} - \bs{\Sigma}_{z,\bs{x}\bs{x}} = \op(1),\quad \hat{\bs{\Sigma}}_{[-q]z,\bs{x}Y(z)} - \bs{\Sigma}_{z,\bs{x}Y(z)} = \op(1).
%\]
Since $\bs{\Sigma}_{z,\bs{x}\bs{x}}$ have invertible limits and $\bs{\beta}^\ast_z$ have finite limits for $z=0,1$, we have
\[
\bs{\beta}_z^\ast - \hat{\bs{\beta}}_{[-q]z}  = \op(1),\quad \hat{\bs{\beta}}_{[-q]z} = \Op(1).
\]

Applying Lemma \ref{lem:property-of-stratified-randomized-experiments}(ii), we have 
\begin{align*}
    & \sum_{k=1}^K \frac{N_{\setk}}{N} \{\bar{Y}_{\setk}(z) - \bar{Y}_{\setk [-q] z}\}^2 = \op(1), \\
    & \sum_{k=1}^K \frac{N_{\setk}}{N} (\bar{\bs{x}}_{\setk[-q] z}-\bar{\bs{x}}_{\setk}) (\bar{\bs{x}}_{\setk[-q] z} -\bar{\bs{x}}_{\setk})^\top = \op(1).
\end{align*}
In light of the above, we have
\begin{align*}
     & \|\hat{f}_{[-q]z}-f^\ast_z\|_N^2 \\
     \leq &  \frac{2}{c}(\bs{\beta}_z^\ast-\hat{\bs{\beta}}_{[-q]z})^\top\bs{\Sigma}_{z,\bs{x}\bs{x}} (\bs{\beta}_z^\ast-\hat{\bs{\beta}}_{[-q]z}) + 4\sum_{k=1}^K \frac{N_{\setk}}{N} \{\bar{Y}_{\setk}(z) - \bar{Y}_{\setk [-q] z}\}^2  \\
    & + 4\hat{\bs{\beta}}_{[-q]z}^\top\Big\{\sum_{k=1}^K \frac{N_{\setk}}{N} (\bar{\bs{x}}_{\setk[-q] z} -\bar{\bs{x}}_{\setk}) (\bar{\bs{x}}_{\setk[-q] z} - \bar{\bs{x}}_{\setk})^\top\Big\} \hat{\bs{\beta}}_{[-q]z} \\
    = & \op(1)  O(1) \op(1) + \op(1) + \Op(1)\op(1)\Op(1) =\op(1).
\end{align*}
The conclusion follows. 
 \end{proof}

\subsection{Proof of \Cref{eq:separate-model-using-optimal-weights}}
\begin{proof}
By definition, we have
    \begin{align*}
     \bs{\beta}_z^\ast = \argmin_{\bs{\beta}} \frac{1}{N}\sum_{k=1}^K \sum_{i: A_i = k} \big\{Y_i(z)-\bar{Y}_{\setk}(z)-(\bs{x}_i-\bar{\bs{x}}_{\setk})^\top \bs{\beta}\big\}^2 \omega_{\setk z}.
\end{align*}
Therefore, 
\begin{align*}
    \tilde{\sigma}^2_{\SR} & = \frac{1}{N}\sum_{z\in \{0,1\}} \sum_{k=1}^K \sum_{i: A_i = k} \big\{Y_i(z)-\bar{Y}_{\setk}(z)-(\bs{x}_i-\bar{\bs{x}}_{\setk})^\top \bs{\beta}^\ast_z \big\}^2 \omega_{\setk z}\\
 &\leq \frac{1}{N} \sum_{z\in \{0,1\}}\sum_{k=1}^K \sum_{i: A_i = k} \{Y_i(z)-\bar{Y}_{\setk}(z)\}^2 \omega_{\setk z} =  \tilde{\sigma}^2_{\textnormal{SR-unadj}}.
\end{align*}
\end{proof}

\subsection{Proof of \Cref{prop:stability-assumption-linear-SR-tom}}
\begin{proof}
\noindent \textbf{(i)} Mimicking the proof of  \Cref{prop:stability-assumption-linear-SR}, we have 
\begin{align*}
     \|\hat{f}_{[-q]z}-f^\ast_z\|_N^2  \leq & \frac{2}{N} \sumi \big\{(\bs{x}_i-\bar{\bs{x}}_{\setk})^\top (\bs{\beta}^\ast-\hat{\bs{\beta}}_{[-q]})\big\}^2 + \sum_{k=1}^K \frac{4N_{\setk}}{N} \{\bar{Y}_{\setk}(z) - \bar{Y}_{\setk [-q] z}\}^2  \\
    & + \sum_{k=1}^K \frac{4N_{\setk}}{N} \big\{(\bar{\bs{x}}_{\setk[-q]z}-\bar{\bs{x}}_{\setk})^\top \hat{\bs{\beta}}_{[-q]}\big\}^2. 
\end{align*}
Let 
\begin{align*}
    & \bs{\Sigma}_{\bs{x}Y} = \frac{1}{N}\sum_{z\in \{0,1\}}\sum_{k=1}^K\sum_{i:A_i=k} \omega_{\setk z} \bs{x}_i \{Y_i(z) - \bar{Y}_{\setk}(z)\}, \\
    & \hat{\bs{\Sigma}}_{[-q],\bs{x}\bs{x}} = \frac{1}{N}\sum_{z\in \{0,1\}}\sum_{k=1}^K\sum_{i:A_i=k,Z_i=z,i\in \ms_{[-q]}} \omega_{\setk [-q] z} (\bs{x}_i- \bar{\bs{x}}_{\setk[-q]z})(\bs{x}_i- \bar{\bs{x}}_{\setk[-q]z})^\top, \\
    &\hat{\bs{\Sigma}}_{[-q],\bs{x}Y} = \frac{1}{N}\sum_{z\in \{0,1\}}\sum_{k=1}^K\sum_{i:A_i=k,Z_i=z,i\in \ms_{[-q]}} \omega_{\setk [-q] z}  (\bs{x}_i- \bar{\bs{x}}_{\setk[-q]z})(Y_i - \bar{Y}_{\setk[-q]z})^\top .
\end{align*}
By FWL theorem, we have $\bs{\beta}^\ast = \bs{\Sigma}_{\bs{x}\bs{x}}^{-1} \bs{\Sigma}_{\bs{x}Y}$ and $\hat{\bs{\beta}}_{[-q]} = \hat{\bs{\Sigma}}_{[-q],\bs{x}\bs{x}}^{-1}\hat{\bs{\Sigma}}_{[-q],\bs{x}Y}.$
%\[
%\bs{\beta}^\ast = \bs{\Sigma}_{\bs{x}\bs{x}}^{-1} \bs{\Sigma}_{\bs{x}Y},\quad \hat{\bs{\beta}}_{[-q]} = \hat{\bs{\Sigma}}_{[-q],\bs{x}\bs{x}}^{-1}\hat{\bs{\Sigma}}_{[-q],\bs{x}Y}.
%\]
Similar to the proof of \Cref{prop:stability-assumption-linear-SR}, we have $\bs{\beta}^\ast-\hat{\bs{\beta}}_{[-q]} = \op(1)$ and $\hat{\bs{\beta}}_{[-q]} = \Op(1)$.
%\[
%\bs{\beta}^\ast-\hat{\bs{\beta}}_{[-q]} = \op(1),\quad \hat{\bs{\beta}}_{[-q]} = \Op(1).
%\]
It follows that $\|\hat{f}_{[-q]z}-f^\ast_z\|_N^2 = \op(1).$

\textbf{(ii)} By definition, we have
  \begin{align*}
     \bs{\beta}^\ast = \argmin_{\bs{\beta}} \frac{1}{N}\sum_{z\in \{0,1\}} \sum_{k=1}^K \sum_{i: A_i = k} \big\{Y_i(z)-\bar{Y}_{\setk}(z)-(\bs{x}_i - \bar{\bs{x}}_{\setk})^\top \bs{\beta}\big\}^2 \omega_{\setk z}.
\end{align*}
Therefore, 
\begin{align*}
    \tilde{\sigma}^2_{\SR} = &\frac{1}{N}\sum_{z\in \{0,1\}} \sum_{k=1}^K \sum_{i: A_i = k} \big\{Y_i(z)-\bar{Y}_{\setk}(z)-(\bs{x}_i-\bar{\bs{x}}_{\setk})^\top \bs{\beta}^\ast\big\}^2 \omega_{\setk z}\\
    \leq & \frac{1}{N}\sum_{z\in \{0,1\}} \sum_{k=1}^K \sum_{i: A_i = k} \{Y_i(z)-\bar{Y}_{\setk}(z)\}^2 \omega_{\setk z} = \tilde{\sigma}^2_{\textnormal{SR-unadj}}.
\end{align*}

Moreover, note that
\[
\Delta_{\SR} = \sum_{k=1}^K \frac{N_{\setk}}{N(N_{\setk}-1)}\sum_{i:A_i=k}(\tau_{\varepsilon,i} - \bar{\tau}_{\varepsilon,\setk})^2 =  \sum_{k=1}^K \frac{N_{\setk}}{N(N_{\setk}-1)}\sum_{i:A_i=k}(\tau_{i} - \bar{\tau}_{\setk})^2.
\]
Therefore,
$
\Delta_{\SR} = \tilde{\sigma}^2_{\textnormal{SR-unadj}} - \sigma^2_{\textnormal{SR-unadj}} = \tilde{\sigma}^2_{\SR} - \sigma^2_{\SR},
$
which yields
$
\sigma^2_{\textnormal{SR-unadj}} \geq \sigma^2_{\SR}.
$

%Hence, we have proven \Cref{prop:stability-assumption-linear-SR-tom} (ii).
\end{proof}

\subsection{Proof of \Cref{prop:stability-assumption-linear-MP}}
\begin{proof}
Note that 
\begin{align*}
    &\|\hat{f}_{[-q]z}-f^\ast_z\|_N^2 = \frac{1}{N} \sumi \big\{\bs{x}_i^\top (\bs{\beta}_z^\ast-\hat{\bs{\beta}}_{[-q]z}) + \alpha^\ast_z-\hat{\alpha}_{[-q]z}\big\}^2 \\
    \leq & \frac{2}{N} \sumi \big\{(\bs{x}_i-\bar{\bs{x}})^\top (\bs{\beta}_z^\ast-\hat{\bs{\beta}}_{[-q]z})\big\}^2 + 2\big\{\bar{\bs{x}}^\top (\bs{\beta}_z^\ast-\hat{\bs{\beta}}_{[-q]z}) +\alpha_{z}^\ast -\hat{\alpha}_{[-q] z}\big\}^2,
\end{align*}
and
\begin{align*}
    &\alpha_{z}^\ast = \bar{Y}(z) - \bar{\bs{x}}^\top \bs{\beta}^\ast_z,\quad \hat{\alpha}_{[-q] z} = \bar{Y}_{[-q] z} - \bar{\bs{x}}_{[-q] z}^\top \hat{\bs{\beta}}_{[-q]z}.
\end{align*}
Therefore,
\begin{align*}
    & \|\hat{f}_{[-q]z}-f^\ast_z\|_N^2 \\
    \leq & \frac{2}{N} \sumi \big\{(\bs{x}_i-\bar{\bs{x}})^\top (\bs{\beta}_z^\ast-\hat{\bs{\beta}}_{[-q]z})\big\}^2 + 2\big\{\bar{\bs{x}}^\top (\bs{\beta}_z^\ast-\hat{\bs{\beta}}_{[-q]z}) +\alpha_{z}^\ast -\hat{\alpha}_{[-q] z}\big\}^2\\
    \leq & \frac{2}{N} \sumi \big\{(\bs{x}_i-\bar{\bs{x}})^\top (\bs{\beta}_z^\ast-\hat{\bs{\beta}}_{[-q]z})\big\}^2 + 4\Big\{\hat{\bs{\beta}}_{[-q]z}(\bar{\bs{x}}- \bar{\bs{x}}_{[-q]z})^\top \Big\}^2  + 4\{\bar{Y}(z)-\bar{Y}_{[-q] z}\}^2.
\end{align*}

Note that $\bs{\beta}^\ast_z = (\bs{\Sigma}_{\bs{x}\bs{x}}^{\MP})^{-1}\bs{\Sigma}_{\bs{x}Y(z)}^{\MP}$ and $\hat{\bs{\beta}}_z = (\hat{\bs{\Sigma}}_{\bs{x}\bs{x}}^{\MP})^{-1}\hat{\bs{\Sigma}}_{\bs{x}Y(z)}^{\MP}$, where
\begin{align*}
    & \bs{\Sigma}_{\bs{x}Y(z)}^{\MP} = \frac{1}{N}\sumi (\bs{x}_i-\bar{\bs{x}})\{Y_i(z) - \bar{Y}(z)\},\\
    & \hat{\bs{\Sigma}}_{[-q]z,\bs{x}\bs{x}}^{\MP} = \frac{1}{N_{[-q]}/2}\sum_{i:Z_i=z,i\in \ms_{[-q]}} (\bs{x}_i-\bar{\bs{x}}_{[-q]z})(\bs{x}_i-\bar{\bs{x}}_{[-q]z})^\top, \\
    & \hat{\bs{\Sigma}}_{[-q]z,\bs{x}Y(z)}^{\MP} = \frac{1}{N_{[-q]}/2}\sum_{i:Z_i=z,i\in \ms_{[-q]}} (\bs{x}_i-\bar{\bs{x}}_{[-q]z})\{Y_i(z) - \bar{Y}_{[-q]z}(z)\}.
\end{align*}

Applying \Cref{lem:variance-of-matched-pair-and-split-by-strutum} and noting that $N_{ [q]}/N \in (c,1-c)$ and $\max\{N^{-1}\sumi \|\bs{x}_i\|^4_\infty$, $N^{-1}\sumi Y_i^4(1)$, $N^{-1}\sumi Y_i^4(0)\} = O(1)$, we have $\bar{\bs{x}}- \bar{\bs{x}}_{[-q]z} = \Op(N^{-1/2})$, $\hat{\bs{\Sigma}}_{[-q]z,\bs{x}\bs{x}}^{\MP} = \bs{\Sigma}_{\bs{x}\bs{x}}^{\MP}  + \Op(N^{-1/2}),$
 $\bar{Y}(z)-\bar{Y}_{[-q] z} = \Op(N^{-1/2})$, and $\hat{\bs{\Sigma}}_{[-q]z,\bs{x}Y(z)}^{\MP} 
 = \bs{\Sigma}_{\bs{x}Y(z)}^{\MP}  + \Op(N^{-1/2})$. Since $\bs{\Sigma}_{\bs{x}\bs{x}}^{\MP}$ has an invertible limit and $\bs{\beta}^\ast_z$, $z=0,1$, have finite limits, it follows that $\bs{\beta}_z^\ast-\hat{\bs{\beta}}_{[-q]z} = \op(1)$ and $\hat{\bs{\beta}}_{[-q]z} = \Op(1)$. Therefore,
$
\|\hat{f}_{[-q]z}-f^\ast_z\|_N^2 = \op(1).
$
\end{proof}

\subsection{Proof of \Cref{mp:regression-of-outcome-difference-on-covariate-difference}}
\begin{proof}
\noindent \textbf{(i)} Note that
\begin{align*}
    &\|\hat{f}_{[-q]z}-f^\ast_z\|_N^2 = \frac{1}{N} \sumi \big\{\bs{x}_i^\top (\bs{\beta}^\ast-\hat{\bs{\beta}}_{[-q]})\big\}^2, 
\end{align*}
\begin{align*}
    \bs{\beta}^\ast = (\bs{\Sigma}_{\bs{x}\bs{x}}^{\MP})^{-1}\bs{\Sigma}_{\bs{x}Y}^{\MP},\quad \hat{\bs{\beta}}_{[-q]} = (\hat{\Sigma}_{[-q],\bs{x}\bs{x}}^{\MP})^{-1}\hat{\Sigma}_{[-q],\bs{x}Y}^{\MP},
\end{align*}
where
\begin{align*}
  &\bs{\Sigma}_{\bs{x}\bs{Y}}^{\MP} = \frac{1}{N}\sum_{k=1}^{N/2} \Big[(\bs{x}_{2k}-\bs{x}_{2k-1})\{Y_{2k}(1)-Y_{2k-1}(0)\} + (\bs{x}_{2k-1}-\bs{x}_{2k})\{Y_{2k-1}(1)-Y_{2k}(0)\}\Big],\\
    & \hat{\Sigma}_{[-q],\bs{x}\bs{x}}^{\MP} = \frac{1}{N_{[-q]}/2}\sum_{k\in \mk_{[-q]}} (\bs{x}_{2k}-\bs{x}_{2k-1})(\bs{x}_{2k}-\bs{x}_{2k-1})^\top,\\    &\hat{\Sigma}_{[-q],\bs{x}\bs{Y}}^{\MP} = \frac{1}{N/2}\sum_{k\in \mk_{[-q]}} \Big\{E_{k} (\bs{x}_{2k}-\bs{x}_{2k-1})(Y_{2k}-Y_{2k-1}) + (1-E_{k}) (\bs{x}_{2k-1}-\bs{x}_{2k})\\
    & \qquad \qquad \qquad (Y_{2k-1}-Y_{2k})\Big\}.
\end{align*}

By \Cref{lem:variance-of-matched-pair-and-split-by-strutum}, we have
\[
\hat{\Sigma}_{[-q],\bs{x}\bs{x}}^{\MP} = \bs{\Sigma}_{\bs{x}\bs{x}}^{\MP}  + \op(1),\quad \hat{\Sigma}_{[-q],\bs{x}\bs{Y}}^{\MP} = \bs{\Sigma}_{\bs{x}\bs{Y}}^{\MP} + \op(1).
\]
Therefore, $   \hat{\bs{\beta}}_{[-q]}- \bs{\beta}^\ast = \op(1).$ As a consequence, we have $\|\hat{f}_{[-q]z}-f^\ast_z\|_N^2  = \op(1)$.

\textbf{(ii)} We have shown in the proof of \Cref{prop:valid-inference-matched-pair} that
\begin{align*}
& \tilde{\sigma}_{\MP}^2 = \frac{1}{N}\sum_{k=1}^{N/2} \{\varepsilon_{2k-1}(1)+ \varepsilon_{2k-1}(0)- \varepsilon_{2k}(1)- \varepsilon_{2k}(0)\}^2+\frac{4}{N}\sum_{k=1}^{N/2} (\bar{\tau}_{\varepsilon,\setk}-\bar{\tau}_{\varepsilon})^2,\\
& \sigma_{\MP}^2 = \frac{1}{N}\sum_{k=1}^{N/2} \{\varepsilon_{2k-1}(1)+\varepsilon_{2k-1}(0)- \varepsilon_{2k}(1)- \varepsilon_{2k}(0)\}^2,\\
& \tilde{\sigma}_{\textnormal{MP-unadj}}^2 = \frac{1}{N}\sum_{k=1}^{N/2} \{Y_{2k-1}(1)+ Y_{2k-1}(0)- Y_{2k}(1)- Y_{2k}(0)\}^2+\frac{4}{N}\sum_{k=1}^{N/2} (\bar{\tau}_{\setk}-\bar{\tau})^2,\\
& \sigma_{\textnormal{MP-unadj}}^2 = \frac{1}{N}\sum_{k=1}^{N/2} \{Y_{2k-1}(1)+Y_{2k-1}(0)- Y_{2k}(1)- Y_{2k}(0)\}^2.
\end{align*}

By the definition of $\bs{\beta}^\ast$, we have
\begin{align*}
    \bs{\beta}^\ast = \argmin_{\bs{\beta}} \frac{1}{N}\sum_{k=1}^{N/2} \{Y_{2k-1}(1)-Y_{2k}(0) - Y_{2k}(1) + Y_{2k-1}(0) - 2(\bs{x}_{2k-1}-\bs{x}_{2k})^\top \bs{\beta}\}^2.
\end{align*}
As a consequence, we have
\begin{align*}
   \sigma_{\textnormal{MP-unadj}}^2 = &\frac{1}{N}\sum_{k=1}^{N/2} \{\varepsilon_{2k-1}(1)+\varepsilon_{2k-1}(0)- \varepsilon_{2k}(1)- \varepsilon_{2k}(0)\}^2 \\
    =& \frac{1}{N}\sum_{k=1}^{N/2} \{Y_{2k-1}(1)+Y_{2k-1}(0)- Y_{2k}(1)- Y_{2k}(0) - 2(\bs{x}_{2k-1}-\bs{x}_{2k})^\top \bs{\beta}^\ast\}^2 \\
    \leq& \frac{1}{N}\sum_{k=1}^{N/2} \{Y_{2k-1}(1)+Y_{2k-1}(0)- Y_{2k}(1)- Y_{2k}(0)\}^2 = \sigma_{\MP}^2.
\end{align*}

By the definition of $f^\ast_z$, we have  $\tau_{\varepsilon, i} = \tau_{i}$. Therefore,
\[
\tilde{\sigma}^2_{\MP} - \sigma^2_{\MP} = \frac{4}{N}\sum_{k=1}^{N/2} (\bar{\tau}_{\varepsilon,\setk}-\bar{\tau}_{\varepsilon})^2 = \frac{4}{N}\sum_{k=1}^{N/2} (\bar{\tau}_{\setk}-\bar{\tau})^2 = \tilde{\sigma}^2_{\textnormal{MP-unadj}} - \sigma^2_{\textnormal{MP-unadj}}.
\]
In light of the above, $\sigma_{\MP}^2 \leq \sigma_{\textnormal{MPE-unadj}}^2$ implies that $\tilde{\sigma}_{\MP}^2 \leq \tilde{\sigma}_{\textnormal{MP-unadj}}^2.$
%\[
%\tilde{\sigma}_{\MP}^2 \leq \tilde{\sigma}_{\textnormal{MP-unadj}}^2.
%\]
This completes the proof.
\end{proof}

\end{document}